\documentclass[submission, Phys]{SciPost}
\PassOptionsToPackage{unicode}{hyperref}
\usepackage{amsthm}
\theoremstyle{definition}
\newtheorem{thm}{Theorem}[section]
\newtheorem{lemma}[thm]{Lemma}
\newtheorem{rmk}[thm]{Remark}
\newtheorem{ex}[thm]{Example}
\newcommand{\Cyc}{\mathrm{HC}}

\newcommand{\PV}{\mathrm{PV}}
\newcommand{\Fuk}{\mathrm{Fuk}}
\newcommand{\Coh}{\mathrm{Coh}}
\newcommand{\Ext}{\mathrm{Ext}}

\newtheorem{conjecture}{Conjecture}

\let\U\relax
\let\C\relax

\usepackage{upgreek}

\usepackage{tikz}
\definecolor{PLUS}{HTML}{0F2080}
\definecolor{MINUS}{HTML}{F5793A}
\definecolor{ZERO}{HTML}{382119}

\usepackage{relsize}

\usetikzlibrary{calc, positioning, arrows.meta,
  decorations.pathreplacing} \tikzset{
  baseline={([yshift=-0.75ex]current bounding box.center)},
  zerosep/.style={inner sep=0pt, outer sep=0pt, minimum size=0pt},
  node distance=8pt, align at top/.style={baseline=(current bounding
    box.north)}, align at bottom/.style={baseline=(current bounding
    box.south)}, }

\newlength{\qsep}
\usetikzlibrary{decorations.markings, arrows, intersections}
\tikzset{
  x=\qsep, y=\qsep,  font=\smaller,
  ->-/.style={decoration={
      markings, mark=at position #1 with
      {\arrow{>}}},postaction={decorate}},
  -<-/.style={decoration={
      markings, mark=at position #1 with
      {\arrow{<}}},postaction={decorate}},
  node/.style={draw, fill=white, shape=circle, minimum size=12pt, inner
    sep=0pt},
  gnode/.style={node},
  dgnode/.style={node, densely dashed},
  ggnode/.style={node, double},
  fnode/.style={node, shape=rectangle},
  tnode/.style={fnode, double, minimum size=12pt},
  q-/.style={-},
  q->/.style={->,  >=stealth, shorten >=1pt, font=\smaller[2]},
  q<-/.style={q->, <-, shorten >=0pt, shorten <=1pt},
  eq-/.style={double, double distance=2pt},
}

\usetikzlibrary{shapes.misc}

\tikzset{cross/.style={cross out, draw=black, minimum size=2*(#1-\pgflinewidth), inner sep=0pt, outer sep=0pt},
cross/.default={1pt}}

\usepackage{upgreek}

\newcommand{\gf}{\mathfrak{g}}

\newcommand{\borel}{\mathfrak{b}}
\newcommand{\nf}{\mathfrak{n}}
\newcommand{\hf}{\mathfrak{h}}
\newcommand{\pf}{\mathfrak{p}}

\newcommand{\del}{\partial}
\newcommand{\delb}{{\bar\partial}}

\newcommand{\id}{\mathop{\mathrm{id}}\nolimits}

\newcommand{\ket}[1]{|#1\rangle}

\newcommand{\vev}[1]{\langle #1 \rangle}

\newcommand{\diag}{\mathop{\mathrm{diag}}\nolimits}
\newcommand{\Hom}{\mathop{\mathrm{Hom}}\nolimits}

\newcommand{\ad}{\mathop{\mathrm{ad}}\nolimits}

\renewcommand{\Im}{\mathop{\mathrm{Im}}\nolimits}
\renewcommand{\Re}{\mathop{\mathrm{Re}}\nolimits}

\newcommand{\tr}{\mathop{\mathrm{tr}}\nolimits}
\newcommand{\End}{\mathop{\mathrm{End}}\nolimits}

\newcommand{\SU}{\mathrm{SU}}

\newcommand{\Spin}{\mathrm{Spin}}

\newcommand{\GL}{\mathrm{GL}}
\newcommand{\glf}{\mathfrak{gl}}

\newcommand{\U}{\mathrm{U}}

\newcommand{\iso}{\cong}

\newcommand{\Z}{\mathbb{Z}}

\newcommand{\R}{\mathbb{R}}
\newcommand{\C}{\mathbb{C}}

\let\nc\newcommand
\let\renc\renewcommand
\nc{\wbar}{\overline}
\let\td\tilde
\let\wtd\widetilde
\let\wht\widehat
\let\mcl\mathcal

\nc{\ab}{{\bar{a}}} \nc{\at}{\tilde{a}} \nc{\ah}{\hat{a}}
\nc{\bb}{{\bar{b}}} \nc{\bt}{\tilde{b}} \nc{\bh}{\hat{b}}
\nc{\cb}{{\bar{c}}} \nc{\ct}{\tilde{c}} %\nc{\ch}{\hat{c}}
\nc{\db}{{\bar{d}}} \nc{\dt}{\tilde{d}} \renc{\dh}{\hat{d}}
\nc{\eb}{{\bar{e}}} \nc{\et}{\tilde{e}} \nc{\eh}{\hat{e}}
\nc{\fb}{{\bar{f}}} \nc{\ft}{\tilde{f}} \nc{\fh}{\hat{f}}
\nc{\gb}{{\bar{g}}} \nc{\gt}{\tilde{g}} \nc{\gh}{\hat{g}}
\nc{\hb}{{\bar{h}}} \nc{\hh}{\hat{h}} %\nc{\ht}{\tilde{h}}
\nc{\ib}{{\bar{\imath}}} \nc{\ih}{\hat{\imath}} %\nc{\it}{\tilde{\imath}}
\nc{\jb}{{\bar{\jmath}}} \nc{\jt}{\tilde{\jmath}} \nc{\jh}{\hat{\jmath}}
\nc{\kb}{{\bar{k}}} \nc{\kt}{\tilde{k}} \nc{\kh}{\hat{k}}
\nc{\lb}{{\bar{l}}} \nc{\lt}{\tilde{l}} \nc{\lh}{\hat{l}}
\nc{\mb}{{\bar{m}}} \nc{\mt}{\tilde{m}} \nc{\mh}{\hat{m}}
\nc{\nb}{{\bar{n}}} \nc{\nt}{\tilde{n}} \nc{\nh}{\hat{n}}
\nc{\ob}{{\bar{o}}} \nc{\ot}{\tilde{o}} \nc{\oh}{\hat{o}}
\nc{\pb}{{\bar{p}}} \nc{\pt}{\tilde{p}} \nc{\ph}{\hat{p}}
\nc{\qb}{{\bar{q}}} \nc{\qt}{\tilde{q}} \nc{\qh}{\hat{q}}
\nc{\rb}{{\bar{r}}} \nc{\rt}{\tilde{r}} \nc{\rh}{{\hat{r}}}
\renc{\sb}{{\bar{s}}} \nc{\st}{\tilde{s}} \nc{\sh}{\hat{s}}
\nc{\tb}{{\bar{t}}} \renc{\th}{\hat{t}} %\nc{\tt}{\tilde{t}}
\nc{\ub}{{\bar{u}}} \nc{\ut}{\tilde{u}} \nc{\uh}{\hat{u}}
\nc{\vb}{{\bar{v}}} \nc{\vt}{\tilde{v}} \nc{\vh}{\hat{v}}
\nc{\wb}{{\bar{w}}} \nc{\wt}{\tilde{w}} \nc{\wh}{\hat{w}}
\nc{\xb}{{\bar{x}}} \nc{\xt}{\tilde{x}} \nc{\xh}{\hat{x}}
\nc{\yb}{{\bar{y}}} \nc{\yt}{\tilde{y}} \nc{\yh}{\hat{y}}
\nc{\zb}{{\bar{z}}} \nc{\zt}{\tilde{z}} \nc{\zh}{\hat{z}}

\nc{\Ab}{{\wbar{A}}} \nc{\At}{{\wtd{A}}} \nc{\Ah}{{\wht{A}}}
\nc{\Bb}{{\wbar{B}}} \nc{\Bt}{{\wtd{B}}} \nc{\Bh}{{\wht{B}}}
\nc{\Cb}{{\wbar{C}}} \nc{\Ct}{{\wtd{C}}} \nc{\Ch}{{\wht{C}}}
\nc{\Db}{{\wbar{D}}} \nc{\Dt}{{\wtd{D}}} \nc{\Dh}{{\wht{D}}}
\nc{\Eb}{{\wbar{E}}} \nc{\Et}{{\wtd{E}}} \nc{\Eh}{{\wht{E}}}
\nc{\Fb}{{\wbar{F}}} \nc{\Ft}{{\wtd{F}}} \nc{\Fh}{{\wht{F}}}
\nc{\Gb}{{\wbar{G}}} \nc{\Gt}{{\wtd{G}}} \nc{\Gh}{{\wht{G}}}
\nc{\Hb}{{\wbar{H}}} \nc{\Ht}{{\wtd{H}}} \nc{\Hh}{{\wht{H}}}
\nc{\Ib}{{\bar{I}}} \nc{\It}{{\wtd{I}}} \nc{\Ih}{{\wht{I}}}
\nc{\Jb}{{\bar{J}}} \nc{\Jt}{{\wtd{J}}} \nc{\Jh}{{\wht{J}}}
\nc{\Kb}{{\wbar{K}}} \nc{\Kt}{{\wtd{K}}} \nc{\Kh}{{\wht{K}}}
\nc{\Lb}{{\wbar{L}}} \nc{\Lt}{{\wtd{L}}} \nc{\Lh}{{\wht{L}}}
\nc{\Mb}{{\wbar{M}}} \nc{\Mt}{{\wtd{M}}} \nc{\Mh}{{\wht{M}}}
\nc{\Nb}{{\wbar{N}}} \nc{\Nt}{{\wtd{N}}} \nc{\Nh}{{\wht{N}}}
\nc{\Ob}{{\wbar{O}}} \nc{\Ot}{{\wtd{O}}} \nc{\Oh}{{\wht{O}}}
\nc{\Pb}{{\wbar{P}}} \nc{\Pt}{{\wtd{P}}} \nc{\Ph}{{\wht{P}}}
\nc{\Qb}{{\wbar{Q}}} \nc{\Qt}{{\wtd{Q}}} \nc{\Qh}{{\wht{Q}}}
\nc{\Rb}{{\wbar{R}}} \nc{\Rt}{{\wtd{R}}} \nc{\Rh}{{\wht{R}}}
\nc{\Sb}{{\wbar{S}}} \nc{\St}{{\wtd{S}}} \nc{\Sh}{{\wht{S}}}
\nc{\Tb}{{\wbar{T}}} \nc{\Tt}{{\wtd{T}}} \nc{\Th}{{\wht{T}}}
\nc{\Ub}{{\wbar{U}}} \nc{\Ut}{{\wtd{U}}} \nc{\Uh}{{\wht{U}}}
\nc{\Vb}{{\wbar{V}}} \nc{\Vt}{{\wtd{V}}} \nc{\Vh}{{\wht{V}}}
\nc{\Wb}{{\wbar{W}}} \nc{\Wt}{{\wtd{W}}} \nc{\Wh}{{\wht{W}}}
\nc{\Xb}{{\wbar{X}}} \nc{\Xt}{{\wtd{X}}} \nc{\Xh}{{\wht{X}}}
\nc{\Yb}{{\wbar{Y}}} \nc{\Yt}{{\wtd{Y}}} \nc{\Yh}{{\wht{Y}}}
\nc{\Zb}{{\wbar{Z}}} \nc{\Zt}{{\wtd{Z}}} \nc{\Zh}{{\wht{Z}}}

\nc{\CA}{{\mcl{A}}} \nc{\CAb}{{\wbar{\CA}}} \nc{\CAt}{{\wtd{\CA}}} \nc{\CAh}{{\wht{\CA}}}
\nc{\CB}{{\mcl{B}}} \nc{\CBb}{{\wbar{\CB}}} \nc{\CBt}{{\wtd{\CB}}} \nc{\CBh}{{\wht{\CB}}}
\nc{\CC}{{\mcl{C}}} \nc{\CCb}{{\wbar{\CC}}} \nc{\CCt}{{\wtd{\CC}}} \nc{\CCh}{{\wht{\CC}}}
\nc{\cD}{{\mcl{D}}} \nc{\cDb}{{\wbar{\cD}}} \nc{\cDt}{{\wtd{\cC}}} \nc{\cDh}{{\wht{\cD}}}
\nc{\CE}{{\mcl{E}}} \nc{\CEb}{{\wbar{\CE}}} \nc{\CEt}{{\wtd{\CE}}} \nc{\CEh}{{\wht{\CE}}}
\nc{\CF}{{\mcl{F}}} \nc{\CFb}{{\wbar{\CF}}} \nc{\CFt}{{\wtd{\CF}}} \nc{\CFh}{{\wht{\CF}}}
\nc{\CG}{{\mcl{G}}} \nc{\CGb}{{\wbar{\CG}}} \nc{\CGt}{{\wtd{\CG}}} \nc{\CGh}{{\wht{\CG}}}
\nc{\CH}{{\mcl{H}}} \nc{\CHb}{{\wbar{\CH}}} \nc{\CHt}{{\wtd{\CH}}} \nc{\CHh}{{\wht{\CH}}}
\nc{\CI}{{\mcl{I}}} \nc{\CIb}{{\wbar{\CI}}} \nc{\CIt}{{\wtd{\CI}}} \nc{\CIh}{{\wht{\CI}}}
\nc{\CJ}{{\mcl{J}}} \nc{\CJb}{{\wbar{\CJ}}} \nc{\CJt}{{\wtd{\CJ}}} \nc{\CJh}{{\wht{\CJ}}}
\nc{\CK}{{\mcl{K}}} \nc{\CKb}{{\wbar{\CK}}} \nc{\CKt}{{\wtd{\CK}}} \nc{\CKh}{{\wht{\CK}}}
\nc{\CL}{{\mcl{L}}} \nc{\CLb}{{\wbar{\CL}}} \nc{\CLt}{{\wtd{\CL}}} \nc{\CLh}{{\wht{\CL}}}
\nc{\CM}{{\mcl{M}}} \nc{\CMb}{{\wbar{\CM}}} \nc{\CMt}{{\wtd{\CM}}} \nc{\CMh}{{\wht{\CM}}}
\nc{\CN}{{\mcl{N}}} \nc{\CNb}{{\wbar{\CN}}} \nc{\CNt}{{\wtd{\CN}}} \nc{\CNh}{{\wht{\CN}}}
\nc{\CO}{{\mcl{O}}} \nc{\COb}{{\wbar{\CO}}} \nc{\COt}{{\wtd{\CO}}} \nc{\COh}{{\wht{\CO}}}
\nc{\CP}{{\mcl{P}}} \nc{\CPb}{{\wbar{\CP}}} \nc{\CPt}{{\wtd{\CP}}} \nc{\CPh}{{\wht{\CP}}}
\nc{\CQ}{{\mcl{Q}}} \nc{\CQb}{{\wbar{\CQ}}} \nc{\CQt}{{\wtd{\CQ}}} \nc{\CQh}{{\wht{\CQ}}}
\nc{\CR}{{\mcl{R}}} \nc{\CRb}{{\wbar{\CR}}} \nc{\CRt}{{\wtd{\CR}}} \nc{\CRh}{{\wht{\CR}}}
\nc{\CS}{{\mcl{S}}} \nc{\CSb}{{\wbar{\CS}}} \nc{\CSt}{{\wtd{\CS}}} \nc{\CSh}{{\wht{\CS}}}
\nc{\CT}{{\mcl{T}}} \nc{\CTb}{{\wbar{\CT}}} \nc{\CTt}{{\wtd{\CT}}} \nc{\CTh}{{\wht{\CT}}}
\nc{\CU}{{\mcl{U}}} \nc{\CUb}{{\wbar{\CU}}} \nc{\CUt}{{\wtd{\CU}}} \nc{\CUh}{{\wht{\CU}}}
\nc{\CV}{{\mcl{V}}} \nc{\CVb}{{\wbar{\CV}}} \nc{\CVt}{{\wtd{\CV}}} \nc{\CVh}{{\wht{\CV}}}
\nc{\CW}{{\mcl{W}}} \nc{\CWb}{{\wbar{\CW}}} \nc{\CWt}{{\wtd{\CW}}} \nc{\CWh}{{\wht{\CW}}}
\nc{\CX}{{\mcl{X}}} \nc{\CXb}{{\wbar{\CX}}} \nc{\CXt}{{\wtd{\CX}}} \nc{\CXh}{{\wht{\CX}}}
\nc{\CY}{{\mcl{Y}}} \nc{\CYb}{{\wbar{\CY}}} \nc{\CYt}{{\wtd{\CY}}} \nc{\CYh}{{\wht{\CY}}}
\nc{\CZ}{{\mcl{Z}}} \nc{\CZb}{{\wbar{\CZ}}} \nc{\CZt}{{\wtd{\CZ}}} \nc{\CZh}{{\wht{\CZ}}}

\let\eps\epsilon
\let\ups\upsilon
\let\veps\varepsilon
\let\vtht\vartheta

\let\vsgm\varsigma
\let\vphi\varphi
\let\vrho\varrho

\nc{\alphab}{{\bar{\alpha}}} \nc{\alphat}{{\td{\alpha}}} \nc{\alphah}{{\hat{\alpha}}}
\nc{\betab}{{\bar{\beta}}}   \nc{\betat}{{\td{\beta}}}   \nc{\betah}{{\hat{\beta}}} 
\nc{\gammab}{{\bar{\gamma}}} \nc{\gammat}{{\td{\gamma}}} \nc{\gammah}{{\hat{\gamma}}} 
\nc{\deltab}{{\bar{\delta}}} \nc{\deltat}{{\td{\delta}}} \nc{\deltah}{{\hat{\delta}}} 
\nc{\epsilonb}{{\bar{\eps}}} \nc{\epsilont}{{\td{\eps}}} \nc{\epsilonh}{{\hat{\eps}}} 
\nc{\vepsb}{{\bar{\veps}}}   \nc{\vepst}{{\td{\veps}}}   \nc{\vepsh}{{\hat{\veps}}} 
\nc{\zetab}{{\bar{\zeta}}}   \nc{\zetat}{{\td{\zeta}}}   \nc{\zetah}{{\hat{\zeta}}} 
\nc{\etab}{{\bar{\eta}}}     \nc{\etat}{{\td{\eta}}}     \nc{\etah}{{\hat{\eta}}} 
\nc{\thetab}{{\bar{\theta}}} \nc{\thetat}{{\td{\theta}}} \nc{\thetah}{{\hat{\theta}}} 
\nc{\vthetab}{{\bar{\vtht}}} \nc{\vthetat}{{\td{\vtht}}} \nc{\vthetah}{{\hat{\vtht}}} 
\nc{\lambdab}{{\bar{\lambda}}} \nc{\lambdat}{{\td{\lambda}}} \nc{\lambdah}{{\hat{\lambda}}} 
\nc{\iotab}{{\bar{\iota}}}   \nc{\iotat}{{\td{\iota}}}   \nc{\iotah}{{\hat{\iota}}} 
\nc{\kappab}{{\bar{\kappa}}} \nc{\kappat}{{\td{\kappa}}} \nc{\kappah}{{\hat{\kappa}}} 
\nc{\lmdb}{{\bar{\lmd}}}     \nc{\lmdt}{{\td{\lmd}}}     \nc{\lmdh}{{\hat{\lmd}}} 
\nc{\mub}{{\bar{\mu}}}       \nc{\mut}{{\td{\mu}}}       \nc{\muh}{{\hat{\mu}}} 
\nc{\nub}{{\bar{\nu}}}       \nc{\nut}{{\td{\nu}}}       \nc{\nuh}{{\hat{\nu}}} 
\nc{\xib}{{\bar{\xi}}}       \nc{\xit}{{\td{\xi}}}       \nc{\xih}{{\hat{\xi}}} 
\nc{\pib}{{\bar{\pi}}}       \nc{\pit}{{\td{\pi}}}       \nc{\pih}{{\hat{\pi}}} 
\nc{\vpib}{{\bar{\vpi}}}     \nc{\vpit}{{\td{\vpi}}}     \nc{\vpih}{{\hat{\vpi}}} 
\nc{\rhob}{{\bar{\rho}}}     \nc{\rhot}{{\td{\rho}}}     \nc{\rhoh}{{\hat{\rho}}} 
\nc{\vrhob}{{\bar{\vrho}}}   \nc{\vrhot}{{\td{\vrho}}}   \nc{\vrhoh}{{\hat{\vrho}}} 
\nc{\sigmab}{{\bar{\sigma}}} \nc{\sigmat}{{\td{\sigma}}} \nc{\sigmah}{{\hat{\sigma}}} 
\nc{\vsigmab}{{\bar{\vsgm}}} \nc{\vsigmat}{{\td{\vsgm}}} \nc{\vsigmah}{{\hat{\vsgm}}} 
\nc{\taub}{{\bar{\tau}}}     \nc{\taut}{{\td{\tau}}}     \nc{\tauh}{{\hat{\tau}}} 
\nc{\upsb}{{\bar{\ups}}} \nc{\upst}{{\td{\ups}}} \nc{\upsh}{{\hat{\ups}}} 
\nc{\phib}{{\bar{\phi}}}     \nc{\phit}{{\td{\phi}}}     \nc{\phih}{{\hat{\phi}}} 
\nc{\varphib}{{\bar{\vphi}}}   \nc{\varphit}{{\td{\vphi}}}   \nc{\varphih}{{\hat{\vphi}}} 
\nc{\chib}{{\bar{\chi}}}     \nc{\chit}{{\td{\chi}}}     \nc{\chih}{{\hat{\chi}}} 
\nc{\psib}{{\bar{\psi}}}     \nc{\psit}{{\td{\psi}}}     \nc{\psih}{{\hat{\psi}}} 
\nc{\omegab}{{\bar{\omega}}} \nc{\omegat}{{\td{\omega}}} \nc{\omegah}{{\hat{\omega}}} 

\nc{\Gammab}{{\wbar{\Gamma}}}     \nc{\Gammat}{{\wtd{\Gamma}}}     \nc{\Gammah}{{\wht{\Gamma}}}
\nc{\Deltab}{{\wbar{\Delta}}}     \nc{\Deltat}{{\wtd{\Delta}}}     \nc{\Deltah}{{\wht{\Delta}}}
\nc{\Thetab}{{\wbar{\Theta}}}     \nc{\Thetat}{{\wtd{\Theta}}}     \nc{\Thetah}{{\wht{\Theta}}}
\nc{\Lambdab}{{\wbar{\Lambda}}}   \nc{\Lambdat}{{\wtd{\Lambda}}}   \nc{\Lambdah}{{\wht{\Lambda}}}
\nc{\Xib}{{\wbar{\Xi}}}           \nc{\Xit}{{\wtd{\Xi}}}           \nc{\Xih}{{\wht{\Xi}}}
\nc{\Pib}{{\wbar{\Pi}}}           \nc{\Pit}{{\wtd{\Pi}}}           \nc{\Pih}{{\wht{\Pi}}}
\nc{\Sigmab}{{\wbar{\Sigma}}}     \nc{\Sigmat}{{\wtd{\Sigma}}}     \nc{\Sigmah}{{\wht{\Sigma}}}
\nc{\Upsilonb}{{\wbar{\Upsilon}}} \nc{\Upsilont}{{\wtd{\Upsilon}}} \nc{\Upsilonh}{{\wht{\Upsilon}}}
\nc{\Phib}{{\wbar{\Phi}}} \nc{\Phit}{{\wtd{\Phi}}} \nc{\Phih}{{\wht{\Phi}}}
\nc{\Psib}{{\wbar{\Psi}}}         \nc{\Psit}{{\wtd{\Psi}}}         \nc{\Psih}{{\wht{\Psi}}}
\nc{\Omegab}{{\wbar{\Omega}}}     \nc{\Omegat}{{\wtd{\Omega}}}     \nc{\Omegah}{{\wht{\Omega}}}

\newcommand{\rmd}{\mathrm{d}}

\newcommand{\iu}{\mathrm{i}}
\newcommand{\str}{\mathop{\mathrm{str}}\nolimits}
\newcommand{\blank}{-}

\newcommand{\Vsf}{\mathsf{V}}
\newcommand{\Psf}{\mathsf{P}}
\newcommand{\Qsf}{\mathsf{Q}}
\newcommand{\Rsf}{\mathsf{R}}
\newcommand{\Ssf}{\mathsf{S}}
\newcommand{\Psft}{\wtd{\mathsf{P}}}
\newcommand{\Qsft}{\wtd{\mathsf{Q}}}
\newcommand{\Rsft}{\wtd{\mathsf{R}}}
\newcommand{\Ssft}{\wtd{\mathsf{S}}}

\newcommand{\QB}{Q_{\mathrm{B}}}

\newcommand{\strans}[1]{#1^{\mathrm{st}}}

\hypersetup{
    colorlinks,
    linkcolor={red!50!black},
    citecolor={blue!50!black},
    urlcolor={blue!80!black}
}

\usepackage[bitstream-charter]{mathdesign}
\DeclareSymbolFont{usualmathcal}{OMS}{cmsy}{m}{n}
\DeclareSymbolFontAlphabet{\mathcal}{usualmathcal}

\begin{document}
\begin{center}{\Large \textbf{
Superspin chains from superstring theory\\
}}\end{center}

% TODO: write the author list here. Use first name (+ other initials) + surname format.
% Separate subsequent authors by a comma, omit comma and use "and" for the last author.
% Mark the corresponding author with a superscript star.
\begin{center}
Nafiz Ishtiaque\textsuperscript{1$\star$},
Seyed F. Moosavian\textsuperscript{2},
Surya Raghavendran\textsuperscript{3} and
Junya Yagi\textsuperscript{4}
%Dee E. Faa\textsuperscript{1},
%Aah B. Cee\textsuperscript{2} and
%Gee K. See\textsuperscript{3$\star$}
\end{center}

\begin{center}
{\bf 1} Institute for Advanced Study, Princeton, NJ 08540 USA
\\
{\bf 2} Department of Physics, McGill University, Montr\'eal, QC H3A 2T8 Canada
\\
{\bf 3} Perimeter Institute for Theoretical Physics, Waterloo, ON N2L 2Y5 Canada
\\
{\bf 4} Yau Mathematical Sciences Center, Tsinghua University, Beijing 100084 China
\\
${}^\star$ {\small \sf nishtiaque@ias.edu}
\end{center}

\begin{center}
\today
\end{center}

% For convenience during refereeing (optional),
% you can turn on line numbers by uncommenting the next line:
%\linenumbers

\section*{Abstract}
{\bf
% TODO: write your abstract here.
We present a correspondence between two-dimensional
  $\CN = (2,2)$ supersymmetric gauge theories and rational integrable
  $\glf(m|n)$ spin chains with spin variables taking values in Verma
  modules.  To explain this correspondence, we realize the gauge
  theories as configurations of branes in string theory and map them
  by dualities to brane configurations that realize line defects in
  four-dimensional Chern--Simons theory with gauge group $\GL(m|n)$.
  The latter configurations embed the superspin chains into
  superstring theory.  We also provide a string theory derivation of a
  similar correspondence, proposed by Nekrasov, for rational
  $\glf(m|n)$ spin chains with spins valued in finite-dimensional
  representations.
}

\vspace{10pt}
\noindent\rule{\textwidth}{1pt}
\tableofcontents\thispagestyle{fancy}
\noindent\rule{\textwidth}{1pt}
\vspace{10pt}

\section{Introduction}

The Bethe/gauge correspondence, discovered by Nekrasov and
Shatashvili~\cite{Nekrasov:2009ui, Nekrasov:2009uh} in 2009, connects
two seemingly unrelated areas of physics.  The Bethe side of the
correspondence refers to one-dimensional integrable quantum spin
chains.  The gauge side is supersymmetric gauge theories.

Arguably the most prominent example of the Bethe/gauge correspondence
involves Heisenberg's XXX spin chain and its generalizations.  In this
example, the eigenvectors of commuting conserved charges of a rational
$\glf(m)$ spin chain are identified with the vacua of a family of
gauge theories whose gauge group is the product of $m-1$ unitary gauge
groups.  The gauge theories have $\CN = (4,4)$ supersymmetry broken to
$\CN = (2,2)$ subgroup by mass deformations, and their gauge and
matter contents are encoded in quiver diagrams whose underlying graphs
contain the Dynkin diagram of type $A_{m-1}$ as a subgraph.

In 2018, Nekrasov~\cite{Nekrasov:2018gne} presented a generalization
of the above correspondence where the relevant spin chains carry
superspins, namely rational $\glf(m|n)$ spin chains.  The
corresponding gauge theories are essentially $\CN = (2,2)$
supersymmetric, as opposed to having softly broken $\CN = (4,4)$
supersymmetry.  One of the main results of this paper is an
explanation of the origin of this correspondence using superstring
theory.

In fact, the goal of the present work is much more ambitious: we wish
to place the Bethe/gauge correspondence for superspin chains into a
large web of dualities that relate diverse phenomena in which the same
superspin chains arise from different supersymmetric gauge theories in
various spacetime dimensions.

Many of the phenomena that are expected to constitute this web of
dualities are yet to be uncovered, but their specializations to the
case of $\glf(m|0) = \glf(m)$ are known and have been studied in
recent years.  Besides the Bethe/gauge correspondence already
described, the structures of rational $\glf(m)$ spin chains (and their
trigonometric and elliptic generalizations) have appeared in
quantization of the Seiberg--Witten geometries of four-dimensional
$\CN = 2$ supersymmetric gauge theories \cite{Nekrasov:2009rc,
  Dorey:2011pa, Chen:2011sj}, the action of surface and line defects
on supersymmetric indices of four-dimensional supersymmetric gauge
theories~\cite{Gaiotto:2012xa, Gadde:2013ftv, Gaiotto:2015usa,
  Maruyoshi:2016caf, Yagi:2017hmj, Maruyoshi:2020cwy}, quantization of
the Coulomb branches of three-dimensional $\CN = 4$ supersymmetric
gauge theories \cite{Bullimore:2015lsa, Braverman:2016pwk}, and
correlation functions of local operators on interfaces in
four-dimensional $\CN = 4$ super Yang--Mills theory
\cite{Dedushenko:2020yzd}, to name a few.

All of these gauge theory setups have realization in string theory,
and one suspects that they are related to each other in one way or
another via string dualities.  This idea has turned out to be true.
It was argued in~\cite{Costello:2018txb} that brane constructions of
these setups (except for the last one which we expect is also related)
are all dual to brane configurations that realize line defects in a
four-dimensional analog of Chern--Simons
theory~\cite{Costello:2013zra, Costello:2013sla, Costello:2017dso}
with gauge group $\GL(m)$.  This theory only has a bosonic gauge
field, but it is secretly supersymmetric.  Indeed, it is equivalent to
a holomorphic--topological twist of six-dimensional $\CN = (1,1)$
super Yang--Mills theory with gauge group $\U(m)$ in the presence of
$\Omega$-deformation \cite{Nekrasov:2002qd, Nekrasov:2003rj,
  Nekrasov:2010ka, Yagi:2014toa, Luo:2014sva}.  The six-dimensional
theory describes the low-energy dynamics of a stack of $m$ D5-branes,
which comprise part of the brane configurations.

Four-dimensional Chern--Simons theory placed on $\R^2 \times \C$ is
topological on the plane $\R^2$ and holomorphic on the complex plane
$\C$.  Due to this holomorphic--topological property, line defects
extending along $\R^2$ automatically satisfy the Yang--Baxter
relation.  Moreover, each line defect carries one complex parameter,
its position in $\C$.  These facts imply that line defects making up a
square lattice in $\R^2$ defines a two-dimensional classical
integrable lattice model; their correlation function equals the
partition function of the lattice model.

Equivalently, with one of the lattice directions regarded as a time
direction, a lattice of line defects in four-dimensional Chern--Simons
theory defines a one-dimensional quantum integrable spin chain.  For
Wilson lines, this spin chain is a rational $\glf(m)$ spin
chain~\cite{Costello:2013zra, Costello:2013sla, Costello:2017dso}.
(If one replaces $\C$ in spacetime with $\C \setminus \{0\}$ or an
elliptic curve, then one obtains trigonometric or elliptic $\glf(m)$
spin chain, respectively.)  Thus, the fact that the gauge theory
setups mentioned above are all dual to line defects in
four-dimensional Chern--Simons theory explains the appearance of
$\glf(m)$ spin chains in these setups.

Now, in view of the Bethe/gauge correspondence for rational
$\glf(m|n)$ spin chains, one wonders how one can incorporate it into
the picture just described.  If one could generalize the string theory
realization of the Bethe/gauge correspondence for bosonic spin chains
to the superspin chain case, one would generalize, implicitly by
string dualities, \emph{all} of the gauge theory phenomena mentioned
above to their $\glf(m|n)$ versions.  This is what we aim to achieve.

In this paper, we provide brane constructions of the gauge theories
pertinent to the Bethe/gauge correspondence for rational $\glf(m|n)$
spin chains, and show that they are related by dualities to line
defects in four-dimensional Chern--Simons theory with gauge group
$\GL(m|n)$.  The latter theory is obtained from two copies of
$\Omega$-deformed six-dimensional $\CN = (1,1)$ super Yang--Mills
theory, with gauge groups $\U(m)$ and $\U(n)$, coupled by a
four-dimensional hypermultiplet in the bifundamental representation of
$\U(m) \times \U(n)$.  In turn, this gauge theory setup arises from a
stack of $m$ D5-branes intersecting a stack of $n$ D5-branes.  This
brane construction is another main result of the paper.

Actually, we present two versions of the Bethe/gauge correspondence,
one for compact spin chains and one for noncompact spin chains.  The
difference is whether spin variables are valued in finite-dimensional
or infinite-dimensional representations.

We introduce the Bethe/gauge correspondence for noncompact rational
$\glf(m|n)$ spin chains in section~\ref{sec:Bethe/gauge-noncompact}.
The brane constructions of the corresponding gauge theories, as well
as the duality relating these theories to line defects in
four-dimensional Chern--Simons theory with gauge group $\GL(m|n)$, are
discussed in section~\ref{sec:string-Bethe/gauge}.  Discussions in
this section provide a string theory explanation for the Bethe/gauge
correspondence.  The case of compact spin chains is treated in
section~\ref{sec:Bethe/gauge-compact}, where we reproduce the
correspondence proposed in~\cite{Nekrasov:2018gne}.

It should be remarked that in an inspiring paper~\cite{Orlando:2010uu}
in 2010, Orlando and Reffert found the Bethe/gauge correspondence for
the rational $\glf(1|2)$ spin chain with spins taking values in the
natural $(1|2)$-dimensional representation $\C^{1|2}$.  Furthermore,
they gave a string theory argument to explain dualities between
different families of gauge theories corresponding to different
choices of Dynkin diagrams of $\glf(1|2)$.  On the spin chain side,
these dualities are known as fermionic dualities.  In
section~\ref{sec:fermionic-dualities} we discuss the fermionic
dualities for rational $\glf(m|n)$ spin chains from a similar point of
view.

As we mentioned above, four-dimensional Chern--Simons theory with
gauge supergroup can be constructed from two copies of six-dimensional
super Yang--Mills theory coupled by four-dimensional matter fields.
There is a related construction in topological string theory, which
may prove useful in future attempts to put some of the physical
arguments given in this work on a rigorous mathematical footing.  We
describe the topological string construction in
appendix~\ref{sec:topstrings}.

The present work unveils only a small part of a collection of
phenomena in which superspin chains emerge from supersymmetric gauge
theories.  It will be extremely interesting to study other, but
ultimately related, phenomena whose existence is predicted by string
dualities and other tools.  We conclude this introduction by stating
mathematical conjectures as examples of such phenomena.

In section~\ref{sec:Bethe/gauge-noncompact}, we define a family of
$\CN = (2,2)$ supersymmetric gauge theories labeled by the set of
$(m+n-1)$-tuples of nonnegative integers $\Z_{\geq0}^{m+n-1}$,
corresponding to a closed rational $\glf(m|n)$ spin chain of length
$L$ with spins valued in Verma modules.  If we turn off all mass
parameters and turn on appropriate Fayet--Iliopoulos (FI) parameters,
these theories are described in the infrared by effective sigma
models.  The target space of the sigma model with label
$\mathbf{M} = (M_1, \dotsc, M_{m+n-1})$ is a Calabi--Yau manifold
$\CM(\mathbf{M})$ with an action of $\GL(L)^{m+n} \times \GL(1)$.  The
topological A-twist of this sigma model, with mass parameters
associated with the maximal torus $T$ of $\GL(L)^{m+n} \times \GL(1)$
turned on, is equivalent to the sector of the spin chain in which
there are $M_r$ magnons of type $r$.  The highest weights of the Verma
modules are determined by the mass parameters.

\begin{conjecture}
  The direct sum of equivariant cohomology groups
  \begin{equation}
    \label{eq:HMM}
    \bigoplus_{\mathbf{M} \in \Z_{\geq0}^{m+n-1}}
    H_T\bigl(\CM(\mathbf{M})\bigr)
  \end{equation}
  is a module over $Y(\glf(m|n))$, isomorphic to the tensor product of
  $L$ evaluation modules obtained from the Verma modules.
\end{conjecture}

\begin{conjecture}
  There is a homomorphism from the Bethe algebra of the Yangian
  $Y(\glf(m|n))$ to the direct sum of equivariant quantum cohomology
  rings
  \begin{equation}
    \label{eq:QHMM}
    \bigoplus_{\mathbf{M} \in \Z_{\geq0}^{m+n-1}}
    \mathit{QH}_T\bigl(\CM(\mathbf{M})\bigr) \,.
  \end{equation}
\end{conjecture}

The first conjecture says that the Hilbert space of states of the
A-model is the same as that of the spin chain.  For $n = 0$ and
$L = 1$, the conjecture is proved in~\cite{MR2827177}.  The second
conjecture means that the algebra of local operators of the A-model
includes the algebra generated by the commuting conserved charges of
the spin chain.  For $n = 0$, this conjecture follows from a result of
\cite{Maulik:2012wi}.

Similar conjectures can be made for the target spaces of effective
sigma models corresponding to compact rational $\glf(m|n)$ spin
chains.  The brane configurations in the compact case have been
recently considered by Rimanyi and Rozansky~\cite{Rimanyi:2021hzq}
from the perspective of geometric construction of
R-matrices~\cite{Maulik:2012wi}, so we expect that the above
conjectures also hold if the target spaces are varieties defined
in~\cite{Rimanyi:2021hzq}.

\section{Bethe/gauge correspondence for noncompact superspin chains}
\label{sec:Bethe/gauge-noncompact}

The Bethe/gauge correspondence for superspin chains relates a closed
spin chain with $\GL(m|n)$ symmetry and two-dimensional gauge theories
with $\CN = (2,2)$ supersymmetry.  In this section we present a
version of the correspondence in which the spin chain consists of
spins taking values in infinite-dimensional highest-weight
representations of $\glf(m|n)$.  After reviewing some basic facts
about $\glf(m|n)$ and its Verma modules, we introduce the spin chain
and its Bethe equations.  Then, we introduce the gauge theories and
their vacuum equations, and explain in what sense the two sides are
equivalent.

\subsection{$\glf(m|n)$ and its Verma modules}
\label{eq:glmn}

To begin with, let us review the structures of $\glf(m|n)$ and its
Verma modules, with emphasis on aspects that are important for the
Bethe/gauge correspondence.

\subsubsection{Lie superalgebra $\glf(m|n)$}

Let $\C^{m|n}$ be the vector space graded by $\Z_2 = \{\bar0, \bar1\}$
whose even subspace $\C^{m|n}_{\bar0} = \C^m$ and odd subspace
$\C^{m|n}_{\bar1} = \C^n$.  Let $(b_1, \dotsc, b_m)$ and
$(f_1, \dotsc, f_n)$ be the standard basis of $\C^m$ and that of
$\C^n$, respectively.  Throughout this section and the next section
except section~\ref{sec:fermionic-dualities}, we fix an ordered basis
$(e_1, \dotsc, e_{m+n})$ of $\C^{m|n}$ that is a permutation of
$(b_1, \dotsc, b_m, f_1, \dotsc, f_n)$.  The corresponding
$\Z_2$-grading $[\blank]\colon \{1, \dotsc, m+n\} \to \Z_2$ is defined
by
\begin{equation}
  [i] =
  \begin{cases}
    \bar{0} & (e_i \in \C^m) \,;
    \\
    \bar{1} & (e_i \in \C^n) \,.
  \end{cases}
\end{equation}

The space of endomorphisms $\End(\C^{m|n})$ of $\C^{m|n}$ is also a
$\Z_2$-graded vector space, with the even subspace
\begin{equation}
  \End(\C^{m|n})_{\bar0}
  =
  \Hom(\C^m, \C^m)
  \oplus \Hom(\C^n, \C^n)
\end{equation}
and the odd subspace
\begin{equation}
  \End(\C^{m|n})_{\bar1}
  =
  \Hom(\C^m,\C^n)
  \oplus \Hom(\C^n, \C^m) \,.
\end{equation}
The elementary matrix $E_{ij}$, which has $1$ in the $(i,j)$th entry
and $0$ elsewhere, has grading $[E_{ij}] = [i] + [j]$.

The Lie superalgebra $\glf(m|n)$ is the $\Z_2$-graded vector space
$\End(\C^{m|n})$, endowed with the graded commutator
$[\blank,\blank]\colon \End(\C^{m|n}) \otimes \End(\C^{m|n})
\to \End(\C^{m|n})$: for elements $a$, $b$ with homogeneous
$\Z_2$-grading,
\begin{equation}
  [a,b] = ab - (-1)^{[a][b]} ba \,.
\end{equation}
We will distinguish the elements of $\glf(m|n)$ from those of
$\End(\C^{m|n})$ by writing them as $\CE_{ij}$ rather than $E_{ij}$.
They satisfy the commutation relations
\begin{equation}
  \label{eq:gl(m|n)-CR}
  [\CE_{ij}, \CE_{kl}]
  = \delta_{jk} \CE_{il} - (-1)^{([i]+[j])([k]+[l])} \delta_{li} \CE_{kj} \,.
\end{equation}

The Cartan subalgebra of $\glf(m|n)$ is generated by
\begin{equation}
  H_r = (-1)^{[r]} \CE_{rr} - (-1)^{[r+1]} \CE_{r+1,r+1} \,,
  \quad
  r = 1 \,, \dotsc, m+n-1 \,,
\end{equation}
and one more diagonal matrix, say $\CE_{11}$.  The elementary matrix
$\CE_{ij}$ has the root $\veps_i - \veps_j$, with
\begin{equation}
  \veps_i = \CE_{ii}^\vee
\end{equation}
being the weight of $e_i$ in the natural $(m+n)$-dimensional
representation.  The positive roots are $\veps_i - \veps_j$ with
$i < j$.  The simple roots are
\begin{equation}
  \alpha_r = \veps_r - \veps_{r+1} \,,
  \quad
  r = 1 \,, \dotsc, m+n-1 \,.
\end{equation}
The elements having the roots $\alpha_r$ and $-\alpha_r$ are
$E_r = \CE_{r,r+1}$ and $F_r = \CE_{r+1,r}$, respectively.  They
satisfy
\begin{align}
  [H_r, E_s] &= a_{rs} E_s \,,
  \\
  [H_r, F_s] &= -a_{rs} F_s \,,
  \\
  [E_r, F_s] &= \delta_{rs} (-1)^{[r]} H_r \,,
\end{align}
where
\begin{equation}
  \label{eq:a_rs}
  a_{rs}
  =
  \alpha_s(H_r)
  =
  \delta_{rs} \bigl((-1)^{[r]} + (-1)^{[r+1]}\bigr)
  - \delta_{r+1,s} (-1)^{[r+1]}
  - \delta_{r,s+1} (-1)^{[r]}
\end{equation}
is the $(r,s)$th entry of the Cartan matrix.

The structure of $\glf(m|n)$ can be encoded in a Dynkin diagram, in
which a simple root $\alpha_r$ is represented by a blank node if
$a_{rr} = \pm 2$ and a crossed node if $a_{rr} = 0$, and two nodes
$\alpha_r$, $\alpha_s$ are connected by an edge if $a_{rs} \neq 0$.
As an example, consider the case with $(m|n) = (3|2)$ and
$(e_1, e_2, e_3, e_4, e_5) = (b_1, b_2, f_1, f_2, b_3)$.  The
associated Dynkin diagram is
\begin{equation}
  \label{eq:Dynkin}
  \begin{tikzpicture}[scale=1.5]
    \draw[-] (0.5,0.5) -- (1.5,0.5);
    \draw[-] (1.5,0.5) -- (2.5,0.5);
    \draw[-] (2.5,0.5) -- (3.5,0.5);

    \node[gnode, minimum size=10pt] at (0.5,0.5) {};
    \node[gnode, minimum size=10pt] at (1.5,0.5) {};
    \node[gnode, minimum size=10pt] at (2.5,0.5) {};
    \node[gnode, minimum size=10pt] at (3.5,0.5) {};

    \draw (1.5,.5) node[cross=4pt] {};
    \draw (3.5,.5) node[cross=4pt] {};

    \node[below=4pt] at (0.5,0.5) {$\eps_1 - \eps_2$};
    \node[below=4pt] at (1.5,0.5) {$\eps_2 - \delta_1$};
    \node[below=4pt] at (2.5,0.5) {$\delta_1 - \delta_2$};
    \node[below=4pt] at (3.5,0.5) {$\delta_2 - \eps_3$};
  \end{tikzpicture}
\end{equation}
where $\eps_i$ and $\delta_i$ are the weights of $b_i$ and $f_i$,
respectively.

Let us give a different presentation of the content of the Dynkin
diagram in terms of a quiver diagram, which makes the connection to
gauge theory transparent.

First, we represent $\veps_i$ by a vertical line, of one of two colors
depending on its grading:
\begin{equation}
  \veps_i
  =
  \begin{cases}
    \
    \begin{tikzpicture}
      \draw[PLUS, very thick] (0,0) -- (0,0.5);
    \end{tikzpicture}
    & ([i] = \bar0) \,;
    \\
    \
    \begin{tikzpicture}
      \draw[MINUS, very thick] (0,0) -- (0,0.5);
    \end{tikzpicture}
    & ([i] = \bar1) \,.
\end{cases}
\end{equation}
The ordered set $(\veps_1, \dotsc, \veps_{m+n})$ is then represented
graphically as $m$ vertical lines of one color and $n$ vertical lines
of the other color, placed in the order specified by the choice of the
$\Z_2$-grading:
\begin{equation}
  \begin{tikzpicture}[scale=1]
    \draw[PLUS, very thick] (0,0) -- (0,1);
    \draw[PLUS, very thick] (1,0) -- (1,1);
    \draw[MINUS, very thick] (2,0) -- (2,1);
    \draw[MINUS, very thick] (3,0) -- (3,1);
    \draw[PLUS, very thick] (4,0) -- (4,1);
  \end{tikzpicture}
\end{equation}

Next, we put a circle node between each pair of adjacent vertical
lines:
\begin{equation}
  \begin{tikzpicture}[scale=1]
    \draw[PLUS, very thick] (0,0) -- (0,1);
    \draw[PLUS, very thick] (1,0) -- (1,1);
    \draw[MINUS, very thick] (2,0) -- (2,1);
    \draw[MINUS, very thick] (3,0) -- (3,1);
    \draw[PLUS, very thick] (4,0) -- (4,1);

    \node[gnode] at (0.5,0.5) {};
    \node[gnode] at (1.5,0.5) {};
    \node[gnode] at (2.5,0.5) {};
    \node[gnode] at (3.5,0.5) {};
  \end{tikzpicture}
\end{equation}
The $r$th node represents the simple root $\alpha_r$.

Finally, for each pair $(r,s)$ with $a_{rs} \neq 0$, we draw an arrow
from the $r$th node to the $s$th node and write the number $a_{rs}$ on
the side.  We can erase the vertical lines at this stage:
\begin{equation}
  \begin{tikzpicture}[scale=1.5]
    \draw[q->] (0.5,0.56) -- node[above]{$-1$} (1.37,0.56);
    \draw[q->] (1.5,0.43) -- node[below]{$-1$} (0.63,0.43);

    \draw[q->] (1.5,0.56) -- node[above]{$+1$} (2.37,0.56);
    \draw[q->] (2.5,0.43) -- node[below]{$+1$} (1.63,0.43);

    \draw[q->] (2.5,0.56) -- node[above]{$+1$} (3.37,0.56);
    \draw[q->] (3.5,0.43) -- node[below]{$+1$} (2.63,0.43);

    \draw[q->] (0.5,0.6) arc (270:-60:0.15) node[shift={(-4pt,17pt)}]{$+2$};
    \draw[q->] (2.5,0.6) arc (270:-60:0.15) node[shift={(-4pt,17pt)}]{$-2$};

    \node[gnode] at (0.5,0.5) {};
    \node[gnode] at (1.5,0.5) {};
    \node[gnode] at (2.5,0.5) {};
    \node[gnode] at (3.5,0.5) {};
  \end{tikzpicture}
\end{equation}

This quiver has the same content as the Dynkin
diagram~\eqref{eq:Dynkin} modulo the action of the Weyl group
$\mathfrak{S}_m \times \mathfrak{S}_n$ which permutes the basis
vectors $(e_1, \dotsc, e_{m+n})$ without changing the $\Z_2$-grading.

\subsubsection{Verma modules of $\glf(m|n)$}

A representation of $\glf(m|n)$ in a $\Z_2$-graded vector space $V$ is
a Lie superalgebra homomorphism $\pi\colon \glf(m|n) \to \End(V)$,
where $\End(V)$ is given the structure of a Lie superalgebra by the
graded commutator.

The Verma module $M(\lambda)\colon \glf(m|n) \to \End(V_\lambda)$,
with highest weight
\begin{equation}
  \label{eq:hw}
  \lambda = \sum_{i=1}^{m+n} \lambda_i \veps_i \,,
  \quad
  \lambda_i \in \C \,,
\end{equation}
is a representation of $\glf(m|n)$ constructed from a highest-weight
vector $\ket{\Omega_\lambda}$ that is an eigenstate of the diagonal
matrices:
\begin{align}
  \CE_{ii} \ket{\Omega_\lambda} &= \lambda_i \ket{\Omega_\lambda} \,,
  \\
  \CE_{ij} \ket{\Omega_\lambda} &= 0 \,,
  \quad i < j \,.
\end{align}
The other vectors in $V_\lambda$ are created by the action of lowering
operators $\{\CE_{ij} \mid i > j\}$ on $\ket{\Omega_\lambda}$, and two
vectors are identified if they are related by the commutation
relations~\eqref{eq:gl(m|n)-CR}.

More explicitly, the Fock space $V_\lambda$ can be described as
follows.  Let us introduce an ordering among all lowering operators
and name them $x_1$, $\dotsc$, $x_p$.  Then, by the
Poincar\'e--Birkhoff--Witt (PBW) theorem, an element of $V_\lambda$ is
a linear combination of states of the form
\begin{equation}
  \label{eq:PBW}
  x_1^{n_1} \dotsm x_p^{n_p} \ket{\Omega_\lambda} \,,
  \quad
  n_i \in
  \begin{cases}
    \Z_{\geq0} & ([a_i] = \bar0) \,;
    \\
    \{0,1\} & ([a_i] = \bar1) \,.
  \end{cases}
\end{equation}
Verma modules are infinite-dimensional unless $(m|n) = (1|1)$, in
which case there is only one lowering operator and it is odd.

Since the lowering operator $\CE_{ij}$ changes the weight by
$\veps_i - \veps_j = -\alpha_j - \alpha_{j+1} - \dotsb -
\alpha_{i-1}$, a state of $M(\lambda)$ has a weight of the form
\begin{equation}
  \label{eq:weight}
  \lambda - \sum_{r=1}^{m+n-1} M_r \alpha_r \,,
  \quad
  M_r \in \Z_{\geq0} \,.
\end{equation}
We can also represent this weight graphically.  To represent the
highest weight $\lambda$, for each vertical line we draw a diagonal
line ending on it and write $\lambda_i$ next to the $i$th diagonal
line; and to represent the weight~\eqref{eq:weight}, we draw $M_r$
horizontal line segments between the $r$th and $(r+1)$st vertical
lines.  Here is an example for $(M_1,M_2,M_3,M_4) = (2,3,2,1)$:
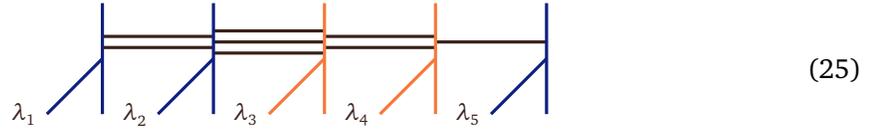
\begin{equation}
  \label{eq:brane-weight}
  \begin{tikzpicture}[scale=1.5]
    \begin{scope}[shift={(0,0.05)}]
      \draw[ZERO, very thick] (0,0.55) -- (1,0.55);
      \draw[ZERO, very thick] (0,0.65) -- (1,0.65);
      
      \draw[ZERO, very thick] (1,0.5) -- (2,0.5);
      \draw[ZERO, very thick] (1,0.6) -- (2,0.6);
      \draw[ZERO, very thick] (1,0.7) -- (2,0.7);
      
      \draw[ZERO, very thick] (2,0.55) -- (3,0.55);
      \draw[ZERO, very thick] (2,0.65) -- (3,0.65);
      
      \draw[ZERO, very thick] (3,0.6) -- (4,0.6);
    \end{scope}

    \draw[PLUS, very thick] (0,0) -- (0,1);
    \draw[PLUS, very thick] (1,0) -- (1,1);
    \draw[MINUS, very thick] (2,0) -- (2,1);
    \draw[MINUS, very thick] (3,0) -- (3,1);
    \draw[PLUS, very thick] (4,0) -- (4,1);

    \draw[PLUS, very thick] (-0.5,0) node[left, ZERO] {$\lambda_1$} -- (0,0.5);
    \draw[PLUS, very thick] (0.5,0) node[left, ZERO] {$\lambda_2$} -- (1,0.5);
    \draw[MINUS, very thick] (1.5,0) node[left, ZERO] {$\lambda_3$} -- (2,0.5);
    \draw[MINUS, very thick] (2.5,0) node[left, ZERO] {$\lambda_4$} -- (3,0.5);
    \draw[PLUS, very thick] (3.5,0) node[left, ZERO] {$\lambda_5$} -- (4,0.5);
  \end{tikzpicture}
\end{equation}

We convert this diagram into a quiver by replacing the diagonal lines
with square nodes and writing $M_r$ inside the $r$th circle node and
$1$ inside the square nodes:
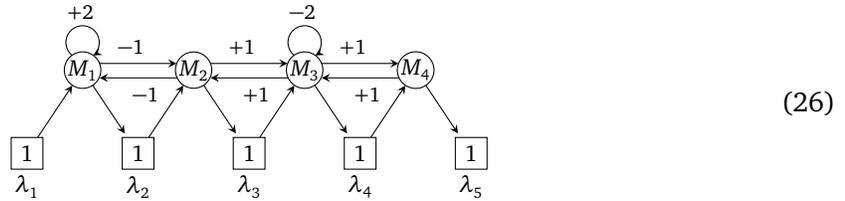
\begin{equation}
  \begin{tikzpicture}[scale=1.5]
    \draw[q->] (0.5,0.56) -- node[above]{$-1$} (1.37,0.56);
    \draw[q->] (1.5,0.43) -- node[below]{$-1$} (0.63,0.43);

    \draw[q->] (1.5,0.56) -- node[above]{$+1$} (2.37,0.56);
    \draw[q->] (2.5,0.43) -- node[below]{$+1$} (1.63,0.43);

    \draw[q->] (2.5,0.56) -- node[above]{$+1$} (3.37,0.56);
    \draw[q->] (3.5,0.43) -- node[below]{$+1$} (2.63,0.43);

    \draw[q->] (0.5,0.6) arc (270:-60:0.15) node[shift={(-4pt,17pt)}]{$+2$};
    \draw[q->] (2.5,0.6) arc (270:-60:0.15) node[shift={(-4pt,17pt)}]{$-2$};

    \draw[q->, shorten >=7pt] (0,-0.25) -- (0.5,0.5);
    \draw[q->, shorten >=9pt] (0.5,0.5) -- (1,-0.25);

    \draw[q->, shorten >=7pt] (1,-0.25) -- (1.5,0.5);
    \draw[q->, shorten >=9pt] (1.5,0.5) -- (2,-0.25);

    \draw[q->, shorten >=7pt] (2,-0.25) -- (2.5,0.5);
    \draw[q->, shorten >=9pt] (2.5,0.5) -- (3,-0.25);

    \draw[q->, shorten >=7pt] (3,-0.25) -- (3.5,0.5);
    \draw[q->, shorten >=9pt] (3.5,0.5) -- (4,-0.25);

    \node[gnode] at (0.5,0.5) {$M_1$};
    \node[gnode] at (1.5,0.5) {$M_2$};
    \node[gnode] at (2.5,0.5) {$M_3$};
    \node[gnode] at (3.5,0.5) {$M_4$};

    \node[fnode] at (0,-0.25) {$1$};
    \node[fnode] at (1,-0.25) {$1$};
    \node[fnode] at (2,-0.25) {$1$};
    \node[fnode] at (3,-0.25) {$1$};
    \node[fnode] at (4,-0.25) {$1$};

    \node at (0,-0.55) {$\lambda_1$};
    \node at (1,-0.55) {$\lambda_2$};
    \node at (2,-0.55) {$\lambda_3$};
    \node at (3,-0.55) {$\lambda_4$};
    \node at (4,-0.55) {$\lambda_5$};
\end{tikzpicture}
\end{equation}
To fix the horizontal positions of the square nodes, we have added
arrows connecting circle and square nodes.

\subsubsection{Tensor products of Verma modules}

If $V_1$ and $V_2$ are $\Z_2$-graded vector spaces, the tensor product
$V_1 \otimes V_2$ is naturally $\Z_2$-graded.  Given two
representations $\pi_1\colon \glf(m|n) \to \End(V_1)$ and
$\pi_2\colon \glf(m|n) \to \End(V_2)$, the tensor product
representation
$\pi_1 \otimes \pi_2 \colon \glf(m|n) \to \End(V_1 \otimes V_2)$ is
defined by
\begin{equation}
  (\pi_1 \otimes \pi_2)(x) (v_1 \otimes v_2)
  =
  \pi_1(x) v_1 \otimes v_2 + (-1)^{[x][v_1]} v_1 \otimes \pi_2(x) v_2 \,,
\end{equation}
where $v_1$, $v_2$ and $x$ are homogeneous in $\Z_2$-grading.  The
tensor products of more than two representations can be defined
recursively.

The tensor product of $L$ Verma modules $M(\lambda^1)$, $\dotsc$,
$M(\lambda^L)$ has a highest-weight vector
$\ket{\Omega_{\lambda^1}} \otimes \dotsb \otimes
\ket{\Omega_{\lambda^L}}$ with highest weight
$\lambda^1 + \dotsb + \lambda^L$.  A weight of
$M(\lambda^1) \otimes \dotsb \otimes M(\lambda^L)$ takes the form
\begin{equation}
  \label{eq:tensor-M-weight}
  \sum_{\ell=1}^L \lambda^\ell - \sum_{r=1}^{m+n-1} M_r \alpha_r \,,
  \quad
  M_r \in \Z_{\geq0} \,.
\end{equation}
Graphically, we represented it by a diagram similar to the diagram
\eqref{eq:brane-weight} for a weight of a single Verma module, but
with $L$ diagonal lines ending on each vertical line.  For example,
the diagram
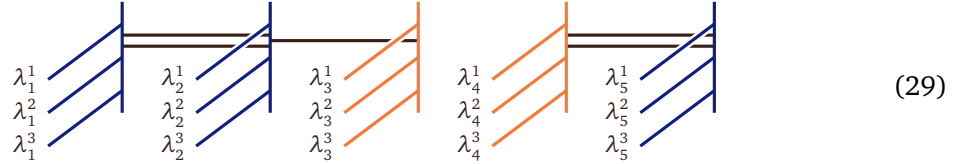
\begin{equation}
  \begin{tikzpicture}[xscale=2, yscale=1.5]
    \begin{scope}[shift={(0,0.05)}]
      \draw[ZERO, very thick] (0,0.55) -- (1,0.55);
      \draw[ZERO, very thick] (0,0.65) -- (1,0.65);
      
      \draw[ZERO, very thick] (1,0.6) -- (2,0.6);
      
      \draw[ZERO, very thick] (3,0.55) -- (4,0.55);
      \draw[ZERO, very thick] (3,0.65) -- (4,0.65);
    \end{scope}

    \draw[white, line width=3pt] (0.5,0.3) -- (1,0.8);
    \draw[white, line width=3pt] (1.5,0.3) -- (2,0.8);
    \draw[white, line width=3pt] (3.5,0) -- (4,0.5);
    \draw[white, line width=3pt] (3.5,0.3) -- (4,0.8);

    \draw[PLUS, very thick] (0,0) -- (0,1);
    \draw[PLUS, very thick] (1,0) -- (1,1);
    \draw[MINUS, very thick] (2,0) -- (2,1);
    \draw[MINUS, very thick] (3,0) -- (3,1);
    \draw[PLUS, very thick] (4,0) -- (4,1);

    \begin{scope}[shift={(0,-0.3)}]
      \draw[PLUS, very thick] (-0.5,0)
      node[left, ZERO] {$\lambda^3_1$} -- (0,0.5);
      \draw[PLUS, very thick] (0.5,0)
      node[left, ZERO] {$\lambda^3_2$} -- (1,0.5);
      \draw[MINUS, very thick] (1.5,0)
      node[left, ZERO] {$\lambda^3_3$} -- (2,0.5);
      \draw[MINUS, very thick] (2.5,0)
      node[left, ZERO] {$\lambda^3_4$} -- (3,0.5);
      \draw[PLUS, very thick] (3.5,0)
      node[left, ZERO] {$\lambda^3_5$} -- (4,0.5);
    \end{scope}

    \begin{scope}[shift={(0,0)}]
      \draw[PLUS, very thick] (-0.5,0)
      node[left, ZERO] {$\lambda^2_1$} -- (0,0.5);
      \draw[PLUS, very thick] (0.5,0)
      node[left, ZERO] {$\lambda^2_2$} -- (1,0.5);
      \draw[MINUS, very thick] (1.5,0)
      node[left, ZERO] {$\lambda^2_3$} -- (2,0.5);
      \draw[MINUS, very thick] (2.5,0)
      node[left, ZERO] {$\lambda^2_4$} -- (3,0.5);
      \draw[PLUS, very thick] (3.5,0)
      node[left, ZERO] {$\lambda^2_5$} -- (4,0.5);
    \end{scope}

    \begin{scope}[shift={(0,0.3)}]
      \draw[PLUS, very thick] (-0.5,0)
      node[left, ZERO] {$\lambda^1_1$} -- (0,0.5);
      \draw[PLUS, very thick] (0.5,0)
      node[left, ZERO] {$\lambda^1_2$} -- (1,0.5);
      \draw[MINUS, very thick] (1.5,0)
      node[left, ZERO] {$\lambda^1_3$} -- (2,0.5);
      \draw[MINUS, very thick] (2.5,0)
      node[left, ZERO] {$\lambda^1_4$} -- (3,0.5);
      \draw[PLUS, very thick] (3.5,0)
      node[left, ZERO] {$\lambda^1_5$} -- (4,0.5);
    \end{scope}
  \end{tikzpicture}
\end{equation}
represents a weight with $(M_1, M_2, M_3, M_4) = (2, 1, 0, 2)$ in the
representation
$M(\lambda^1) \otimes M(\lambda^2) \otimes M(\lambda^3)$ of
$\glf(3|2)$.

The corresponding quiver diagram is the same as before, except that
the $i$th square node is now labeled $L$ and accompanied by the
$L$-tuple $\vec\lambda_i = (\lambda^1_i, \dotsc, \lambda^L_i)$:
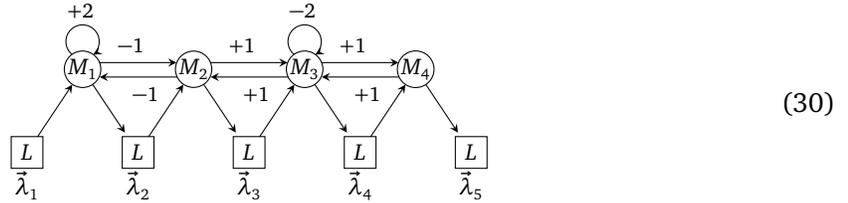
\begin{equation}
  \begin{tikzpicture}[scale=1.5]
    \draw[q->] (0.5,0.56) -- node[above]{$-1$} (1.37,0.56);
    \draw[q->] (1.5,0.43) -- node[below]{$-1$} (0.63,0.43);

    \draw[q->] (1.5,0.56) -- node[above]{$+1$} (2.37,0.56);
    \draw[q->] (2.5,0.43) -- node[below]{$+1$} (1.63,0.43);

    \draw[q->] (2.5,0.56) -- node[above]{$+1$} (3.37,0.56);
    \draw[q->] (3.5,0.43) -- node[below]{$+1$} (2.63,0.43);

    \draw[q->] (0.5,0.6) arc (270:-60:0.15) node[shift={(-4pt,17pt)}]{$+2$};
    \draw[q->] (2.5,0.6) arc (270:-60:0.15) node[shift={(-4pt,17pt)}]{$-2$};

    \draw[q->, shorten >=7pt] (0,-0.25) -- (0.5,0.5);
    \draw[q->, shorten >=9pt] (0.5,0.5) -- (1,-0.25);

    \draw[q->, shorten >=7pt] (1,-0.25) -- (1.5,0.5);
    \draw[q->, shorten >=9pt] (1.5,0.5) -- (2,-0.25);

    \draw[q->, shorten >=7pt] (2,-0.25) -- (2.5,0.5);
    \draw[q->, shorten >=9pt] (2.5,0.5) -- (3,-0.25);

    \draw[q->, shorten >=7pt] (3,-0.25) -- (3.5,0.5);
    \draw[q->, shorten >=9pt] (3.5,0.5) -- (4,-0.25);

    \node[gnode] at (0.5,0.5) {$M_1$};
    \node[gnode] at (1.5,0.5) {$M_2$};
    \node[gnode] at (2.5,0.5) {$M_3$};
    \node[gnode] at (3.5,0.5) {$M_4$};

    \node[fnode] at (0,-0.25) {$L$};
    \node[fnode] at (1,-0.25) {$L$};
    \node[fnode] at (2,-0.25) {$L$};
    \node[fnode] at (3,-0.25) {$L$};
    \node[fnode] at (4,-0.25) {$L$};

    \node at (0,-0.55) {$\vec\lambda_1$};
    \node at (1,-0.55) {$\vec\lambda_2$};
    \node at (2,-0.55) {$\vec\lambda_3$};
    \node at (3,-0.55) {$\vec\lambda_4$};
    \node at (4,-0.55) {$\vec\lambda_5$};
  \end{tikzpicture}
\end{equation}

The above quiver diagram will be identified with a quiver describing
a two-dimensional $\CN = (2,2)$ supersymmetric gauge theory that
appears on the gauge theory side of the Bethe/gauge correspondence.
The graphical representation using lines will be interpreted as a
diagram depicting a brane configuration in string theory.

\subsection{Bethe side}

Now we explain the Bethe side of the Bethe/gauge correspondence.  The
spin chains we consider in this paper are rational $\glf(m|n)$ spin
chains, for which spins take values in representations of $\glf(m|n)$.
More generally, spins in rational $\glf(m|n)$ spin chains are valued
in representations of the Yangian $Y(\glf(m|n))$ of $\glf(m|n)$.

\subsubsection{Yangian}

The \emph{Yangian} $Y(\glf(m|n))$ is a $\Z_2$-graded Hopf algebra,
which in particular is a unital associative $\Z_2$-graded algebra.  It
is generated by elements
\begin{equation}
  T_{ij}^{(l)} \,,
  \quad
  i, j = 1, \dotsc, m+n \,,
  \quad
  l \in \Z_{> 0} \,,
\end{equation}
with grading $[T_{ij}^{(l)}] = [i] + [j]$.  The level-$1$ generators
$T_{ij}^{(1)}$ span a subalgebra isomorphic to $\glf(m|n)$, with the
identification of generators being
$\CE_{ij} = (-1)^{[j]} T_{ji}^{(1)}$.

To describe the algebra relations for all generators in a compact
form, let us introduce a formal variable $\sigma$ and combine the
generators into a single $\End(\C^{m|n}) \otimes Y(\glf(m|n))$-valued
power series in $\sigma^{-1}$:
\begin{equation}
  T(\sigma)
  =
  \sum_{i,j=1}^{m+n} E_{ij} \otimes T_{ij}(\sigma)
  =
  \sum_{i,j=1}^{m+n} \sum_{l=0}^\infty
  \frac{\hbar^l}{\sigma^l} E_{ij} \otimes T_{ij}^{(l)} \,.
\end{equation}
Here $T_{ij}^{(0)} = \delta_{ij}$ and $\hbar$ is a complex parameter.
We can think of $T(\sigma)$ as an $(m+n) \times (m+n)$ matrix whose
entries are elements of $Y(\glf(m|n))[[\sigma^{-1}]]$; it is called
the \emph{monodromy matrix} and is a function of \emph{spectral
  parameter} $\sigma$.  In terms of the monodromy matrix, the algebra
relations for $Y(\glf(m|n))$ are encoded in the \emph{RTT relation}
\begin{equation}
  \label{eq:RTT}
  R_{12}(\sigma_1 - \sigma_2) T_1(\sigma_1) T_2(\sigma_2)
  = T_2(\sigma_2) T_1(\sigma_1) R_{12}(\sigma_1 - \sigma_2) \,.
\end{equation}
This is a relation between elements in
$\End(\C^{m|n}) \otimes \End(\C^{m|n}) \otimes
Y(\glf(m|n))[[\sigma^{-1}]]$, and the subscript(s) on an operator
indicate which factor(s) of $\C^{m|n}$ the operator acts on.

The operator
$R_{12}(\sigma) \in \End(\C^{m|n}) \otimes \End(\C^{m|n})$ that
appears in the RTT relation is the \emph{rational $\glf(m|n)$
  R-matrix}.  It is given by
\begin{equation}
  \label{eq:R}
  R_{12}(\sigma) = \sigma I \otimes I + \hbar P_{12} \,,
\end{equation}
where $I$ is the identity matrix and
\begin{equation}
  P_{12} = \sum_{i,j=1}^{m+n} (-1)^{[j]} E_{ij} \otimes E_{ji} \,.
\end{equation}
The permutation operator $P_{12}$ swaps tensor factors as
\begin{equation}
  P_{12} (e_i \otimes e_j) = (-1)^{[i][j]} e_j \otimes e_i \,.
\end{equation}
The R-matrix commutes with the automorphism group $\GL(m|n)$ of
$\C^{m|n}$:
\begin{equation}
  \label{eq:R-GL-symmetry}
  [g \otimes g, R_{12}(\sigma)] = 0 \,,
  \quad
  g \in \GL(m|n) \,.
\end{equation}

The dynamics of a closed rational $\glf(m|n)$ spin chain is generated
by the \emph{transfer matrix}
\begin{equation}
  t(g, \sigma)
  = \str_{\C^{m|n}}\bigl(g T(\sigma)\bigr)
  = \sum_{i=1}^{m+n} (-1)^{[i]} g_{ij} T_{ji}(\sigma) \,,
  \quad
  g \in \GL(m|n) \,.
\end{equation}
The supertrace taken over $\C^{m|n}$ corresponds to the topology of
the spin chain which is closed, and $g$ twists the periodic boundary
condition.

Multiplying both sides of the RTT relation \eqref{eq:RTT} by
$g \otimes g \otimes 1$ from the left and
$R_{12}(\sigma_1 - \sigma_2)^{-1}$ from the right, then using the
symmetry \eqref{eq:R-GL-symmetry} of the R-matrix and taking the
supertrace over $\C^{m|n} \otimes \C^{m|n}$, we see that transfer
matrices for a fixed $g$ commute with each other:
\begin{equation}
  [t(g,\sigma_1), t(g,\sigma_2)] = 0 \,.
\end{equation}
Therefore, if we expand $t(g,\sigma)$ in powers of $\sigma^{-1}$, the
coefficients are mutually commuting elements of $Y(\glf(m|n))$.  They
generate a commutative subalgebra called the \emph{Bethe algebra} (or
the Baxter algebra) of $Y(\glf(m|n))$.

\subsubsection{Representations of $Y(\glf(m|n))$}

While the Yangian $Y(\glf(m|n))$ and its Bethe algebra are the
algebraic structures underlying rational $\glf(m|n)$ spin chains, to
get a concrete physical realization of a spin chain we need to specify
a representation of $Y(\glf(m|n))$.

A representation $\rho\colon Y(\glf(m|n)) \to \End(V)$ of
$Y(\glf(m|n))$ maps the monodromy matrix to an $\End(V)$-valued matrix
\begin{equation}
  \rho\bigl(T(\sigma)\bigr)
  =
  \sum_{i,j=1}^{m+n} \sum_{l=0}^\infty
  \frac{\hbar^l}{\sigma^l}
  E_{ij} \otimes \rho(T_{ij}^{(l)}) \,.
\end{equation}
Conversely, an $\End(V)$-valued matrix satisfying the RTT relation
determines a representation of $Y(\glf(m|n))$.

Given a representation $\pi\colon \glf(m|n) \to \End(V)$ of
$\glf(m|n)$, we obtain a one-parameter family of representations
$\pi_\zeta\colon Y(\glf(m|n)) \to \End(V)$, $\zeta \in \C$, of
$Y(\glf(m|n))$ by
\begin{equation}
  \pi_\zeta\bigl(T(\sigma)\bigr)
  =
  I \otimes \id_V
  + \frac{\hbar}{\sigma-\zeta} \sum_{i,j=1}^{m+n}
  (-1)^{[j]} E_{ij} \otimes \pi(\CE_{ji}) \,.
\end{equation}
This is known as the \emph{evaluation module} for $\pi$, and $\zeta$
is called the \emph{inhomogeneity parameter}.  Note that we have
\begin{equation}
  \pi_\zeta\bigl((-1)^{[j]} T_{ji}^{(1)}\bigr) = \pi(\CE_{ij}) \,.
\end{equation}

For a representation $\pi\colon \glf(m|n) \to \End(V)$ of $\glf(m|n)$,
let us define a one-parameter family of representations
$\pi^c\colon \glf(m|n) \to \End(V)$, $c \in \C$, by
\begin{equation}
  \pi^c(\CE_{ij})
  =
  \pi(\CE_{ij}) + (-1)^{[i]} c\delta_{ij}\id_V \,.
\end{equation}
In the associated Yangian representations, the parameter $c$ is
related to a shift in the inhomogeneity parameter.  Suppose that
$\rho\colon Y(\glf(m|n)) \to \End(V)$ is a representation of
$Y(\glf(m|n))$.  Then, $\rho(T(\sigma))$ satisfies the RTT equation,
and for any function $f$ of $\sigma$, the RTT equation is still
satisfied when $\rho(T(\sigma))$ is replaced by
$f(\sigma) \rho(T(\sigma))$.  Therefore, if $f(\sigma)$ can be
expanded in a power series in $\sigma^{-1}$ starting from $1$, then
$f(\sigma) \rho(T(\sigma))$ defines a new representation of
$Y(\glf(m|n))$.  Since
\begin{equation}
  \label{eq:piczeta}
  \pi^c_\zeta\bigl(T(\sigma)\bigr)
  =
  \pi_\zeta(T(\sigma))
  + \frac{c\hbar}{\sigma - \zeta} I \otimes \id_V
  =
  \frac{\sigma - \zeta + c\hbar}{\sigma - \zeta}
  \pi_{\zeta - c\hbar}\bigl(T(\sigma)\bigr)
  \,,
\end{equation}
we see that $\pi^c_\zeta$ and $\pi_{\zeta - c\hbar}$ are related in
this manner.

To construct tensor product representations of $Y(\glf(m|n))$, we use
the coproduct
$\Delta\colon Y(\glf(m|n)) \linebreak[1] \to Y(\glf(m|n)) \otimes Y(\glf(m|n))$.
The map $\Delta$ is defined by the formula
\begin{equation}
  \Delta\bigl(T(\sigma)\bigr)
  =
  \sum_{i,j,k=1}^{m+n} (-1)^{([i] + [k])([k] + [j])}
  E_{ij} \otimes T_{ik}(\sigma) \otimes T_{kj}(\sigma) \,.
\end{equation}
Given two representations $\rho_1\colon Y(\glf(m|n)) \to \End(V_1)$
and $\rho_2\colon Y(\glf(m|n)) \to \End(V_2)$, the tensor product
representation
$\rho_1 \dot\otimes \rho_2\colon Y(\glf(m|n)) \to \End(V_1 \otimes
V_2)$ is defined by
\begin{equation}
  \rho_1 \dot\otimes \rho_2 = (\rho_1 \otimes \rho_2) \circ \Delta \,.
\end{equation}
A calculation shows that $(\rho_1 \dot\otimes \rho_2)(T(\sigma))$
satisfies the RTT relation.

\subsubsection{The spin chain}

Now, fix a positive integer $L$, and choose $L$ highest weights
\begin{equation}
  \vec\lambda = (\lambda^1, \dotsc, \lambda^L)
\end{equation}
of $\glf(m|n)$ and $L$ inhomogeneity parameters
\begin{equation}
  \vec\zeta = (\zeta^1, \dotsc, \zeta^L) \,.
\end{equation}
Also, choose a diagonal element%
\footnote{The spin chain can be defined for any choice of $g$, not
  necessarily diagonal ones. However, nondiagonalizable choices of $g$
  do not appear to have a clear interpretation on the gauge theory
  side of the Bethe/gauge correspondence.}
of $\GL(m|n)$:
\begin{equation}
  g = \diag(e^{\phi_1}, \dotsc, e^{\phi_{m+n}}) \,,
  \quad
  \phi_i \in \C \,.
\end{equation}
We consider the closed rational $\glf(m|n)$ spin chain of length $L$,
with the spin at the $\ell$th site valued in the evaluation module
$M(\lambda^\ell)_{\zeta^\ell}$ for the Verma module $M(\lambda^\ell)$
and the periodic boundary condition twisted by $g$.

The Hilbert space of states of this spin chain is the tensor product
\begin{equation}
  \label{eq:Hilbert}
  V_{\vec\lambda} = \bigotimes_{\ell=1}^L V_{\lambda^\ell} \,,
\end{equation}
and the $L$ spins can be thought of as a single spin in the tensor
product representation
\begin{equation}
  M(\vec\lambda)_{\vec\zeta}
  =
  M(\lambda^1)_{\zeta^1} \dot\otimes \dotsb \dot\otimes
  M(\lambda^L)_{\zeta^L}
  \,.
\end{equation}
The transfer matrix of the spin chain
\begin{equation}
  \label{eq:t-spin-chain}
  M(\vec\lambda)_{\vec\zeta} \bigl(t(g,\sigma)\bigr)
\end{equation}
generates commuting conserved charges acting on $V_{\vec\lambda}$,
making the spin chain integrable.  The Hamiltonian is a linear
combination of these charges.

\subsubsection{Bethe equations}

The Hilbert space of the spin chain is spanned by vectors that
simultaneously diagonalize the commuting conserved charges, or
equivalently, diagonalize the transfer matrix~\eqref{eq:t-spin-chain}
for all values of the spectral parameter $\sigma$.  These
eigenvectors, referred to as \emph{Bethe vectors}, are the main
characters from the Bethe side of the Bethe/gauge correspondence.

The Bethe vectors for the rational $\glf(m|n)$ spin chain have been
constructed by Bethe ansatz methods~\cite{Ragoucy:2007kg,
  Belliard:2008di}.  The construction starts with the highest-weight
vector $\ket{\Omega_{\vec\lambda}}$ of the tensor product
representation
\begin{equation}
  M(\vec\lambda) = \bigotimes_{\ell=1}^L M(\lambda^\ell) \,.
\end{equation}
This state is called the \emph{pseudovacuum} and satisfies
\begin{align}
  M(\vec\lambda)_{\vec\zeta}\bigl(T_{ii}(\sigma)\bigr)
  \ket{\Omega_{\vec\lambda}}
  &=
    \Biggl(\prod_{\ell=1}^L 
    \frac{\sigma - \zeta^\ell + (-1)^{[i]} \lambda_i^\ell \hbar}
         {\sigma - \zeta^\ell}\Biggr)
    \ket{\Omega_{\vec\lambda}} \,,
  \\
  M(\vec\lambda)_{\vec\zeta}\bigl(T_{ij}(\sigma)\bigr)
  \ket{\Omega_{\vec\lambda}}
  &=
    0 \,,
  \quad
  i > j \,.
\end{align}
According to our graphical notation, the pseudovacuum is represented
by a diagram with no horizontal segments:
\begin{equation}
  \ket{\Omega_{\vec\lambda}}
  =
  \begin{tikzpicture}[xscale=2, yscale=1.5]
    \draw[PLUS, very thick] (0,0) -- (0,1);
    \draw[PLUS, very thick] (1,0) -- (1,1);
    \draw[MINUS, very thick] (2,0) -- (2,1);
    \draw[MINUS, very thick] (3,0) -- (3,1);
    \draw[PLUS, very thick] (4,0) -- (4,1);

    \begin{scope}[shift={(0,-0.3)}]
      \draw[PLUS, very thick] (-0.5,0)
      node[left, ZERO] {$\lambda^3_1$} -- (0,0.5);
      \draw[PLUS, very thick] (0.5,0)
      node[left, ZERO] {$\lambda^3_2$} -- (1,0.5);
      \draw[MINUS, very thick] (1.5,0)
      node[left, ZERO] {$\lambda^3_3$} -- (2,0.5);
      \draw[MINUS, very thick] (2.5,0)
      node[left, ZERO] {$\lambda^3_4$} -- (3,0.5);
      \draw[PLUS, very thick] (3.5,0)
      node[left, ZERO] {$\lambda^3_5$} -- (4,0.5);
    \end{scope}

    \begin{scope}[shift={(0,0)}]
      \draw[PLUS, very thick] (-0.5,0)
      node[left, ZERO] {$\lambda^2_1$} -- (0,0.5);
      \draw[PLUS, very thick] (0.5,0)
      node[left, ZERO] {$\lambda^2_2$} -- (1,0.5);
      \draw[MINUS, very thick] (1.5,0)
      node[left, ZERO] {$\lambda^2_3$} -- (2,0.5);
      \draw[MINUS, very thick] (2.5,0)
      node[left, ZERO] {$\lambda^2_4$} -- (3,0.5);
      \draw[PLUS, very thick] (3.5,0)
      node[left, ZERO] {$\lambda^2_5$} -- (4,0.5);
    \end{scope}

    \begin{scope}[shift={(0,0.3)}]
      \draw[PLUS, very thick] (-0.5,0)
      node[left, ZERO] {$\lambda^1_1$} -- (0,0.5);
      \draw[PLUS, very thick] (0.5,0)
      node[left, ZERO] {$\lambda^1_2$} -- (1,0.5);
      \draw[MINUS, very thick] (1.5,0)
      node[left, ZERO] {$\lambda^1_3$} -- (2,0.5);
      \draw[MINUS, very thick] (2.5,0)
      node[left, ZERO] {$\lambda^1_4$} -- (3,0.5);
      \draw[PLUS, very thick] (3.5,0)
      node[left, ZERO] {$\lambda^1_5$} -- (4,0.5);
    \end{scope}
  \end{tikzpicture}
\end{equation}

Excited states are obtained from the pseudovacuum by the action of
creation operators $M(\vec\lambda)_{\vec\zeta}(T_{ij}(\sigma))$,
$i < j$.  The operator $M(\vec\lambda)_{\vec\zeta}(T_{ij}(\sigma))$
contains $M(\vec\lambda)(\CE_{ji})$ and changes the $\glf(m|n)$ weight
by
$\veps_j - \veps_i = -\alpha_i - \alpha_{i+1} - \dotsb -
\alpha_{j-1}$.  Roughly speaking, we can interpret this action as
creating a single quasi-particle, or a \emph{magnon}, of rapidity
$\sigma$ and type $r$ for each $r = i$, $i+1$, $\dotsc$, $j-1$.
Graphically, we think of it as creating a horizontal line connecting
the $i$th and $j$th vertical lines:
\begin{equation}
  T_{13}(\sigma_1) T_{25}(\sigma_2) T_{45}(\sigma_3)
  \ket{\Omega_{\vec\lambda}}
  \sim
  \begin{tikzpicture}[xscale=1.5, yscale=1.5]
    \begin{scope}[shift={(0,0.175)}]
      \draw[ZERO, very thick] (0,0.5) -- node[above, xshift=-2em, yshift=-2pt] {$\sigma_1$} (2,0.5);
    \end{scope}

    \begin{scope}[shift={(1,0.05)}]
      \draw[ZERO, very thick] (0,0.5) -- node[above, yshift=-2pt] {$\sigma_2$} (3,0.5);
    \end{scope}

    \begin{scope}[shift={(3,-0.275)}]
      \draw[ZERO, very thick] (0,0.5) -- node[above left, yshift=-2pt] {$\sigma_3$} (1,0.5);
    \end{scope}

    \draw[white, line width=3pt] (1,0.65) -- (1,0.7);
    \draw[white, line width=3pt] (2,0.53) -- (2,0.57);
    \draw[white, line width=3pt] (3,0.53) -- (3,0.57);

    \draw[white, line width=3pt] (0.5,0.3) -- (1,0.8);
    \draw[white, line width=3pt] (1.5,0.3) -- (2,0.8);
    \draw[white, line width=3pt] (2.5,0.3) -- (3,0.8);
    \draw[white, line width=3pt] (3.5,0) -- (4,0.5);
    \draw[white, line width=3pt] (3.5,0.3) -- (4,0.8);

    \draw[PLUS, very thick] (0,0) -- (0,1);
    \draw[PLUS, very thick] (1,0) -- (1,1);
    \draw[MINUS, very thick] (2,0) -- (2,1);
    \draw[MINUS, very thick] (3,0) -- (3,1);
    \draw[PLUS, very thick] (4,0) -- (4,1);

    \begin{scope}[shift={(0,-0.3)}]
      \draw[PLUS, very thick] (-0.5,0) -- (0,0.5);
      \draw[PLUS, very thick] (0.5,0) -- (1,0.5);
      \draw[MINUS, very thick] (1.5,0) -- (2,0.5);
      \draw[MINUS, very thick] (2.5,0) -- (3,0.5);
      \draw[PLUS, very thick] (3.5,0) -- (4,0.5);
    \end{scope}

    \begin{scope}[shift={(0,0)}]
      \draw[PLUS, very thick] (-0.5,0) -- (0,0.5);
      \draw[PLUS, very thick] (0.5,0) -- (1,0.5);
      \draw[MINUS, very thick] (1.5,0) -- (2,0.5);
      \draw[MINUS, very thick] (2.5,0) -- (3,0.5);
      \draw[PLUS, very thick] (3.5,0) -- (4,0.5);
    \end{scope}

    \begin{scope}[shift={(0,0.3)}]
      \draw[PLUS, very thick] (-0.5,0) -- (0,0.5);
      \draw[PLUS, very thick] (0.5,0) -- (1,0.5);
      \draw[MINUS, very thick] (1.5,0) -- (2,0.5);
      \draw[MINUS, very thick] (2.5,0) -- (3,0.5);
      \draw[PLUS, very thick] (3.5,0) -- (4,0.5);
    \end{scope}
  \end{tikzpicture}
\end{equation}
This is, however, not a precise correspondence because the left-hand
side depends on the ordering of creation operators.

The operator $T_{ji}(\sigma)$, $i < j$, changes the weight by
$\alpha_i + \alpha_{i+1} + \dotsb + \alpha_{j-1}$, so annihilates
magnons of type $m = i$, $i+1$, $\dotsc$, $j-1$.  It removes one
horizontal line from each of the intervals between the $i$th and $j$th
vertical lines.  If there is no horizontal line to remove, then the
state is annihilated.

Eigenvectors of the transfer matrix are excited states constructed by
certain linear combinations of creation operators.  It turns out that
a Bethe vector with the $\glf(m|n)$ weight~\eqref{eq:tensor-M-weight}
is specified by a \emph{Bethe root}
\begin{equation}
  (\{\sigma_1^1, \dotsc, \sigma_1^{M_1}\},
  \{\sigma_2^1, \dotsc, \sigma_2^{M_2}\},
  \dotsc,
  \{\sigma_{m+n-1}^1, \dotsc, \sigma_{m+n-1}^{M_{m+n-1}}\}) \,,
\end{equation}
which is a solution of the \emph{Bethe equations}
\begin{multline}
  \label{eq:Bethe}
  e^{\tau_r} \prod_{s=1}^{m+n-1} \prod_{b_s = 1}^{M_s}
  \frac{\sigma_r^{a_r} - \sigma_s^{b_s} + \frac12 a_{rs} \hbar}
       {\sigma_r^{a_r} - \sigma_s^{b_s} - \frac12 a_{rs} \hbar}
  \\
  =
  (-1)^{\delta_{[r],[r+1]}}
  \prod_{\ell=1}^L
  \frac{\sigma_r^{a_r} - \zeta^\ell
    + (-1)^{[r]} \lambda_r^\ell\hbar - \frac12 c_r \hbar}
      {\sigma_r^{a_r} - \zeta^\ell
        + (-1)^{[r+1]} \lambda_{r+1}^\ell \hbar - \frac12 c_r \hbar}
  \,,
  \\
  a_r = 1, \dotsc, M_r \,,
  \quad
  r = 1, \dotsc, m+n-1 \,.
\end{multline}
Here we have defined
\begin{equation}
  \tau_r = (-1)^{[r+1]} \phi_{r+1} - (-1)^{[r]} \phi_r
\end{equation}
and
\begin{equation}
  c_i = \sum_{j=1}^i (-1)^{[j]} \,.
\end{equation}

We represent this Bethe vector by a diagram with $M_r$ horizontal
lines between the $r$th and $(r+1)$st vertical lines, with labels
$\{\sigma_r^1, \dotsc, \sigma_r^{M_r}\}$.  Therefore, the diagram
\begin{equation}
  \label{eq:Bethe-vector-diagram}
  \begin{tikzpicture}[scale=2]
    \begin{scope}[shift={(0,0.15)}]
      \draw[ZERO, very thick] (0,0.5) -- node[above left=-2pt] {$\sigma_1^1$} (1,0.5);
    \end{scope}

    \begin{scope}[shift={(2,0.1)}]
      \draw[ZERO, very thick] (0,0.5) -- node[above left=-2pt] {$\sigma_3^1$} (1,0.5);
    \end{scope}

    \begin{scope}[shift={(3,0)}]
      \draw[ZERO, very thick] (0,0.4) -- node[above left=-2pt] {$\sigma_4^2$} (1,0.4);
      \draw[ZERO, very thick] (0,0.85) -- node[above left=-2pt] {$\sigma_4^1$} (1,0.85);
    \end{scope}

    \begin{scope}[shift={(1,-0.1)}]
      \draw[ZERO, very thick] (0,0.8) -- node[above left=-2pt] {$\sigma_2^1$} (1,0.8);
      \draw[ZERO, very thick] (0,0.45) -- node[above left=-2pt] {$\sigma_2^2$} (1,0.45);
    \end{scope}

    \draw[white, line width=3pt] (0.5,0.3) -- (1,0.8);
    \draw[white, line width=3pt] (1.5,0) -- (2,0.5);
    \draw[white, line width=3pt] (1.5,0.3) -- (2,0.8);
    \draw[white, line width=3pt] (2.5,0.3) -- (3,0.8);
    \draw[white, line width=3pt] (3.5,-0.3) -- (4,0.2);
    \draw[white, line width=3pt] (3.5,0) -- (4,0.5);
    \draw[white, line width=3pt] (3.5,0.3) -- (4,0.8);

    \draw[PLUS, very thick] (0,0) -- (0,1);
    \draw[PLUS, very thick] (1,0) -- (1,1);
    \draw[MINUS, very thick] (2,0) -- (2,1);
    \draw[MINUS, very thick] (3,0) -- (3,1);
    \draw[PLUS, very thick] (4,0) -- (4,1);

    \begin{scope}[shift={(0,-0.3)}]
      \draw[PLUS, very thick] (-0.5,0)
      node[below left=-4pt, ZERO] {$\lambda^3_1$} -- (0,0.5);
      \draw[PLUS, very thick] (0.5,0)
      node[below left=-4pt, ZERO] {$\lambda^3_2$} -- (1,0.5);
      \draw[MINUS, very thick] (1.5,0)
      node[below left=-4pt, ZERO] {$\lambda^3_3$} -- (2,0.5);
      \draw[MINUS, very thick] (2.5,0)
      node[below left=-4pt, ZERO] {$\lambda^3_4$} -- (3,0.5);
      \draw[PLUS, very thick] (3.5,0)
      node[below left=-4pt, ZERO] {$\lambda^3_5$} -- (4,0.5);
    \end{scope}

    \begin{scope}[shift={(0,0)}]
      \draw[PLUS, very thick] (-0.5,0)
      node[below left=-4pt, ZERO] {$\lambda^2_1$} -- (0,0.5);
      \draw[PLUS, very thick] (0.5,0)
      node[below left=-4pt, ZERO] {$\lambda^2_2$} -- (1,0.5);
      \draw[MINUS, very thick] (1.5,0)
      node[below left=-4pt, ZERO] {$\lambda^2_3$} -- (2,0.5);
      \draw[MINUS, very thick] (2.5,0)
      node[below left=-4pt, ZERO] {$\lambda^2_4$} -- (3,0.5);
      \draw[PLUS, very thick] (3.5,0)
      node[below left=-4pt, ZERO] {$\lambda^2_5$} -- (4,0.5);
    \end{scope}

    \begin{scope}[shift={(0,0.3)}]
      \draw[PLUS, very thick] (-0.5,0)
      node[below left=-4pt, ZERO] {$\lambda^1_1$} -- (0,0.5);
      \draw[PLUS, very thick] (0.5,0)
      node[below left=-4pt, ZERO] {$\lambda^1_2$} -- (1,0.5);
      \draw[MINUS, very thick] (1.5,0)
      node[below left=-4pt, ZERO] {$\lambda^1_3$} -- (2,0.5);
      \draw[MINUS, very thick] (2.5,0)
      node[below left=-4pt, ZERO] {$\lambda^1_4$} -- (3,0.5);
      \draw[PLUS, very thick] (3.5,0)
      node[below left=-4pt, ZERO] {$\lambda^1_5$} -- (4,0.5);
    \end{scope}
  \end{tikzpicture}
\end{equation}
represents a Bethe vector of the closed $\glf(3|2)$ rational spin
chain of length $L = 3$ that belongs to the magnon sector
$(M_1,M_2,M_3,M_4) = (1,2,1,2)$ and corresponds to the Bethe root
$(\{\sigma_1^1\}, \{\sigma^1_2, \sigma^2_2\}, \{\sigma_3^1\},
\{\sigma_4^1, \sigma_4^2\})$.

Note that the Bethe equations are invariant under the shift
\begin{equation}
  \zeta^\ell \mapsto \zeta^\ell + c^\ell \hbar \,,
  \quad
  \lambda^\ell_i \mapsto \lambda^\ell_i + (-1)^{[i]} c^\ell \,,
  \quad
  c^\ell \in \C \,.
\end{equation}
This is a consequence of the relation~\eqref{eq:piczeta} between a
representation with shifted highest weight and a representation with
shifted inhomogeneity parameter.  Since multiplying the transfer
matrix by a function of the spectral parameter does not change its
eigenvectors, the Bethe equations remain the same if we shift the
highest weights and inhomogeneity parameters as above.

\subsection{Gauge side}
\label{sec:gauge-side}

Now we turn to the gauge side of the Bethe/gauge correspondence.  The
closed rational $\glf(m|n)$ spin chain discussed above corresponds to
a family of two-dimensional $\CN = (2,2)$ supersymmetric gauge
theories whose field contents are described by quivers.  These
theories have supersymmetric vacua that are in one-to-one
correspondence with the Bethe vectors of the spin chain.  For
background knowledge on $\CN = (2,2)$ supersymmetric gauge theories,
we refer the reader to \cite{Hori:2003ic}.

\subsubsection{The gauge theories}

The magnon sectors of the spin chain (the weight spaces of the Hilbert
space~\eqref{eq:Hilbert}) are labeled by $(m+n-1)$-tuples of
nonnegative integers.  The sector with magnon numbers
$(M_1, \dotsc, M_{m+n-1})$ corresponds to a theory with the product
gauge group
\begin{equation}
  \U(M_1) \times \dotsb \times \U(M_{m+n-1}) \,.
\end{equation}
Correspondingly, the theory has vector multiplets $\Vsf_r$, $r = 1$,
$\dotsc$, $m+n-1$, one for each unitary gauge group factor.

In addition, the theory has various chiral multiplets.  If
$[r] = [r+1]$, then there is one chiral multiplet transforming in the
adjoint representation of $\U(M_r)$:%
\footnote{Here and thereafter, statements about fields such as the one
  that follows only indicate the representations in which they are
  valued.}
\begin{equation}
  \upphi_r
  \in \Hom(\C^{M_r},\C^{M_r}) \,,
  \quad
  [r] = [r+1] \,,
  \quad
  r = 1, \dotsc, m+n-1 \,.
\end{equation}
There are also chiral multiplets
\begin{alignat}{2}
  \Psf_i &\in \Hom(\C^{M_{i-1}}, \C^{M_i}) \,,
  & \quad &
  i = 2, \dotsc, m+n-1 \,,
  \\
  \Psft_i &\in \Hom(\C^{M_i},\C^{M_{i-1}}) \,,
  & \quad &
  i = 2, \dotsc, m+n-1 \,,
  \\
  \Qsf_i &\in \Hom(\C^L,\C^{M_{i-1}}) \,,
  & \quad &
  i = 2, \dotsc, m+n \,,
  \\
  \Qsft_i &\in \Hom(\C^{M_i},\C^L) \,,
  & \quad &
  i = 1, \dotsc, m+n-1 \,.
\end{alignat}
It is convenient to introduce the notations
\begin{equation}
  \upphi_r = 0 \,,
  \quad
  [r] \neq [r+1] \,,
\end{equation}
and
\begin{equation}
  \Psf_1 = \Psft_1 = \Psf_{m+n} = \Psft_{m+n} = \Qsf_1 = \Qsft_{m+n} = 0 \,.
\end{equation}

These chiral multiplets are coupled by the superpotential
\begin{multline}
  \label{eq:W}
  W
  =
  \sum_{r=1}^{m+n-1}
  \tr_{\C^{M_r}} \Bigl(
  \upphi_r \Psft_{r+1} \Psf_{r+1}
  -
  \upphi_r \Psf_r \Psft_r
  +
  \Psf_r \Qsf_r \Qsft_r
  \\
  +
  \bigl((-1)^{[r]} - (-1)^{[r+1]}\bigr)
  \Psft_{r+1} \Psf_{r+1} \Psf_r \Psft_r \Bigr)
  \,.
\end{multline}
The terms involving adjoint chiral multiplets are the cubic
superpotentials required for $\CN = (4,4)$ supersymmetry, which the
theory possesses if either $m = 0$ or $n = 0$.  The last quartic terms
are present only for the gauge group factors without adjoint chiral
multiplets~\cite{Uranga:1998vf}.%
\footnote{The quartic terms can be understood as follows.  Suppose
  that $[r] \neq [r+1]$, say $[r] = 0$ and $[r+1] = 1$, and introduce
  a pair of chiral multiplets $\upphi_r^\pm$ in the adjoint
  representation of $\U(M_r)$.  These multiplets are massive and
  couple with bifundamental chiral multiplets via superpotential terms
  of the form
  $\tr_{\C^{M_r}}(\upphi_r^- \Psft_{r+1} \Psf_{r+1} - \upphi_r^+
  \Psf_r \Psft_r + m \upphi_r^+ \upphi_r^-)$.  Integrating out
  $\upphi_r^\pm$, we get the quartic term.  From the point of view of
  the brane construction discussed in section~\ref{sec:brane}, we
  imagine the situation in which $\mathrm{NS5}_r$ and
  $\mathrm{NS5}_{r+1}$ are almost orthogonal but not quite.  The
  adjoint chiral multiplets $\upphi_r^\pm$ correspond to the positions
  of D2-branes along $\R^2_{\pm\hbar}$.}

The field content of the theory can be encoded in a quiver diagram.
This is the same quiver as the one that specifies a weight in the
tensor product of $L$ Verma modules of $\glf(m|n)$.  Here is the
quiver for the now-familiar $\glf(3|2)$ example, with the arrows
labeled with the corresponding chiral multiplets:
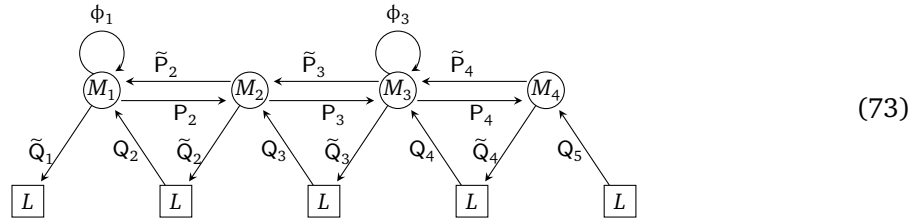
\begin{equation}
  \begin{tikzpicture}[scale=2]
    \draw[q->, shorten >=9pt] (1.37,0.56)
    -- node[above=-2pt]{$\Psft_2$} (0.5,0.56);
    \draw[q->, shorten >=9pt] (0.63,0.43)
    -- node[below=-2pt]{$\Psf_2$} (1.5,0.43);

    \draw[q->, shorten >=9pt] (2.37,0.56)
    -- node[above=-2pt]{$\Psft_3$} (1.5,0.56);
    \draw[q->, shorten >=9pt] (1.63,0.43)
    -- node[below=-2pt]{$\Psf_3$} (2.5,0.43);

    \draw[q->, shorten >=9pt] (3.37,0.56)
    -- node[above=-2pt]{$\Psft_4$} (2.5,0.56);
    \draw[q->, shorten >=9pt] (2.63,0.43)
    -- node[below=-2pt]{$\Psf_4$} (3.5,0.43);

    \draw[q->] (0.5,0.6) arc (270:-60:0.15) node[shift={(-4pt,22pt)}]{$\upphi_1$};
    \draw[q->] (2.5,0.6) arc (270:-60:0.15) node[shift={(-4pt,22pt)}]{$\upphi_3$};

    \draw[q->, shorten >=9pt] (0.5,0.5)
    -- node[left, shift={(0pt,-2pt)}]{$\Qsft_1$} (0,-0.25);
    \draw[q->, shorten >=9pt] (1,-0.25)
    -- node[left, shift={(4pt,-2pt)}]{$\Qsf_2$} (0.5,0.5);

    \draw[q->, shorten >=9pt] (1.5,0.5)
    -- node[left, shift={(0pt,-2pt)}]{$\Qsft_2$} (1,-0.25);
    \draw[q->, shorten >=9pt] (2,-0.25)
    -- node[left, shift={(4pt,-2pt)}]{$\Qsf_3$} (1.5,0.5);

    \draw[q->, shorten >=9pt] (2.5,0.5)
    -- node[left, shift={(0pt,-2pt)}]{$\Qsft_3$} (2,-0.25);
    \draw[q->, shorten >=9pt] (3,-0.25)
    -- node[left, shift={(4pt,-2pt)}]{$\Qsf_4$} (2.5,0.5);

    \draw[q->, shorten >=9pt] (3.5,0.5)
    -- node[left, shift={(0pt,-2pt)}]{$\Qsft_4$} (3,-0.25);
    \draw[q->, shorten >=9pt] (4,-0.25)
    -- node[left, shift={(4pt,-2pt)}]{$\Qsf_5$} (3.5,0.5);

    \node[gnode] at (0.5,0.5) {$M_1$};
    \node[gnode] at (1.5,0.5) {$M_2$};
    \node[gnode] at (2.5,0.5) {$M_3$};
    \node[gnode] at (3.5,0.5) {$M_4$};

    \node[fnode] at (0,-0.25) {$L$};
    \node[fnode] at (1,-0.25) {$L$};
    \node[fnode] at (2,-0.25) {$L$};
    \node[fnode] at (3,-0.25) {$L$};
    \node[fnode] at (4,-0.25) {$L$};
\end{tikzpicture}
\end{equation}
A circle node labeled $M$ is a $\U(M)$ gauge group.  The theory has
$m + n$ copies of $\U(L)$ flavor groups, denoted here by square nodes.
We name them $\U(L)_1$, $\U(L)_2$, $\dotsc$, $\U(L)_{m+n}$ from left
to right.  The terms in the superpotential~\eqref{eq:W} correspond to
closed paths of length three and four in the quiver.

Apart from the $\U(L)^{m+n}$ flavor symmetry, the theory has an
important $\U(1)$ global symmetry preserved by the superpotential.  We
call it $\U(1)_{\hbar}$.  The charges of the chiral multiplets under
$\U(1)_{\hbar}$ are
\begin{align}
  \upphi_r&\colon 2(-1)^{[r]} \,,
  \\
  \Psf_i&\colon -(-1)^{[i]} \,,
  \\
  \Psft_i&\colon -(-1)^{[i]} \,,
  \\
  \Qsf_i&\colon \frac12 (-1)^{[i]} \,,
  \\
  \Qsft_i&\colon \frac12 (-1)^{[i]} \,.
\end{align}
For $\upphi_r$ and $\Psf_i$, $\Psft_i$, their charges coincide with
the corresponding Cartan matrix elements.

Most parameters of the spin chain correspond in the gauge theory to
the twisted masses with respect to the global symmetry
$\U(L)^{m+n} \times \U(1)_{\hbar}$.  We can turn on the twisted masses
as follows.  First, we couple vector multiplets for
$\U(L)^{m+n} \times \U(1)_{\hbar}$ to the chiral multiplets and gauge
the global symmetry.  Then, we give vacuum expectation values to the
adjoint scalar fields in these vector multiplets.  Finally, we take
the limit in which the gauge couplings for
$\U(L)^{m+n} \times \U(1)_{\hbar}$ go to zero, thereby freezing the
vector multiplets just added.  The vacuum expectation values of the
scalar fields appear as complex mass parameters for the chiral
multiplets, which are the twisted masses in question.

To the scalar field for $\U(L)_i$, we give the vacuum expectation
value
\begin{equation}
  \diag(\mu_i^1, \dotsc, \mu_i^L) \,.
\end{equation}
This yields twisted masses $-\mu_i^\ell$ to $\Qsf_i$ and
$+\mu_i^\ell$ to $\Qsft_i$.  To the scalar field for
$\U(1)_{\hbar}$, we give the vacuum expectation value $\hbar/2$.  This
yields a twisted mass $q\hbar/2$ to a chiral multiplet that has charge
$q$ under $\U(1)_{\hbar}$.  To summarize, the twisted masses of the
chiral multiplets are
\begin{align}
  (\upphi_r)^{a_r}{}_{b_r} &\colon (-1)^{[r]} \hbar\,,
  \\
  (\Psf_i)^{a_i}{}_{b_{i-1}} &\colon  -\frac12 (-1)^{[i]} \hbar \,,
  \\
  (\Psft_i)^{a_{i-1}}{}_{b_i} &\colon  -\frac12 (-1)^{[i]} \hbar \,,
  \\
  (\Qsf_i)^{a_{i-1}}{}_\ell &\colon
           -\mu_i^\ell + \frac14 (-1)^{[i]} \hbar \,,
  \\
  (\Qsft_i)^\ell{}_{a_i} &\colon
            +\mu_i^\ell + \frac14 (-1)^{[i]} \hbar \,.
\end{align}

Lastly, we need FI parameters and theta angles in order to account for
the twist parameters for the periodic boundary condition of the spin
chain.  To do so, from each vector multiplet $\Vsf_r$ we construct an
adjoint twisted chiral multiplet $\Upsigma_r$ whose lowest component
is the vector multiplet scalar $\sigma_r$.  Then, we choose
complexified FI parameters
\begin{equation}
  t_1, \dotsc, t_{m+n-1} \in \C/2\pi\iu\Z \,,
\end{equation}
and turn on the twisted superpotential
\begin{equation}
  \label{eq:Wt-tree}
  \Wt = -\sum_{r=1}^{m+n-1} t_r \tr\Upsigma_r \,.
\end{equation}
The real and imaginary parts of $t_r$ are related to the FI parameter
$r_r$ and the theta angle $\theta_r$ for $\U(M_r)$ as
$t_r = r_r - \iu\theta_r$.

\subsubsection{Vacuum equations}

We are interested in the vacua of this theory when it is defined on a
periodic space.  For $\CN = (2,2)$ supersymmetric gauge theories, the
exact low-energy effective descriptions are known and we can use these
descriptions to determine their vacua.  See~\cite{Hori:2003ic} for
detailed discussions in the abelian case.  Nonabelian examples are
treated in \cite{Hanany:1997vm}.

Let $\Upsigma_r^{a_r}$, $a_r = 1$, $\dotsc$, $M_r$, be the diagonal
components of $\Upsigma_r$, and let $\Upsigma$ and $\sigma$
collectively denote $\{\Upsigma_r^{a_r}\}$ and their scalar components
$\{\sigma_r^{a_r}\}$, respectively.

If the fields $\sigma$ take generic large values and are slowly
varying, the chiral multiplets and the off-diagonal components of the
vector multiplets (or equivalently the corresponding twisted chiral
multiplets) can be considered as having large masses due to higgsing.
Their masses are
\begin{align}
  (\Upsigma_r)^{a_r}{}_{b_r} &\colon
  \sigma_r^{a_r} - \sigma_r^{b_r} \,,
  \\    
  (\upphi_r)^{a_r}{}_{b_r} &\colon
  \sigma_r^{a_r} - \sigma_r^{b_r} + (-1)^{[r]} \hbar \,,
  \\
  (\Psf_i)^{a_i}{}_{b_{i-1}}&\colon
  \sigma_i^{a_i} - \sigma_{i-1}^{b_{i-1}} - \frac12 (-1)^{[i]} \hbar \,,
  \\
  (\Psft_i)^{a_{i-1}}{}_{b_i}&\colon
  \sigma_{i-1}^{a_{i-1}} - \sigma_i^{b_i} - \frac12 (-1)^{[i]} \hbar \,,
  \\
  (\Qsf_i)^{a_{i-1}}{}_\ell&\colon
  \sigma_{i-1}^{a_{i-1}} - \mu_i^\ell + \frac14 (-1)^{[i]} \hbar \,,
  \\
  (\Qsft_i)^\ell{}_{a_i}&\colon
  \mu_i^\ell - \sigma_i^{a_i} + \frac14 (-1)^{[i]} \hbar \,,
\end{align}
where $a_r$, $b_r$ are indices for the $\U(M_r)$ gauge group factor.
After integrating out these heavy fields, we are left with an
effective description of the theory that involves only $\Upsigma$.

The effective theory is determined solely by a single holomorphic
function of $\sigma$, the \emph{effective twisted superpotential}
$\Wt_{\mathrm{eff}}$, and it can be calculated exactly at one-loop
order.  Integrating out a chiral multiplet whose mass due to the
higgsing is $m(\sigma)$ contributes to $\Wt_{\mathrm{eff}}(\sigma)$ by
the term
\begin{equation}
  -m(\sigma) \bigl(\log m(\sigma) - 1\bigr) \,.
\end{equation}
An off-diagonal component of a vector multiplet also contributes in
the same way~\cite{Hori:2011pd}.  Integrating out high-energy modes of
$\Upsigma$ does not alter the form of $\Wt_{\mathrm{eff}}$
\cite{Hori:2003ic}.

The vacua of the theory are the solutions of the vacuum equations
\begin{equation}
  \exp\biggl(\frac{\del\Wt_{\mathrm{eff}}(\sigma)}{\del\sigma_r^{a_r}}\biggr)
  = 1 \,,
  \quad
  a_r = 1, \dotsc, M_r \,,
  \quad
  r = 1, \dotsc, m+n-1 \,.
\end{equation}
These equations are invariant under shifts of the exponent by integer
multiples of $2\pi\iu$, reflecting the fact that the imaginary part of
the exponent is the effective theta angle.  In the case at hand, the
vacuum equations read
\begin{multline}
  e^{t_r}
  (-1)^{M_r+1}
  \Biggl(\prod_{b_r=1}^{M_r}
  \frac{\sigma_r^{a_r} - \sigma_r^{b_r} + (-1)^{[r]} \hbar}
       {\sigma_r^{b_r} - \sigma_r^{a_r} + (-1)^{[r]} \hbar}
  \Biggr)^{\delta_{[r],[r+1]}}
  \prod_{\ell=1}^L
  \frac{\sigma_r^{a_r} - \mu_{r+1}^\ell + \frac14 (-1)^{[r+1]} \hbar}
       {\mu_r^\ell - \sigma_r^{a_r} + \frac14 (-1)^{[r]} \hbar}
  \\
  \times
  \prod_{b_{r-1}=1}^{M_{r-1}}
  \frac{\sigma_r^{a_r} - \sigma_{r-1}^{b_{r-1}} - \frac12 (-1)^{[r]} \hbar}
       {\sigma_{r-1}^{b_{r-1}} - \sigma_r^{a_r} - \frac12 (-1)^{[r]} \hbar}
  \prod_{b_{r+1}=1}^{M_{r+1}}
  \frac{\sigma_r^{a_r} - \sigma_{r+1}^{b_{r+1}} - \frac12 (-1)^{[r+1]} \hbar}
       {\sigma_{r+1}^{b_{r+1}} - \sigma_r^{a_r} - \frac12 (-1)^{[r+1]} \hbar}
  =
  1 \,,
  \\
  a_r = 1, \dotsc, M_r \,,
  \quad
  r = 1, \dotsc, m+n-1 \,,
\end{multline}
with $M_0 = M_{m+n} = 0$.  The factor $e^{t_r}$ comes from the
tree-level twisted superpotential~\eqref{eq:Wt-tree}, and the factor
$(-1)^{M_r+1}$ comes from the off-diagonal components of the vector
multiplets.  Using the Cartan matrix~\eqref{eq:a_rs}, we can rewrite
these equations as
\begin{equation}
  \label{eq:vacuum}
  e^{\tau_r}
  \prod_{s=1}^{m+n-1} \prod_{b_s = 1}^{M_s}
  \frac{\sigma_r^{a_r} - \sigma_s^{b_s} + \frac12 a_{rs} \hbar}
       {\sigma_r^{a_r} - \sigma_s^{b_s} - \frac12 a_{rs} \hbar}
  \prod_{\ell=1}^L
  \frac{\sigma_r^{a_r} - \mu_{r+1}^\ell + \frac14 (-1)^{[r+1]} \hbar}
       {\sigma_r^{a_r} - \mu_r^\ell - \frac14 (-1)^{[r]} \hbar}
  =
  (-1)^{\delta_{[r],[r+1]}} \,,
\end{equation}
where
\begin{equation}
  \tau_r = t_r + \iu\pi \bigl((1 - \delta_{[r],[r+1]})(M_r + 1) + M_{r-1} + M_{r+1} + L\bigr) \,.
\end{equation}

\subsubsection{Twisted chiral ring}

The above two-dimensional gauge theory has $\CN = (2,2)$
supersymmetry, generated by four supercharges $Q_\pm$, $\Qb_\pm$.
Under a vector $\U(1)$ R-symmetry, $\Qb_\pm$ and $Q_\pm$ have charges
$+1$ and $-1$, while under an axial $\U(1)$ R-symmetry, $\Qb_+$, $Q_-$
have charge $+1$ and $\Qb_-$, $Q_+$ have charge $-1$.  The theory is
unitary and the supercharges satisfy the reality conditions
$Q_\pm^* = \Qb_\pm$.  It turns out that the theory has unbroken vector
$\U(1)$ R-symmetries, and this fact implies the absence of certain
central charges $Z$, $Z^*$ in the $\CN = (2,2)$ supersymmetry algebra.

The linear combination $Q = \Qb_+ + Q_-$ satisfies
\begin{equation}
  Q^2 = \Zt \,,
\end{equation}
with $\Zt$ being another central charge.  In the gauge theory that we
are considering,
\begin{equation}
  \Zt
  =
  \hbar F_{\hbar} + \sum_{i=1}^{m+n} \sum_{\ell=1}^L \mu_i^\ell F_i^\ell \,,
\end{equation}
where $F_{\hbar}$ is the generator of $\U(1)_{\hbar}$ and $F_i^\ell$,
$\ell = 1$, $\dotsc$, $L$, are the generators of the maximal torus of
$\U(L)_i$.  Since $Q$ squares to zero in the sector in which
$\Zt = 0$, we can define the $Q$-cohomology in the space of
$\Zt$-invariant states and in the algebra of $\Zt$-invariant
operators.  The subalgebra of the latter consisting of the elements
represented by local operators is the \emph{twisted chiral ring} of
the theory.

The $\CN = (2,2)$ supersymmetry algebra with $Z = Z^* = 0$ says that
the Hamiltonian $H$ satisfies $\{Q,Q^*\} = 2H$.  Therefore, $H$ is
positive semidefinite and vacuum states are annihilated by $Q$ and
$Q^*$.  In particular, vacua have $\Zt = 0$.  According to Hodge
theory, the $Q$-cohomology of states is isomorphic to the space of
vacua.

Besides the Hamiltonian, the momentum $P$ is also $Q$-exact:
$\{Q, Q_+ - \Qb_-\} = 2P$.  It follows that translations act trivially
in the $Q$-cohomology.  In particular, the twisted chiral ring is
commutative since we can switch the order of two local $Q$-cohomology
classes along the time axis by moving them around continuously inside
the two-dimensional spacetime.

In fact, for the theory considered here, not just the Hamiltonian and
the momentum but the entire stress tensor is $Q$-exact.  As a
consequence, the $Q$-cohomology of states and the twisted chiral ring
are topological, and there is a state--operator correspondence between
them: the two are isomorphic as vector spaces.

Being topological, the twisted chiral ring can be computed in the
effective theory.  As a vector space, it is the space of polynomials
in the scalar fields $\{\sigma_r^{a_r}\}$ modulo the action of the
Weyl group of the gauge group and the relations imposed by the vacuum
equations~\eqref{eq:vacuum}.  On a vacuum state, specified by a
solution of the vacuum equations, an element of the twisted chiral
ring acts by evaluation on the solution.  Therefore, a vacuum is a
simultaneous eigenstate of the elements of the twisted chiral ring.

\subsection{The correspondence}

The Bethe equations~\eqref{eq:Bethe} and the vacuum
equations~\eqref{eq:vacuum} coincide under the identification
\begin{equation}
  \label{eq:mu-lambda}
  \mu_i^\ell
  =
  \zeta^\ell
  - (-1)^{[i]} \biggl(\lambda^\ell_i + \frac14\biggr) \hbar
  + \frac12 c_i \hbar \,,
\end{equation}
together with the obvious identification between parameters for which
we have been using the same symbols.  (We are measuring twisted masses
in an appropriate unit so that they are numbers here.)

Thus, the vacua of the gauge theory are identified with the Bethe
vectors of the corresponding magnon sector of the spin chain.  Under
this identification, the elements of the twisted chiral ring are
identified with the commuting conserved charges of the spin chain.
This is the statement of the Bethe/gauge correspondence.

One conclusion we can immediately draw from the Bethe/gauge
correspondence is that the gauge theory has no supersymmetric vacuum
unless the assignment $(M_1, \dotsc, M_{m+n-1})$ of the ranks of the
unitary gauge groups corresponds to a weight of
$M(\lambda^1) \otimes \dotsb \otimes M(\lambda^L)$.  For example, for
$(m|n) = (1|1)$, supersymmetry is broken if and only if
\begin{equation}
  M_1 > L
\end{equation}
because the fermionic lowering operator can be applied at most $L$
times, at which point all spin sites are occupied by fermionic
excitations.  This is consistent with the known result that
supersymmetry is broken in a two-dimensional $\CN = (2,2)$
supersymmetric gauge theory with gauge group $\U(M)$, $L_f$
fundamental chiral multiplets and $L_a$ antifundamental chiral
multiplets if $M > \max(L_f, L_a)$ \cite{Benini:2014mia}.

\section{String theory realization of the Bethe/gauge correspondence}
\label{sec:string-Bethe/gauge}

Although we have presented the Bethe/gauge correspondence for
noncompact rational $\glf(m|n)$ superspin chains, we have not yet
explained why such a correspondence should exist.  In this section we
provide an explanation using string theory.

We will discuss how to construct the vacua of the relevant gauge
theories using branes, and how to map these brane configurations to
other ones that realize configurations of line defects in
four-dimensional Chern--Simons theory with gauge group $G = \GL(m|n)$.
The emergence of integrable spin chains is understood naturally in the
latter setup.

Moreover, we will give an explanation of fermionic dualities known in
the literature of integrable superspin chains.

\subsection{Brane construction of the gauge theory vacua}
\label{sec:brane}

The gauge theory and its vacua described in
sections~\ref{sec:gauge-side} can be constructed with branes in string
theory.  In fact, we have already represented the corresponding Bethe
vectors graphically in a way that makes the connection to the brane
construction transparent.

\subsubsection{Semiclassical type IIA configuration}

The construction uses NS5-branes
\begin{equation}
  \mathrm{NS5}_i \,,
  \quad
  i = 1, \dotsc, m+n \,,
\end{equation}
D4-branes
\begin{equation}
  \mathrm{D4}_i^\ell \,,
  \quad
  i = 1, \dotsc, m+n,
  \quad \ell = 1, \dotsc, L \,,
\end{equation}
and D2-branes
\begin{equation}
  \mathrm{D2}_r^{a_r} \,,
  \quad
  r = 1, \dotsc, m+n-1,
  \quad a_r = 1, \dotsc, M_r \,,
\end{equation}
in type IIA superstring theory.  The indices $i$ and $r$ are
$\Z_2$-graded as before.

First, let us consider the case in which all FI parameters are zero.
In this case, a semiclassical brane configuration for a vacuum state
of the gauge theory is summarized as follows:
\begin{equation}
  \label{eq:brane-gauge}
  \begin{array}{r@{\colon\ }c@{\ \times\ }c@{\ \times\ }c@{\ \times\ }c@{\ \times\ }c@{\ \times\ }c@{\ \times\ }c}
    \mathrm{Spacetime}
    & \R & \check S^1 & \C & \R_X & \R_Y
    & \R^2_{+\hbar} & \R^2_{-\hbar}
    \\
    \mathrm{NS5}_i \ ([i]=\bar0)
    & \R & \check S^1 & \C
    & \{X_i\} & \{Y_i\} & \R^2_{+\hbar} & \{0\}
    \\
    \mathrm{NS5}_i \ ([i]=\bar1)
    & \R & \check S^1 & \C
    & \{X_i\} & \{Y_i\} & \{0\} & \R^2_{-\hbar}
    \\
    \mathrm{D4}_i^\ell \ ([i]=\bar0)
    & \R & \check S^1 & \{\mu_i^\ell\}
    & \{X_i\} & [Y_i, \infty) & \R^2_{+\hbar} & \{0\}
    \\
    \mathrm{D4}_i^\ell \ ([i]=\bar1)
    & \R & \check S^1 & \{\mu_i^\ell\}
    & \{X_i\} & [Y_i, \infty) & \{0\} & \R^2_{-\hbar}
    \\
    \mathrm{D2}_r^{a_r}
    & \R & \check S^1 & \{\sigma_r^{a_r}\}
    & [X_r,X_{r+1}] & \{\Yt_r\} & \{0\} & \{0\}
  \end{array}
\end{equation}
In the spacetime, $\check S^1$ is a circle, $\R_X$ and $\R_Y$ are
lines, and $\R^2_{+\hbar}$ and $\R^2_{-\hbar}$ are planes.
Corresponding to the vanishing FI parameters, we have
\begin{equation}
  Y_i = \Yt_r = 0
\end{equation}
for all $i$ and $r$.

All of these branes wrap the cylinder $\R \times \check S^1$, which is
the spacetime of the gauge theory.  The branes $\mathrm{NS5}_i$ and
$\mathrm{D4}_i^\ell$ extend over $\R^2_{(-1)^{[i]}\hbar}$ and are
located at the origin of $\R^2_{(-1)^{[i]+\bar1}\hbar}$.  Moreover,
$\mathrm{NS5}_i$ extends over $\C$, whereas $\mathrm{D4}_i^\ell$
extends along $\R_X$ and ends on $\mathrm{NS5}_i$.  Along $\R_X$, the
NS5-branes are ordered according to the ordered basis of $\C^{m|n}$
specifying the Dynkin diagram of $\glf(m|n)$:
\begin{equation}
  \label{eq:X-ordering}
  X_1 < X_2 < \dotsb < X_{m+n} \,.
\end{equation}

The graphical representation~\eqref{eq:Bethe-vector-diagram} for a
Bethe vector can be reinterpreted as the above brane configuration.
In that picture, the vertical direction is the direction of $\C$ and
the horizontal direction is $\R_X$; the vertical lines are the
NS5-branes.  The diagonal lines ending on the vertical ones are the
D4-branes.  The horizontal line segments between the $r$th and
$(r+1)$st vertical lines are the D2-branes $\mathrm{D2}_r^{a_r}$,
$a_r = 1$, $\dotsc$, $M_r$, suspended between the two NS5-branes:
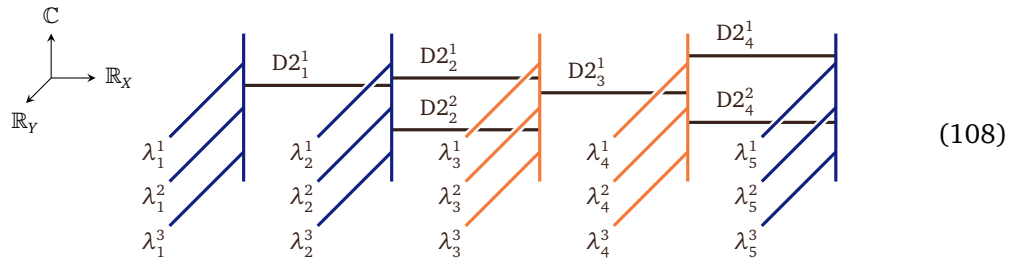
\begin{equation}
  \begin{tikzpicture}[scale=2]
    \begin{scope}[shift={(-1.3,0.7)}, scale=0.3]
      \draw[->, >=stealth] (0,0) -- (1,0) node[right] {$\R_X$};
      \draw[->, >=stealth] (0,0) -- (0,1) node[above] {$\C$};
      \draw[->, >=stealth] (0,0) -- (45:-0.8) node[below] {$\R_Y$};
    \end{scope}

    \begin{scope}[shift={(0,0.15)}]
      \draw[ZERO, very thick] (0,0.5) -- node[above left=-2pt] {$\mathrm{D2}_1^1$} (1,0.5);
    \end{scope}

    \begin{scope}[shift={(2,0.1)}]
      \draw[ZERO, very thick] (0,0.5) -- node[above left=-2pt] {$\mathrm{D2}_3^1$} (1,0.5);
    \end{scope}

    \begin{scope}[shift={(3,0)}]
      \draw[ZERO, very thick] (0,0.4) -- node[above left=-2pt] {$\mathrm{D2}_4^2$} (1,0.4);
      \draw[ZERO, very thick] (0,0.85) -- node[above left=-2pt] {$\mathrm{D2}_4^1$} (1,0.85);
    \end{scope}

    \begin{scope}[shift={(1,-0.1)}]
      \draw[ZERO, very thick] (0,0.8) -- node[above left=-2pt] {$\mathrm{D2}_2^1$} (1,0.8);
      \draw[ZERO, very thick] (0,0.45) -- node[above left=-2pt] {$\mathrm{D2}_2^2$} (1,0.45);
    \end{scope}

    \draw[white, line width=3pt] (0.5,0.3) -- (1,0.8);
    \draw[white, line width=3pt] (1.5,0) -- (2,0.5);
    \draw[white, line width=3pt] (1.5,0.3) -- (2,0.8);
    \draw[white, line width=3pt] (2.5,0.3) -- (3,0.8);
    \draw[white, line width=3pt] (3.5,-0.3) -- (4,0.2);
    \draw[white, line width=3pt] (3.5,0) -- (4,0.5);
    \draw[white, line width=3pt] (3.5,0.3) -- (4,0.8);

    \draw[PLUS, very thick] (0,0) -- (0,1);
    \draw[PLUS, very thick] (1,0) -- (1,1);
    \draw[MINUS, very thick] (2,0) -- (2,1);
    \draw[MINUS, very thick] (3,0) -- (3,1);
    \draw[PLUS, very thick] (4,0) -- (4,1);

    \begin{scope}[shift={(0,-0.3)}]
      \draw[PLUS, very thick] (-0.5,0)
      node[below left=-4pt, ZERO] {$\lambda^3_1$} -- (0,0.5);
      \draw[PLUS, very thick] (0.5,0)
      node[below left=-4pt, ZERO] {$\lambda^3_2$} -- (1,0.5);
      \draw[MINUS, very thick] (1.5,0)
      node[below left=-4pt, ZERO] {$\lambda^3_3$} -- (2,0.5);
      \draw[MINUS, very thick] (2.5,0)
      node[below left=-4pt, ZERO] {$\lambda^3_4$} -- (3,0.5);
      \draw[PLUS, very thick] (3.5,0)
      node[below left=-4pt, ZERO] {$\lambda^3_5$} -- (4,0.5);
    \end{scope}

    \begin{scope}[shift={(0,0)}]
      \draw[PLUS, very thick] (-0.5,0)
      node[below left=-4pt, ZERO] {$\lambda^2_1$} -- (0,0.5);
      \draw[PLUS, very thick] (0.5,0)
      node[below left=-4pt, ZERO] {$\lambda^2_2$} -- (1,0.5);
      \draw[MINUS, very thick] (1.5,0)
      node[below left=-4pt, ZERO] {$\lambda^2_3$} -- (2,0.5);
      \draw[MINUS, very thick] (2.5,0)
      node[below left=-4pt, ZERO] {$\lambda^2_4$} -- (3,0.5);
      \draw[PLUS, very thick] (3.5,0)
      node[below left=-4pt, ZERO] {$\lambda^2_5$} -- (4,0.5);
    \end{scope}

    \begin{scope}[shift={(0,0.3)}]
      \draw[PLUS, very thick] (-0.5,0)
      node[below left=-4pt, ZERO] {$\lambda^1_1$} -- (0,0.5);
      \draw[PLUS, very thick] (0.5,0)
      node[below left=-4pt, ZERO] {$\lambda^1_2$} -- (1,0.5);
      \draw[MINUS, very thick] (1.5,0)
      node[below left=-4pt, ZERO] {$\lambda^1_3$} -- (2,0.5);
      \draw[MINUS, very thick] (2.5,0)
      node[below left=-4pt, ZERO] {$\lambda^1_4$} -- (3,0.5);
      \draw[PLUS, very thick] (3.5,0)
      node[below left=-4pt, ZERO] {$\lambda^1_5$} -- (4,0.5);
    \end{scope}
  \end{tikzpicture}
\end{equation}

Strings stretched between $\mathrm{D2}_r^{a_r}$ and
$\mathrm{D2}_r^{b_r}$ produce the components $(\Vsf_r)^{a_r}{}_{b_r}$
and $(\Vsf_r)^{b_r}{}_{a_r}$ of the vector multiplet $\Vsf_r$ for the
gauge group factor $\U(M_r)$.  The D2-branes can move along $\C$, and
the position of $\mathrm{D2}_r^{a_r}$ in $\C$ determines the scalar
field $\sigma_r^{a_r}$ of the twisted chiral multiplet
$\Upsigma_r^{a_r}$.

If $[r] = [r+1]$, the D2-branes suspended between $\mathrm{NS5}_r$ and
$\mathrm{NS5}_{r+1}$ can also move along $\R^2_{(-1)^{[r]} \hbar}$,
over which the two NS5-branes extend.  Accordingly, in this case
strings with both ends attached on these D2-branes give rise to an
additional chiral multiplet, namely the adjoint chiral multiplet
$\upphi_r$.  The positions of the D2-branes in
$\R^2_{(-1)^{[r]} \hbar}$ are the diagonal components of the scalar
field in $\upphi_r$.

Strings stretched between $\mathrm{D2}_{r-1}^{a_{r-1}}$ and
$\mathrm{D2}_r^{b_r}$ yield the components
$(\Psf_r)_{a_{r-1}}{}^{b_r}$ and $(\Psft_r)^{b_r}{}_{a_{r-1}}$ of the
bifundamental chiral multiplets between $\U(M_{r-1})$ and $\U(M_r)$.
Strings from $\mathrm{D2}_i^{a_i}$ to $\mathrm{D4}_i^\ell$ are
responsible for $(\Qsft_i)^\ell{}_{a_i}$ and those from
$\mathrm{D4}_i^\ell$ to $\mathrm{D2}_{i-1}^{a_{i-1}}$ give
$(\Qsf_i)^{a_{i-1}}{}_\ell$.

Various parameters of the gauge theory are identified as follows.  The
gauge coupling for $\U(M_r)$ is $\sqrt{g_s/l_s(X_{r+1} - X_r)}$,
where $g_s$ is the string coupling and $l_s$ is the string length.
The position of $\mathrm{D4}_i^\ell$ in $\C$ determines the twisted
mass $\mu_i^\ell$.  The FI parameter for $\U(M_r)$ is
$(Y_{r+1} - Y_r)/g_s l_s$, while the theta angle is given by the
difference in the periodic scalars on $\mathrm{NS5}_r$ and
$\mathrm{NS5}_{r+1}$ (up to a shift by $\iu\pi$ which we will explain
shortly).  Since we are taking $Y_i = 0$ for all $i$, all FI
parameters are zero.  Introducing the twisted masses proportional to
$\hbar$ requires turning on a nontrivial B-field.  For the moment we
take $\hbar = 0$.

The rotation symmetry of the directions orthogonal to the D2-brane
worldvolumes becomes a global symmetry of the gauge theory.  The
rotation symmetry $\U(1)_\C$ of $\C$ is an axial R-symmetry, under
which the vector multiplet scalars have charge $2$.  The rotation
symmetries $\U(1)_{\R^2_{\pm\hbar}}$ of $\R^2_{\pm\hbar}$ are vector
R-symmetries.  The adjoint chiral multiplet $\upphi_r$ has charge
$(2,0)$ or $(0,2)$ under
$\U(1)_{\R^2_{+\hbar}} \times \U(1)_{\R^2_{-\hbar}}$, depending on
whether $[r] = \bar0$ or $\bar1$.

Now, let us turn on FI parameters by displacing the NS5-branes along
$\R_Y$ by different amounts.  As we vary their positions, the
D2-branes suspended between them get rotated in $\R_X \times \R_Y$ by
various angles.  Such a configuration no longer preserves
supersymmetry.  If the twisted masses are generic, the D2-branes
cannot stretch between D4-branes without breaking supersymmetry
either.  Moreover, if $M_r > L$ and $[r] \neq [r+1]$, suspending $M_r$
D2-branes between $\mathrm{NS5}_r$ and the $L$ D4-branes ending on
$\mathrm{NS5}_{r+1}$ (or between $\mathrm{NS5}_{r+1}$ and the $L$
D4-branes ending on $\mathrm{NS5}_r$) breaks supersymmetry by the
s-rule.  It appears that there are no supersymmetric vacua for generic
FI parameters, twisted masses and magnon numbers.

This analysis is semiclassical, however.  Quantum mechanically,
D4-branes bend NS5-branes on which they end and the conclusion is
altered.

\subsubsection{Lift to M-theory}

Important aspects of the quantum corrections to the above brane
configuration can be understood by uplift to M-theory.  The M-theory
spacetime contains an additional compact direction $S^1_{\mathrm{M}}$.
Let $\vartheta$ be its coordinate with period $2\pi$, and introduce a
complex coordinate $Y + \iu\vartheta$ for the cylinder
$\R_Y \times S^1_{\mathrm{M}}$.  Further introducing
$w = e^{-(Y + \iu\vartheta)}$, we map the cylinder to the punctured complex
plane $\C^\times$.

All NS5-branes and D4-branes are lifted to M5-branes in M-theory.  For
each~$i$, $\mathrm{NS5}_i$ and $\mathrm{D4}_i^\ell$, $\ell = 1$,
$\dotsc$, $L$, merge into a single M5-brane $\mathrm{M5}_i$, wrapping
a Riemann surface $\Sigma_i$ in $\C \times \C^\times$.  The D2-brane
$\mathrm{D2}_r^{a_r}$ is lifted to an M2-brane $\mathrm{M2}_r^{a_r}$
stretched between $\mathrm{M5}_r$ and $\mathrm{M5}_{r+1}$.  Hence, the
brane configuration in M-theory is as follows:
\begin{equation}
  \begin{array}{r@{\colon\ }c@{\ \times\ }c@{\ \times\ }c@{\ \times\ }c@{\ \times\ }c@{\ \times\ }c}
    \mathrm{Spacetime}
    & \R & \check S^1 & \C \times \C^\times & \R_X & \R^2_{+\hbar} & \R^2_{-\hbar}
    \\
    \mathrm{M5}_i \ ([i]=\bar0)
    & \R & \check S^1 & \Sigma_i & \{X_i\} & \R^2_{+\hbar} & \{0\}
    \\
    \mathrm{M5}_i \ ([i]=\bar1)
    & \R & \check S^1 & \Sigma_i & \{X_i\} & \{0\} & \R^2_{-\hbar}
    \\
    \mathrm{M2}_r^{a_r}
    & \R & \check S^1& \{(\sigma_r^{a_r}, w_r^{a_r})\}
   & [X_r, X_{r+1}] & \{0\} & \{0\}
  \end{array}
\end{equation}

In terms of the coordinates $(z,w)$ of $\C \times \C^\times$, the
Riemann surface $\Sigma_i$ is defined by the equation
\begin{equation}
  w = q_i \prod_{\ell=1}^L (z - \mu_i^\ell) \,,
\end{equation}
where $q_i$ is a constant.  The zero of $w$ at $z = \mu_i^\ell$
describes $\mathrm{D4}_i^\ell$, which extends to $+\infty$ in $\R_Y$.
If $\Sigma_r$ and $\Sigma_{r+1}$ intersect in $\C \times \C^\times$,
then $\mathrm{M2}_r^{a_r}$ can be placed at an intersection point so
that its worldvolume is orthogonal to the M5-branes.

Therefore, the M2-branes can be suspended between the M5-branes in a
manner that preserves supersymmetry if $\mathrm{M2}_r^{a_r}$ is placed
at $z = \sigma_r^{a_r}$ and the coordinate $\sigma_r^{a_r}$, for
each $r$ and $a_r$, satisfies the condition
\begin{equation}
  q_r \prod_{\ell=1}^L (\sigma_r^{a_r} - \mu_r^\ell)
  =
  w_r
  =
  q_{r+1} \prod_{\ell=1}^L (\sigma_r^{a_r} - \mu_{r+1}^\ell) \,.
\end{equation}
Comparing these equations with the vacuum equations~\eqref{eq:vacuum}
for $\hbar = 0$,
\begin{equation}
  e^{\tau_r}
  \prod_{\ell=1}^L
  \frac{\sigma_r^{a_r} - \mu_{r+1}^\ell}{\sigma_r^{a_r} - \mu_r^\ell}
  =
  (-1)^{\delta_{[r],[r+1]}} \,,
\end{equation}
we see that they coincide if we identify
\begin{equation}
  (-1)^{\delta_{[r],[r+1]}} e^{\tau_r}
  =
  \frac{q_{r+1}}{q_r} \,.
\end{equation}
Since $\vartheta_i = (-1)^{[i]} \arg q_i$ is the classical value of
the periodic scalar field on $\mathrm{NS5}_i$, the difference
$(-1)^{[r+1]} \vartheta_{r+1} - (-1)^{[r]} \vartheta_r$ is indeed
equal to $\theta_r$, up to a shift by $\iu\pi$.

\subsubsection{Turning on $\hbar$}

Finally, we explain how to make $\hbar \neq 0$.  The global symmetry
$\U(1)_{\hbar}$ is the antidiagonal subgroup of
$\U(1)_{\R^2_{+\hbar}} \times \U(1)_{\R^2_{-\hbar}}$.  To turn on the
twisted masses for $\U(1)_{\hbar}$, we follow the fluxtrap procedure
\cite{Hellerman:2011mv, Reffert:2011dp}.  This is done as follows.

First, we compactify $\C$ to a torus $T^2 \iso \C/(R_1\Z + \iu R_2\Z)$
in the type IIA setup and apply T-duality on both directions of $T^2$.
The D2-branes become D4-branes wrapping the dual torus
$\check T^2 \iso \C/(\check R_1\Z + \iu\check R_2\Z)$.  Next, we twist
the product between $\check T^2$ and
$\R^2_{+\hbar} \times \R^2_{-\hbar}$ by the action of $\U(1)_{\hbar}$.
(More precisely, we replace
$\check T^2 \times \R^2_{+\hbar} \times \R^2_{-\hbar}$ with the
quotient of $\C \times \R^2_{+\hbar} \times \R^2_{-\hbar}$ such that
translations on $\C$ by $\check R_1$ and $\iu \check R_2$ are
accompanied by the action of the elements $\exp(\Re\hbar)$ and
$\exp(\Im\hbar)$ of $\U(1)_{\hbar}$, respectively.)  Last, we apply
T-duality on $\check T^2$ and decompactify $T^2$ to $\C$ by taking
$R_1$, $R_2 \to \infty$.  This last T-duality yields a certain B-field
due to the twist in the product between $\check T^2$ and
$\R^2_{+\hbar} \times \R^2_{-\hbar}$ introduced earlier.

From the point of view of the gauge theory, the first step amounts to
lifting the two-dimensional theory on $\R \times \check S^1$ to a
four-dimensional theory on $\R \times \check S^1 \times \check T^2$.  Then,
the second step turns on a holonomy for the background gauge field for
$\U(1)_{\hbar}$.  The last step dimensionally reduces the
four-dimensional theory back to two dimensions.  Since the components
of a four-dimensional gauge field along $\check T^2$ become the complex
scalar field for the corresponding two-dimensional gauge field, this
procedure induces the twisted masses for $\U(1)_{\hbar}$.

\subsection{Four-dimensional Chern--Simons theory}

We can convert the brane configuration~\eqref{eq:brane-gauge} to a
configuration realizing line defects in four-dimensional Chern--Simons
theory with gauge group $\GL(m|n)$.  To do so, we apply T-duality
along $\check S^1$ and then S-duality.

Under these dualities, $\check S^1$ is mapped to the dual circle
$S^1$, the NS5-branes are mapped to D5-branes, the D4-branes are
mapped to D3-branes, and the D2-branes are mapped to F1-branes
(fundamental strings).  Thus we obtain the following type IIB setup:
\begin{equation}
  \label{eq:brane-4dCS}
  \begin{array}{r@{\colon\ }c@{\ \times\ }c@{\ \times\ }c@{\ \times\ }c@{\ \times\ }c@{\ \times\ }c@{\ \times\ }c}
    \mathrm{Spacetime}
    & \R & S^1 & \C & \R_X & \R_Y
    & \R^2_{+\hbar} & \R^2_{-\hbar}
    \\
    \mathrm{D5}_i \ ([i]=\bar0)
    & \R & S^1 & \C
    & \{X_i\} & \{Y_i\} & \R^2_{+\hbar} & \{0\}
    \\
    \mathrm{D5}_i \ ([i]=\bar1)
    & \R & S^1 & \C
    & \{X_i\} & \{Y_i\} & \{0\} & \R^2_{-\hbar}
    \\
    \mathrm{D3}_i^\ell \ ([i]=\bar0)
    & \R & \{y_i^\ell\} & \{\mu_i^\ell\}
    & \{X_i\} & [Y_i, \infty) & \R^2_{+\hbar} & \{0\}
    \\
    \mathrm{D3}_i^\ell \ ([i]=\bar1)
    & \R & \{y_i^\ell\} & \{\mu_i^\ell\}
    & \{X_i\} & [Y_i, \infty) & \{0\} & \R^2_{-\hbar}
    \\
    \mathrm{F1}_r^{a_r}
    & \R & \{\yt_r^{a_r}\} & \{\sigma_r^{a_r}\}
    & [X_r, X_{r+1}] & \{\Yt_r\} & \{0\} & \{0\}
  \end{array}
\end{equation}
The positions of the D3-branes and the F1-branes on $S^1$ are given by
the holonomies around $\check S^1$ of the gauge fields on their
counterparts in the type IIA setup.  The B-field inducing the twisted
masses for $\U(1)_{\hbar}$ becomes a Ramond--Ramond (RR) two-form field
in the new setup.

As in the two-dimensional theory discussed before, the vacuum sector
of the theory on the D5-branes, with $\R$ taken to be the time
direction, is captured by the cohomology with respect to a certain
supercharge.  This supercharge is dual to the supercharge $Q$ of the
two-dimensional theory, and we will use the same symbol to denote it.

We claim that the $Q$-invariant sector of the theory, which governs
the $Q$-cohomology, is equivalent to four-dimensional Chern--Simons
theory with gauge group $\GL(m|n)$.

Before demonstrating this equivalence, let us remark that related
brane constructions have appeared in the literature.%
\footnote{To relate these brane constructions to ours, we endow
  $\R^2_{+\hbar} \times \R^2_{-\hbar}$ with a Taub--NUT metric.  (The
  $Q$-invariant sector is independent of the choice of metric as long
  as it preserves the rotational symmetries of $\R^2_{+\hbar}$ and
  $\R^2_{-\hbar}$.)  If we regard the Taub--NUT space as a circle
  fibration over $\R^3$, then $\R^2_{+\hbar} \times \{0\}$ and
  $\{0\} \times \R^2_{-\hbar}$ are two semi-infinite cigar-shaped
  subspaces extending in the opposite directions such that their tips
  touch at the origin of $\R^3$; see \cite{Hellerman:2012zf}, appendix
  B.  T-duality in the direction of the circle fibers produces an
  NS5-brane which sits at the origin of $\R^3$ and extends in the
  directions transverse to the Taub--NUT space.  The D5-branes
  wrapping the two cigars are turned into D4-branes ending on the
  NS5-brane from two sides.  Considering the case with $n = 0$, we
  reproduce the construction of \cite{Ashwinkumar:2018tmm}.  The field
  theory counterpart of this T-duality was analyzed
  in~\cite{Ishtiaque:2020ufn}.  From the D4--NS5 brane configuration
  we obtain the brane configuration of \cite{Mikhaylov:2014aoa},
  roughly speaking, by further replacing $\C$ with a cylinder, taking
  T-duality in the circumferential direction of the cylinder, and
  decompactifying the dual cylinder.  (Such T-duality was considered
  in \cite{Yamazaki:2019prm}.)}
In \cite{Mikhaylov:2014aoa}, Mikhaylov and Witten gave a brane
construction of three-dimensional Chern--Simons theory with gauge
group $\GL(m|n)$, extending the construction for gauge group $\GL(m)$
given in~\cite{Witten:2011zz}.  Their construction uses $m$ D4-branes
and $n$ D4-branes ending on an NS5-brane from opposite sides.  The
supergroup Chern--Simons theory appears at the intersection of the
three kinds of branes.  In~\cite{Ashwinkumar:2018tmm}, a construction
of four-dimensional Chern--Simons theory with gauge group $\GL(m)$ was
proposed.  In this construction, $m$ D4-branes end on an NS5-brane.

\subsubsection{Case with $m = 0$ or $n = 0$}

In the case in which all D5-branes are of even type ($n = 0$) or of
odd type ($m=0$), the result just described was derived in
\cite{Costello:2018txb}.  Let us briefly review the derivation in
\cite{Costello:2018txb}.

For $n = 0$, the worldvolume theory on the D5-branes is a deformation
of six-dimensional $\CN = (1,1)$ super Yang--Mills theory with gauge
group $\U(m)$, placed on
$\R \times S^1 \times \C \times \R^2_{+\hbar}$.  The deformation is
what is often called an \emph{$\Omega$-deformation} and controlled by
$\hbar$: we have
\begin{equation}
  Q^2 = \hbar F_{\hbar} \,,
\end{equation}
where $F_{\hbar}$ is the generator of $\U(1)_{\hbar}$.  (Here we are
considering the situation in which there are no D3-branes, hence no
$\U(L)$ flavor symmetries.)  In the six-dimensional theory,
$\U(1)_{\hbar}$ is the antidiagonal subgroup of the rotation group
$\U(1)_{\R^2_{+\hbar}}$ on $\R^2_{+\hbar}$ and the subgroup
$\U(1)_{\R^2_{-\hbar}}$ of the R-symmetry group $\Spin(4)$ coming from
the rotation symmetry of $\R^2_{-\hbar}$.

Six-dimensional (Euclidean) $\CN = (1,1)$ super Yang--Mills theory on
$\R \times S^1 \times \C \times \R^2_{+\hbar}$ reduces to
two-dimensional $\CN = (8,8)$ super Yang--Mills theory on
$\R^2_{+\hbar}$ by dimensional reduction.  In the undeformed case
(when $\hbar = 0$), the supercharge $Q$ belongs to an $\CN = (2,2)$
subalgebra of the $\CN = (8,8)$ supersymmetry algebra.  Accordingly,
$\CN = (1,1)$ super Yang--Mills theory on
$\R \times S^1 \times \C \times \R^2_{+\hbar}$ may be thought of as an
$\CN = (2,2)$ supersymmetric gauge theory on $\R^2_{+\hbar}$, with
infinite-dimensional target space and gauge group.  The
$\Omega$-deformation of the six-dimensional theory induces an
$\Omega$-deformation of the two-dimensional theory.

In general, the $Q$-invariant sector of an $\Omega$-deformed
$\CN = (2,2)$ supersymmetric gauge theory on $\R^2$ is equivalent to a
zero-dimensional theory \cite{Yagi:2014toa, Luo:2014sva,
  Costello:2018txb}.  Let $\CG$ be the gauge group of the theory and
$\CG_\C$ be its complexification.  By $\CN = (2,2)$ supersymmetry, the
chiral multiplets take values in a K\"ahler manifold $\CX$ with
$\CG_\C$-action.  The superpotential is a $\CG_\C$-invariant
holomorphic function $\CW$ on $\CX$.  The path integral with insertion
of $Q$-invariant observables localizes to a $\CG_\C$-invariant
submanifold $\gamma$ of $\CX$.  This submanifold is essentially a
Lefschetz thimble: $\gamma$ is the union of all gradient flows
generated by the real part of $\CW/\hbar$, terminating on the
$\CG_\C$-orbit of a chosen critical point of $\CW$. (For simplicity,
we assume that the critical points of $\CW$ are nondegenerate up to
the $\CG_\C$-action.)  The localized path integral takes the form
\begin{equation}
  \label{eq:localized-PI}
  \int_{\gamma/\CG_\C} \exp\biggl(\frac{2\pi}{\hbar} \CW\biggr) \CO \,,
\end{equation}
where $\CO$ descends from the $Q$-invariant observables inserted in
the path integral.  This is the path integral for a zero-dimensional
gauge theory with gauge group $\CG_\C$ and target space $\gamma$.  The
action functional is $-2\pi\CW/\hbar$.

The remarkable aspect of this localization phenomenon is that the
gauge group gets complexified.  In the localization process, some
fermionic fields have zero modes.  They may be regarded as ghost
fields for partial gauge fixing that breaks $\CG_\C$ down to $\CG$.
Even though the action functional is holomorphic and its real part is
not bounded from below, the integral~\eqref{eq:localized-PI} can
converge since $\Re(\CW/\hbar)$ gets smaller and smaller along the
backward gradient flows in $\gamma$.

For the six-dimensional $\CN = (1,1)$ super Yang--Mills theory on
$\R \times S^1 \times \C \times \R^2_{+\hbar}$, the gauge group $\CG$
is the space of maps from $\R \times S^1 \times \C$ to $\U(m)$.  In
addition to the vector multiplet, the theory has three chiral
multiplets in the adjoint representation of $\CG$.  Their scalar
fields are $Q$-invariant and can be combined into a one-form on
$\R \times S^1 \times \C$:
\begin{equation}
  \label{eq:A-gl(m)}
  \CA
  =
  (A_x + \iu X) \rmd x
  + (A_y + \iu Y) \rmd y
  + A_\zb \, \rmd\zb \,.
\end{equation}
Here, $A_x$, $A_y$ are the components of the gauge field along
$\R \times S^1$, $A_\zb$ is the antiholomorphic component of the gauge
field along $\C$, and $X$, $Y$ are two of the four scalar fields of
the six-dimensional theory associated to motions along $\R_X$ and
$\R_Y$, respectively.  The superpotential is given by
\begin{equation}
  \CW
  =
  -\frac{\iu}{e^2} \int_{\R \times S^1 \times \C} \rmd z \wedge
  \tr\biggl(\CA \wedge \rmd \CA
  + \frac23 \CA \wedge \CA \wedge \CA\biggr) \,,
\end{equation}
where $e$ is the gauge coupling and $\tr$ is an invariant symmetric
bilinear form on the Lie algebra of $\U(m)$, which we can take to be
the trace in the defining representation.

According to the localization argument, the $\Omega$-deformation of
the six-dimensional theory is equivalent to a zero-dimensional gauge
theory.  This zero-dimensional theory has infinite-dimensional target
space and gauge group, and can be more naturally interpreted as a
four-dimensional gauge theory.  Its action is
\begin{equation}
  -\frac{2\pi\iu}{\hbar e^2} \int_{\Sigma \times \C} \rmd z \wedge
  \tr_{\C^m}\biggl(\CA^{00} \wedge \rmd \CA^{00}
  + \frac23 \CA^{00} \wedge \CA^{00} \wedge \CA^{00}\biggr)
  \,.
\end{equation}
Here, we have written the partial $\glf(m)$
connection~\eqref{eq:A-gl(m)} as $\CA^{00}$ to emphasize its place in
the Lie superalgebra $\glf(m|n)$ that will arise later.  This is the
action for four-dimensional Chern--Simons theory.  Thus we conclude
that the $Q$-invariant sector of the $\Omega$-deformed six-dimensional
$\CN = (1,1)$ super Yang--Mills theory on
$\R \times S^1 \times \C \times \R^2_{+\hbar}$ with gauge group
$\U(m)$ is equivalent to four-dimensional Chern--Simons theory on
$\R \times S^1 \times \C$ with gauge group $\GL(m)$ and coupling given
by $\hbar$.

Similarly, if we consider the case $m = 0$, the worldvolume theory on
the D5-branes is an $\Omega$-deformed six-dimensional $\CN = (1,1)$
super Yang--Mills theory on $\R \times S^1 \times \C \times \R^2_{-\hbar}$
with gauge group $\U(n)$.  Its $Q$-invariant sector is equivalent to
four-dimensional Chern--Simons theory on $\R \times S^1 \times \C$ with gauge
group $\GL(n)$ and action
\begin{equation}
  +\frac{2\pi\iu}{\hbar e^2} \int_{\R \times S^1 \times \C} \rmd z \wedge
  \tr_{\C^n}\biggl(\CA^{11} \wedge \rmd \CA^{11}
  + \frac23 \CA^{11} \wedge \CA^{11} \wedge \CA^{11}\biggr)
  \,,
\end{equation}
with the partial $\glf(n)$ connection $\CA^{11}$ defined in the same
way as $\CA^{00}$.

\subsubsection{Case with nonzero $m$ and $n$}

Let us turn to the case in which $m$ and $n$ are both nonzero.  In
this case, the two sets of D5-branes lead to two copies of
four-dimensional Chern--Simons theory on $\R \times S^1 \times \C$, one with
gauge group $\GL(m)$ and the other with gauge group $\GL(n)$, with
opposite couplings.  Arranging $\CA^{00}$ and $\CA^{11}$ into a
matrix
\begin{equation}
  \CA^0
  =
  \begin{pmatrix}
    \CA^{00} & 0 \\
    0 & \CA^{11}
  \end{pmatrix} \,,
\end{equation}
we can write the sum of their action functionals as
\begin{equation}
  \label{eq:4dCS-0}
  -\frac{2\pi\iu}{\hbar e^2} \int_{\R \times S^1 \times \C} \rmd z \wedge
  \str_{\C^{m|n}}\biggl(\CA^0 \wedge \rmd \CA^0
  + \frac23 \CA^0 \wedge \CA^0 \wedge \CA^0\biggr)
  \,.
\end{equation}

The two copies are coupled through strings stretched between the two
sets of D5-branes.  These strings produce a four-dimensional $\CN = 2$
hypermultiplet on $\R \times S^1 \times \C$ in the bifundamental
representation of $\GL(m) \times \GL(n)$.  It consists of bosonic
complex scalars
\begin{align}
  q &\in \Hom(\C^m,\C^n) \,,
  \\
  \qt^\dagger &\in \Hom(\C^m,\C^n) \,,
  \\
  q^\dagger &\in \Hom(\C^n,\C^m) \,,
  \\
  \qt &\in \Hom(\C^n,\C^m) \,,
\end{align}
and fermionic Weyl spinors
\begin{align}
  \psi &\in \Hom(\C^m,\C^n) \,,
  \\
  \psit^\dagger &\in \Hom(\C^m,\C^n) \,,
  \\
  \psi^\dagger &\in \Hom(\C^n,\C^m) \,,
  \\
  \psit &\in \Hom(\C^n,\C^m) \,.
\end{align}

In the absence of coupling to the two copies of four-dimensional
Chern--Simons theory, the bifundamental hypermultiplet preserves eight
supercharges.  The supercharge $Q$ is a linear combination of these
supercharges such that the generators of translations on
$\R \times S^1$ and antiholomorphic translations on $\C$ are
$Q$-exact.  By redefining fields if necessary, we can take $Q$ to be
the supercharge used in the holomorphic--topological twist studied in
\cite{Kapustin:2006hi}, with the parameter $t = i$.

It turns out that most of the action for the hypermultiplet is
$Q$-exact.  The remaining part of the action can be expressed in a
suggestive form.  Endow the cylinder $\R \times S^1$ with a complex
coordinate $w$, and define
\begin{align}
  \CA^{10}
  &=
    -\psit^\dagger_- \rmd w + \psi_- \rmd\wb
    - \frac12(\psit_+^\dagger - \psi_+) \rmd\zb \,,
  \\
  \CA^{01}
  &=
    \psi^\dagger_- \rmd w + \psit_- \rmd\wb
    + \frac12(\psit_+^\dagger + \psi^\dagger_+) \rmd\zb
\end{align}
and
\begin{align}
  c^{10} &= 4\iu q^\dagger \,,
  \\
  c^{01} &= 4\qt^\dagger \,,
  \\
  b^{10} &= \qt \,,
  \\
  b^{01} &= -\iu q \,,
  \\
  B^{10} &= \iu(\psit_+ - \psi^\dagger_+) \,,
  \\
  B^{01} &= -\iu(\psi_+ + \psit^\dagger_+) \,.
\end{align}
We introduce a matrix
\begin{equation}
  \CA^1
  =
  \begin{pmatrix}
    0 & \CA^{01} \\
    \CA^{10} & 0
  \end{pmatrix}
\end{equation}
and matrices $c^1$, $b^1$, $B^1$ defined likewise.  On these
matrices $Q$ acts by
\begin{align}
  Q \cdot \CA^1 &= -\rmd'c^1 \,,
  \\
  Q \cdot c^1 &= 0 \,,
  \\
  Q \cdot b^1 &= B^1 \,,
  \\
  Q \cdot B^1 &= 0 \,,
\end{align}
where
\begin{equation}
  \rmd'
  = \rmd - \rmd z \, \del_z
  = \rmd w \, \del_w + \rmd\wb \, \del_\wb + \rmd\zb \, \del_\zb \,.
\end{equation}
The non-$Q$-exact part of the action is
\begin{equation}
  \label{eq:Kapustin-action}
  -\frac{2\pi\iu}{\hbar e^2} \int_{\R \times S^1 \times \C} \rmd z \wedge
  \str_{\C^{m|n}}(\CA^1 \wedge \rmd\CA^1) \,.
\end{equation}
(Since this is quadratic in fermions, the prefactor is inessential.)

To describe the intersecting D5-branes, we couple this bifundamental
hypermultiplet to the two copies of four-dimensional Chern--Simons
theory by identifying the flavor groups $\GL(m)$ and $\GL(n)$ with the
gauge groups of the latter.  Concretely, we replace the de Rham
differential that appears in the above formulas with the
gauge-covariant differential
\begin{equation}
  \rmd_{\CA^0} = \rmd + \CA^0 \,.
\end{equation}
Thus, the action of $Q$ on the fields is modified to
\begin{align}
  Q \cdot \CA^1 &= -\rmd'_{\CA^0} c^1 \,,
  \\
  Q \cdot c^1 &= 0 \,,
  \\
  Q \cdot b^1 &= B^1 \,,
  \\
  Q \cdot B^1 &= 0 \,,
\end{align}
and the action functional for the bifundamental hypermultiplet becomes
\begin{equation}
  \label{eq:4dCS-1}
  -\frac{2\pi\iu}{\hbar e^2} \int_{\R \times S^1 \times \C} \rmd z \wedge
  \str_{\C^{m|n}}(\CA^1 \wedge \rmd_{\CA^0} \CA^1) \,.
\end{equation}

Combining $\CA^0$ and $\CA^1$ into a single matrix
\begin{equation}
  \CA
  =
  \CA^0 + \CA^1
  =
  \begin{pmatrix}
    \CA^{00} & \CA^{01} \\
    \CA^{10} & \CA^{11}
  \end{pmatrix}
  \,,
\end{equation}
we can write the total action, which is the sum of the
actions~\eqref{eq:4dCS-0} and~\eqref{eq:4dCS-1}, as
\begin{equation}
  -\frac{2\pi\iu}{\hbar e^2} \int_{\R \times S^1 \times \C} \rmd z \wedge
  \str_{\C^{m|n}}\biggl(\CA \wedge \rmd\CA
  + \frac23 \CA \wedge \CA \wedge \CA\biggr) \,.
\end{equation}
This is the action for four-dimensional Chern--Simons theory with
gauge group $\GL(m|n)$.

Before concluding that we have obtained the desired theory, we need to
solve two problems.  First, although the above action is invariant
under $\GL(m|n)$ gauge transformation, the gauge group of the theory
is still $\GL(m) \times \GL(n)$, not $\GL(m|n)$.  Second, the
gauge-invariant action~\eqref{eq:4dCS-1} for the bifundamental
hypermultiplet is not $Q$-invariant due to the coupling to $\CA^0$.
Its $Q$-variation gives
\begin{equation}
  -\frac{4\pi\iu}{\hbar e^2} \int_{\R \times S^1 \times \C} \rmd z \wedge
  \str_{\C^{m|n}}(\CA^1 \wedge [\CF^0, c^1]) \,,
\end{equation}
where $\CF^0$ is the curvature of $\CA^0$.

The two problems are solved simultaneously if we correct the
$Q$-action on $\CA^0$ and $B^1$ to
\begin{align}
  Q \cdot \CA^0 &= \{\CA^1, c^1\} \,,
  \\
  Q \cdot B^1 &= \frac12 [\{c^1,c^1\},b^1] \,.
\end{align}
With this modification, the $Q$-variation of the bosonic
action~\eqref{eq:4dCS-0} cancels that of the fermionic
action~\eqref{eq:4dCS-1}.  At the same time, $c^1$, $b^1$ and $B^1$
can now be interpreted as a ghost, an antighost and an auxiliary field
used in the Becchi--Rouet--Stora--Tyutin (BRST) procedure for partial
gauge fixing of $\GL(m|n)$ down to $\GL(m) \times \GL(n)$
\cite{Kapustin:2009cd}.

To make the last point more explicit, let us introduce a ghost $c^0$,
an antighost $b^0$ and an auxiliary field $B^0$ for gauge fixing of
$\GL(m) \times \GL(n)$.  The BRST charge $\QB$ acts on the fields by
\begin{align}
  \QB \cdot \CA^0 &= -\rmd_{\CA_0}' c^0 \,,
  \\
  \QB \cdot c^0 &= \frac12 \{c^0,c^0\} \,,
  \\
  \QB \cdot b^0 &= B^0 \,,
  \\
  \QB \cdot B^0 &= 0
\end{align}
and
\begin{align}
  \QB \cdot \CA^1 &= \{c^0,\CA^1\} \,,
  \\
  \QB \cdot c^1 &= [c^0,c^1] \,,
  \\
  \QB \cdot b^1 &= [c^0,b^1] \,,
  \\
  \QB \cdot B^1 &= \{c^0,B^1\} \,.
\end{align}
Let us postulate that
\begin{align}
  Q \cdot c^0 &= -\frac12 \{c^1,c^1\} \,,
  \\
  Q \cdot b^0 &= 0 \,,
  \\
  Q \cdot B^0 &= 0 \,.
\end{align}
Then, the modified BRST charge
\begin{equation}
  \Qh = \QB + Q
\end{equation}
satisfies $\Qh^2 = 0$ and 
\begin{align}\label{brsttrans}
  \Qh \cdot \CA &= -\rmd_\CA' c \,,
  \\
  \Qh \cdot c &= \frac12 \{c^0,c^0\} - \frac12 \{c^1,c^1\} + [c^0,c^1] \,,
  \\
  \Qh \cdot b &= B \,,
  \\
  \Qh \cdot B &= 0 \,,
\end{align}
where
\begin{align}
  c &= c^0 + c^1 \,,
  \\
  b &= b^0 + b^1 \,,
  \\
  B &= B^0 + B^1 + [c^0,b^1] \,.
\end{align}
The $\Qh$-cohomology computes the mixed Lie superalgebra cohomology
defined in~\cite{MR4050665}.

Thus, we conclude that the $Q$-invariant sector of the theory on the
intersecting D5-branes is four-dimensional Chern--Simons theory with
gauge group $\GL(m|n)$.

\subsection{Emergence of the spin chain}

The D3-branes and the F1-branes in the type IIB
setup~\eqref{eq:brane-4dCS} intersect the D5-branes along lines in
$\R \times S^1 \times \C$.  As such, they create line defects in the
four-dimensional Chern--Simons theory on $\R \times S^1 \times \C$,
extending along $\R$ and supported at points in $\C$.  Such a
configuration of line defects in four-dimensional Chern--Simons theory
is naturally identified with an integrable spin chain
\cite{Costello:2013zra, Costello:2013sla, Costello:2017dso}.  We now
show that this spin chain is precisely the one that appears in the
Bethe/gauge correspondence.

\subsubsection{Line defects and spin chains}

Let us first explain the relation between line defects in
four-dimensional Chern--Simons theory and integrable spin chains.

Consider four-dimensional Chern--Simons theory on
$\R \times \R \times \C$, with gauge group $G$ which we take to be a
complex simple Lie supergroup.  Its field is a partial $G$-connection
of the form
\begin{equation}
  \CA = \CA_x \, \rmd x + \CA_y \, \rmd y + \CA_\zb \, \rmd\zb \,.
\end{equation}
We insert line defects
\begin{equation}
  \CL^\ell \,, \quad \ell = 1, \dotsc, L \,,
\end{equation}
extending in the $x$-direction, which we regard as the time direction.
Along the $y$-axis, we arrange $\CL^1$, $\dotsc$, $\CL^L$ in the
ascending order.  They are supported at points $\zeta^1$, $\dotsc$,
$\zeta^L$ in $\C$.

Solutions of the equation of motion for four-dimensional Chern--Simons
theory, away from the line defects, are flat connections.  Away from
the line defects, flat connections can be gauged away.  Then, all
information about the state of the theory is localized in the
neighborhoods of the line defects, and the Hilbert space $V$
factorises into the tensor product of the spaces attached to the line
defects:
\begin{equation}
  V = V^1 \otimes \dotsb \otimes V^L \,.
\end{equation}
This is identified with the Hilbert space of an open spin chain with $L$
sites.  The space $V^\ell$ supported on $\CL^\ell$ is the state space
for the $\ell$th spin.

The four-dimensional Chern--Simons theory is topological on
$\R \times \R$ and holomorphic on $\C$.  Due to the topological
invariance on $\R \times \R$, the Hamiltonian is zero.  To change the
state, we can insert a Wilson line extending in the $y$-direction,
crossing the $L$ line defects introduced earlier.  This Wilson line is
a non-gauge-invariant operator acting on $V$ and interpreted as a
monodromy matrix $T(\sigma)$ in the spin chain.  The spectral
parameter $\sigma$ is the position of the Wilson line in $\C$, and the
holomorphy on $\C$ implies that $T(\sigma)$ is holomorphic in $\sigma$.

If we introduce two Wilson lines and make them intersect in
$\R \times \R$, we get an R-matrix at the intersection.  The two sides
of the RTT relation \eqref{eq:RTT} correspond to two different
configurations of two open Wilson lines crossing each other and the
line defects $\CL^1$, $\dotsc$, $\CL^L$.  The topological invariance
on $\R \times \R$ and the existence of the extra dimensions $\C$ imply
the equivalence of the two configurations.

This R-matrix can be computed by perturbation theory
\cite{Costello:2017dso}, and it was found to be the R-matrix for the
rational spin chain with $G$ symmetry.  Therefore, this setup produces
an open rational spin chain.  For $G = \GL(m|n)$, the R-matrix is the
one given in~\eqref{eq:R}, up to some equivalence relations.

To obtain a closed spin chain, we simply compactify the $y$-axis $\R$
to $S^1$.  Now, flat connections have global gauge-invariant
information, namely the holonomy around $S^1$.  The holonomy is fixed
by the boundary condition at infinity and becomes a parameter of the
spin chain.  Flat connections can still be gauged away almost
everywhere.  Away from the line defects, we can make them vanish
except on a single line parallel to the line defects and placed
between $\CL^L$ and $\CL^1$, say.  The holonomy is then identified
with the twist parameter $g$ of the periodic boundary condition in the
spin chain.  A Wilson loop winding around $S^1$ gives a transfer
matrix $t(g, \sigma)$ evaluated in the representation of the Wilson
loop.

\subsubsection{Line defects created by D3-branes}

Let us return to the setup for the Bethe/gauge correspondence.

By the topological invariance on $\R \times S^1$, the positions of the
D3-branes on $S^1$ do not matter.  For each $\ell$, we gather the
$m+n$ D3-branes $\mathrm{D3}_i^\ell$, $i = 1$, $\dotsc$, $m+n$, to the
same position $y^\ell$ on $S^1$:
\begin{equation}
  y_1^\ell = \dotsb = y_{m+n}^\ell = y^\ell \,.
\end{equation}
Since they are also located at the same point $\zeta^\ell$ in $\C$ up
to first order in $\hbar$, we can regard them as creating a single
line defect $\CL^\ell$ supported on the line
$\R \times \{y^\ell\} \times \{\zeta^\ell\}$ in
$\R \times S^1 \times \C$, treating the differences
$\mu_i^\ell - \zeta^\ell$, $i = 1$, $\dotsc$, $m+n$, in the positions
in $\C$ as parameters of the line defect.  The D3-branes thus create
$L$ line defects $\CL^1$, $\dotsc$, $\CL^L$.

From the Bethe/gauge correspondence we know what the Hilbert space
$V^\ell$ of $\CL^\ell$ must be: it is the evaluation module of the
Yangian $Y(\glf(m|n))$ with spectral parameter $\zeta^\ell$,
obtained from the Verma module $V_{\lambda^\ell}$ of $\glf(m|n)$.
The F1-branes represent excitations in this Hilbert space.

Let us derive this Hilbert space from the point of view of brane
dynamics.  In the four-dimensional Chern--Simons theory on
$\R \times S^1 \times \C$, the positions of the D5-branes in $\R_X$
and $\R_Y$ parametrize the vacuum expectation values of the gauge
fields along the topological directions.  For the purpose of
identifying the Hilbert space of the line defect, we can consider the
situation in which all D5-branes are coincident, say $X_i = Y_i = 0$
for all~$i$.

The $Q$-invariant sector of the theory living on the D3-branes that
create $\CL^\ell$ is the BF theory with gauge group $G = \GL(m|n)$,
defined on $\R \times [0,\infty)$.  This theory can be obtained from
four-dimensional Chern--Simons theory on
$\R \times [0,\infty) \times \C$ by dimensional reduction on $\C$.  It
has the action
\begin{equation}
  \label{eq:BF}
  \frac{1}{\hbar} \int_{\R \times [0,\infty)}
  \sigma (\rmd\eta + \eta \wedge \eta) \,,
\end{equation}
where $\eta$ is the gauge field and $\sigma$ is a scalar field valued
in $\gf^*$, the dual of the Lie algebra $\gf = \glf(m|n)$.  The field
$\sigma$ comes from the reduction of the antiholomorphic component
$A_\zb$ of the four-dimensional gauge field on $\C$.

One way to see that this is the right theory is to note that upon
exchanging the directions of $S^1$ and $\R_Y$, the D3-branes are
T-dual to D5-branes on
$\R \times S^1 \times \C \times \R^2_{\pm\hbar}$ (after compactifying
$\C$ to a torus).  Since the D5-branes are described by
four-dimensional Chern--Simons theory with gauge group $\GL(m|n)$, the
D3-branes are described by the BF theory.  The exchange of two
directions amounts to swapping the role of the real and imaginary
parts of $A_y$, and does not alter the analysis in any essential way.

We view the BF theory as a Poisson sigma model \cite{Ikeda:1993fh,
  Schaller:1994es} with target space $\gf^*$.  The space of functions
on $\gf^*$ is the symmetric algebra $S(\gf)$ of $\gf$.  Linear
functions are elements of $\gf$, and the Poisson bracket between them
is given by the Lie bracket.  Extending the Poisson bracket to
$S(\gf)$ by the Leibniz rule, we endow $\gf^*$ with the Poisson
structure.  The action~\eqref{eq:BF} of the BF theory is related to
the action of the Poisson sigma model by integration by parts.

There are two boundaries in the spacetime
$\R \times [0, \infty) \subset \R \times \R_Y$, one at $Y = 0$ and the
other at $Y = \infty$.  It is more convenient to think of the
spacetime of the theory as the limit of
\begin{equation}
  \R \times [0,r]
\end{equation}
as $r \to \infty$.  Physically, we can realize this setup by
making the D3-brane $\mathrm{D3}_i^\ell$ end on an NS5-brane
$\mathrm{NS5}_i^\ell$ with worldvolume
\begin{equation}
  \R \times \{y^\ell\} \times \{\mu^\ell_i\}
  \times \R_X \times \{r\} \times \R^2_{+\hbar} \times \R^2_{-\hbar}
  \subset
  \R \times S^1 \times \C
  \times \R_X \times \R_Y \times \R^2_{+\hbar} \times \R^2_{-\hbar} \,.
\end{equation}
Since the BF theory is topological, the value of $r$ does not
matter.  In particular, we can take the limit $r \to 0$.  In
this limit the theory reduces to a one-dimensional quantum mechanical
system.  This quantum mechanical system describes the line defect
after coupling to the four-dimensional Chern--Simons theory.

The Hilbert space of the BF theory depends on the boundary conditions
imposed on the two boundaries $\R \times \{0\}$ and
$\R \times \{r\}$.  On each boundary, we impose a boundary
condition that defines what is known as a coisotropic brane in the
context of Poisson sigma models \cite{Cattaneo:2003dp}.

The boundary condition on $\R \times \{0\}$ is simple.  The imaginary
part of the component $\eta_x$ of the gauge field along $\R$
parametrizes the positions of the D3-branes in $\R_X$.  These are
necessarily fixed to the positions of the D5-branes on the boundary
where the D3-branes end on the D5-branes.  By holomorphy, the real
part must also obey the Dirichlet boundary condition.  Thus we have
\begin{equation}
  \eta|_{\R \times \{0\}} = 0 \,.
\end{equation}
The field $\sigma$ is unconstrained on the boundary.  This
boundary condition completely breaks the gauge symmetry.  The global
symmetry $G$ on the boundary is used for coupling to the
four-dimensional Chern--Simons theory by gauging.

We propose that the boundary condition on $\R \times \{r\}$ is
determined by the positions of the NS5-branes in $\C$ as follows.  The
diagonal part of $\sigma$ parametrizes these positions.  Let
\begin{equation}
  \gf = \mathfrak{n_-} \oplus \mathfrak{h} \oplus \mathfrak{n_+}
\end{equation}
be the triangular decomposition of $\gf$ with respect to the chosen
basis; thus $\hf$ is spanned by diagonal matrices, and $\nf_+$ and
$\nf_-$ are spanned by strictly upper triangular matrices and strictly
lower triangular matrices, respectively.  Then, the boundary condition
is
\begin{align}
  \eta|_{\R \times \{r\}} &\in \borel \,,
  \\
  \sigma|_{\R \times \{r\}}
  &\in \nf_-^* + \underline{\lambda}^\ell \,,
\end{align}
where $\borel = \hf \oplus \nf_+$ is the Borel subalgebra and
$\underline{\lambda}^\ell$ is an element of $\hf^*$.  This condition
breaks the gauge group on the boundary to the Borel subgroup $B$ whose
Lie algebra is $\borel$.%
\footnote{The choice of Borel subgroup is determined by the positions
  of the D5-branes in $\R_X$, which are in turn given by the vacuum
  expectation value of the time component $\CA_x = A_x + \iu X$ of the
  gauge field; the $\mathfrak{u}(m|n)$-valued field $X$ has the vacuum
  expectation value $\vev{X} = \sum_{i=1}^{m+n} \iu X_i \CE_{ii}$.  In
  this background, a state evolving for duration $T$ is scaled by the
  factor $\exp(T\vev{\CA_x}) = \exp(-T\sum_{i=1}^{m+n} X_i \CE_{ii})$.
  (We have set $\vev{A_x} = 0$ for simplicity.)  Therefore, if we
  compactify the time direction to a circle (say, of radius $1$), as
  one does when computing the partition function of the lattice model
  equivalent to the spin chain, then the periodic boundary condition
  is twisted by the action of $\exp(-\sum_{i=1}^{m+n} X_i \CE_{ii})$.
  In the magnon sector $(M_1, \dotsc, M_{m+n-1})$, this action is
  multiplication by
  $\exp(-\sum_{\ell=1}^L \sum_{i=1}^{m+n} \lambda_i^\ell X_i +
  \sum_{r=1}^{m+n-1} M_r (X_r - X_{r+1}))$.  We must have the
  ordering~\eqref{eq:X-ordering} for the partition function to be a
  power series in small variables.}

By the state--operator correspondence, the Hilbert space is isomorphic
to the space of observables supported at the junction of the above two
coisotropic branes.  This is a bimodule over the algebras of local
observables on the two boundaries.

Let $\{T_\alpha\}_{\alpha=1}^{\dim\nf_-}$ be a basis of $\nf_-$ and
extend it a basis $\{T_a\}_{a=1}^{\dim\gf}$ of $\gf$.  Let
$\sigma_a = \langle T_a, \sigma\rangle$.  Classically, the algebra of
local observables is the algebra of gauge-invariant polynomials in
$\{\sigma_a\}$.

On the boundary $\R \times \{0\}$, local observables are simply
polynomials in $\{\sigma_a\}$ since there is no gauge symmetry there.
The algebra of local observables on the boundary is therefore $S(\gf)$
at the classical level.  Quantum corrections lead to a noncommutative
deformation of $S(\gf)$.  At the quantum level, the algebra is
isomorphic to the universal enveloping algebra of $\gf$ with bracket
$\hbar[\blank,\blank]$; if
$[T_a,T_b] = \sum_{c=1}^{\dim\gf} f_{ab}{}^c T_c$, then
$[\sigma_a, \sigma_b] = \hbar \sum_{c=1}^{\dim\gf} f_{ab}{}^c
\sigma_c$ quantum mechanically~\cite{Kontsevich:1997vb,
  Cattaneo:1999fm}.  In our setup $\hbar$ is a complex parameter
rather than a formal parameter, so the algebra is isomorphic to
$U(\gf)$, with $\sigma_a$ mapped to $\hbar T_a$.  (The quantization
map from $S(\gf)$ to $U(\gf)$ is complicated for polynomials of higher
degree.)

On the boundary $\R \times \{r\}$, the algebra of local observables is
trivial.  The boundary condition says that local operators are
constructed entirely from $\{\sigma_\alpha\}$.  Nontrivial polynomials
cannot commute with $\hf$ and, in particular, cannot be $B$-invariant.
The only local observables are multiples of the identity operator.

At the junction, the two boundary conditions combined imply that
observables are polynomials in $\{\sigma_\alpha\}$.  As a vector
space, the space of observables is generated from the ``highest-weight
vector'' $1$ by the action of ``creation operators''
$\{\sigma_\alpha\}$.  On this vector space the algebra $U(\gf)$ acts.
Thus, the Hilbert space is a Verma module of $U(\gf)$, as expected.

It remains to show that the highest weight of the Verma module is
determined by the positions of the D3-branes.  Classically, the
boundary condition on $\R \times \{r\}$ implies that the highest
weight of the module is $\underline{\lambda}^\ell$.  There is,
however, a quantum correction which shifts the highest weight.%
\footnote{This shift can be understood as originating from normal
  ordering of creation and annihilation operators, which correspond to
  the positive and negative roots.  We will see a similar shift in
  section~\ref{sec:cov-contra}.}
The highest weight is actually
\begin{equation}
  \label{eq:lambda-lambdat}
  \lambda^\ell = \underline{\lambda}^\ell - \rho \,,
\end{equation}
where $\rho$ is the Weyl vector defined in terms of the character of
the $\borel$-module $\gf/\borel$ as
\begin{equation}
  \rho(\blank) = -\frac12 \str_{\gf/\borel} \ad(\blank) \,.
\end{equation}
Since $\rho(\nf_+) = \rho([\borel,\borel]) = 0$, $\rho$ can be
regarded as an element of $\hf^*$.  This is the graded half sum of
positive roots.  When $\gf$ is an ordinary Lie algebra, $\rho$ is the
ordinary Weyl vector and the above shift was derived
in~\cite{MR3305442} based on results from \cite{MR2504434,
  MR3118583}.

For $\gf = \glf(m|n)$, we have
\begin{equation}
  \label{eq:Weyl-vector-glmn}
  \rho
  = \sum_{\substack{k,l=1 \\ k < l}}^{m+n}
    \frac12 (-1)^{[k]+[l]} (\veps_k - \veps_l)
  = \sum_{i=1}^{m+n} \frac12  (-1)^{[i]} (m-n + (-1)^{[i]} - 2 c_i) \veps_i \,.
\end{equation}
Comparing the relations~\eqref{eq:lambda-lambdat}
and~\eqref{eq:mu-lambda}, we find
\begin{equation}
  \label{eq:lambda-zeta-mu}
  (-1)^{[i]} 
  \biggl(\underline{\lambda}_i^\ell - \frac14\biggr)
  =
  \frac{1}{\hbar} (\zeta^\ell -\mu_i^\ell)
  + \frac12 (m-n - c_i) \,.
\end{equation}

\subsubsection{Line defects for parabolic Verma modules of scalar
  type}

There is a generalization of the above brane construction which
produces line defects in parabolic Verma modules of scalar type.

Let $(l_1, \dotsc, l_s)$ be an ordered partition of $m + n$:
$\sum_{\alpha=1}^s l_\alpha = m + n$.  The partition specifies a
parabolic subalgebra $\pf$ of $\glf(m|n)$, namely the subalgebra
consisting of upper-triangular block-diagonal matrices with diagonal
blocks of orders $l_1$, $\dotsc$, $l_s$.  A character $\chi$ of $\pf$
is determined by an $s$-tuple of complex numbers
$(\chi_1, \dotsc, \chi_s)$ as
\begin{equation}
  \chi(\blank) = \str(\chi^\vee \blank) \,,
\end{equation}
where the matrix $\chi^\vee$ is given by
\begin{equation}
  \chi^\vee
  =
  \diag(\chi_1, \dotsc, \chi_1, \dotsc,
  \underbrace{\chi_\alpha, \dotsc, \chi_\alpha}_{\text{$l_\alpha$ times}},
  \dotsc, \chi_s, \dotsc, \chi_s) \,.
\end{equation}

For each $\alpha$, we take $l_\alpha$ D3-branes and make them end on a
single NS5-brane on one side.  On the other side, they end on separate
D5-branes as in the previous construction.  In total, we have $m + n$
D3-branes suspended between $m+n$ D5-branes and $s$ NS5-branes.

On the D3-branes we get the BF theory with gauge group $\GL(m|n)$.
The boundary condition on the D5-brane side is the same as before.  On
the NS5-brane side, the boundary condition is
\begin{align}
  \eta|_{\R \times \{r\}} &\in \pf \,,
  \\
  \sigma|_{\R \times \{r\}}
  &\in \pf^\perp + \underline{\lambda}^\ell \,,
\end{align}
where $\pf^\perp$ is the annihilator of $\pf$ in $\gf^*$ and
$\underline{\lambda}^\ell$ is a character of $\pf$.  Classically (that
is to say, when $\underline{\lambda}^\ell$ is of order $\hbar^{-1}$
and quantities of order $\hbar^0$ are ignored), the value of
$\underline{\lambda}^\ell_\alpha$ is the position of the $\alpha$th
NS5-brane in $\C$.

We expect that the Hilbert space of the BF theory with these boundary
conditions is a parabolic Verma module of scalar type
\begin{equation}
  U(\gf) \otimes_{U(\pf)} \C_{\underline{\lambda}^\ell - \rho} \,.
\end{equation}
The character $\rho$ of $\pf$ is defined by
\begin{equation}
  \rho = -\frac12 \str_{\gf/\pf} \ad_\pf
\end{equation}
and $\C_{\underline{\lambda}^\ell - \rho}$ is the one-dimensional $U(\pf)$-module
determined by the character $\underline{\lambda}^\ell - \rho$.

\subsection{Fermionic Dualities}
\label{sec:fermionic-dualities}

As we have emphasized in our discussions, the Lie superalgebra
$\glf(m|n)$ does not possess a unique Dynkin diagram.  A Dynkin
diagram is specified by a choice of ordered basis of $\C^{m|n}$ (or a
choice of $\Z_2$-grading if we identify Dynkin diagrams related by the
action of the Weyl group), and different choices are related by a
series of certain adjacent transpositions, called odd reflections.
Under odd reflections, a highest-weight representation is mapped to a
highest-weight representation, but the highest weight is not preserved
because the definition of raising and lowering operators is altered.

Odd reflections change how we describe representations of $\glf(m|n)$,
and the description of the Bethe vectors of the superspin chain is
changed accordingly.  The map from the Bethe vectors for one choice of
ordered basis to another is known as a \emph{fermionic duality}.

Fermionic dualities have been studied before purely from an algebraic
perspective~\cite{
  doi:10.1142/9789812798268_0007,
  Essler:1992he,
  Tsuboi:1998ne,
  MR2074080,
  MR2042980,
  Volin:2010xz}, with a notable exception of the work by Orlando and
Reffert~\cite{Orlando:2010uu} where they employed the point of view of
string theory to discuss the fermionic dualities for the
supersymmetric $t$-$J$ model, which is the rational $\glf(1|2)$ spin
chain with spins valued in the natural representation $\C^{1|2}$.
Here we offer a string theory explanation for important aspects of
fermionic dualities for the rational $\glf(m|n)$ spin chain with Verma
modules, namely their action on highest weights and magnon numbers.

\subsubsection{Odd reflections and fermionic dualities}

Recall from section~\ref{eq:glmn} that the definitions of positive and
simple roots depend on a choice of ordered basis
$(e_1, \dotsc, e_{m+n})$ of $\C^{m|n}$, which is a permutation of
$(b_1, \dotsc, b_m, f_1, \dotsc, f_n)$ such that $(b_1, \dotsc, b_m)$
and $(f_1, \dotsc, f_n)$ are the standard bases of $\C^m$ and $\C^n$,
respectively.  There is a natural identification between these basis
vectors and their weights:
\begin{align}
  (e_1, \dotsc, e_{m+n}) &\leftrightarrow (\veps_1, \dotsc, \veps_{m+n}) \,,
  \\
  (b_1, \dotsc, b_m) &\leftrightarrow (\eps_1, \dotsc, \eps_m) \,,
  \\
  (f_1, \dotsc, f_n) &\leftrightarrow (\delta_1, \dotsc, \delta_n) \,.
\end{align}
In the following discussion we will consider permutations of
$(\veps_1, \dotsc, \veps_{m+n})$ induced by those of
$(e_1, \dotsc, e_{m+n})$.

For a given choice of ordered basis $(\veps_1, \dotsc, \veps_{m+n})$
of the dual of the Cartan subalgebra of $\glf(m|n)$, the set of
positive roots is
\begin{equation}
  \Phi^+ = \{\veps_i - \veps_j \mid i < j \}
\end{equation}
and the set of simple roots is
\begin{equation}
  \label{+roots}
  \Pi = \{\veps_r - \veps_{r+1} \mid r = 1, \dotsc, m+n-1\} \,.
\end{equation}
A root $\veps_i - \veps_j$ is said to be \emph{even} if $[i] = [j]$
and \emph{odd} if $[i] \neq [j]$.

Pick an odd simple root $\alpha_s = \veps_s - \veps_{s+1}$ and apply
to the ordered basis the adjacent transposition
$\sigma_s\colon \{1, \dotsc, m+n\} \to \{1, \dotsc, m+n\}$
interchanging $\veps_s$ and $\veps_{s+1}$:
\begin{multline}
  (\veps_{\sigma_s(1)}, \dotsc, \veps_{\sigma_s(s-1)},
  \veps_{\sigma_s(s)}, \veps_{\sigma_s(s+1)}, \veps_{\sigma_s(s+2)},
  \dotsc, \veps_{\sigma_s(m+n)})
  \\
  =
  (\veps_1, \dotsc, \veps_{s-1},
  \veps_{s+1}, \veps_s, \veps_{s+2},
  \dotsc, \veps_{m+n}) \,.
\end{multline}
The adjacent transposition alters the notion of positive and simple
roots.  For the new ordered basis
$(\veps_{\sigma_s(1)}, \dotsc, \veps_{\sigma_s(m+n)})$, the set of
positive roots is
\begin{equation}
  \label{eq:OR-Phi+}
  \Phi^+_{\alpha_s}
  =
  \{\veps_{\sigma_s(i)} - \veps_{\sigma_s(j)} \mid i < j\}
  =
  \{-\alpha_s\} \cup \Phi^+ \setminus \{\alpha_s\}
\end{equation}
and the set of simple roots is
\begin{equation}
  \begin{split}
    \Pi_{\alpha_s}
    &=
    \{\veps_{\sigma_s(r)} - \veps_{\sigma_s(r+1)} \mid r = 1, \dotsc, m+n-1\}
    \\
    &=
    \{\veps_{s-1} - \veps_{s+1}, \veps_{s+1} - \veps_s, \veps_s - \veps_{s+2}\}
    \cup
    \Pi \setminus \{\alpha_{s-1}, \alpha_s, \alpha_{s+1}\}
    \,.
  \end{split}
\end{equation}
This automorphism of the root system which transforms the positive and
simple roots is called the \emph{odd reflection} with respect to the
odd simple root $\alpha_s$.

As an example, take $(m|n)=(3|2)$ and
$(\veps_1, \veps_2, \veps_3, \veps_4, \veps_5) = (\eps_1, \eps_2,
\delta_1, \delta_2, \eps_3)$.  This choice of ordered basis gives the
Dynkin diagram~\eqref{eq:Dynkin}.  There are two odd simple roots,
$\alpha_2 = \eps_2 - \delta_1$ and $\alpha_4 = \delta_2 - \eps_3$,
represented by the crossed nodes.  Reflection with respect to
$\alpha_2$ swaps $\eps_2$ and $\delta_1$, leading to the new ordered
basis $(\eps_1, \delta_1, \eps_2, \delta_2, \eps_3)$.  The Dynkin
diagram corresponding to the reflected simple roots is
\begin{equation}
  \begin{tikzpicture}[scale=1.5]
    \draw[-] (0.5,0.5) -- (1.5,0.5);
    \draw[-] (1.5,0.5) -- (2.5,0.5);
    \draw[-] (2.5,0.5) -- (3.5,0.5);

    \node[gnode, minimum size=10pt] at (0.5,0.5) {};
    \node[gnode, minimum size=10pt] at (1.5,0.5) {};
    \node[gnode, minimum size=10pt] at (2.5,0.5) {};
    \node[gnode, minimum size=10pt] at (3.5,0.5) {};

    \draw (0.5,.5) node[cross=4pt] {};
    \draw (1.5,.5) node[cross=4pt] {};
    \draw (2.5,.5) node[cross=4pt] {};
    \draw (3.5,.5) node[cross=4pt] {};

    \node[below=4pt] at (0.5,0.5) {$\eps_1 - \delta_1$};
    \node[below=4pt] at (1.5,0.5) {$\delta_1 - \eps_2$};
    \node[below=4pt] at (2.5,0.5) {$\eps_2 - \delta_2$};
    \node[below=4pt] at (3.5,0.5) {$\delta_2 - \eps_3$};
  \end{tikzpicture}
\end{equation}
We see that all simple roots are now odd.  Reflection of the original
ordered basis with respect to $\alpha_4$ results in the ordered basis
$(\eps_1, \eps_2, \delta_1, \eps_3, \delta_2)$ and the Dynkin diagram
\begin{equation}
  \begin{tikzpicture}[scale=1.5]
    \draw[-] (0.5,0.5) -- (1.5,0.5);
    \draw[-] (1.5,0.5) -- (2.5,0.5);
    \draw[-] (2.5,0.5) -- (3.5,0.5);

    \node[gnode, minimum size=10pt] at (0.5,0.5) {};
    \node[gnode, minimum size=10pt] at (1.5,0.5) {};
    \node[gnode, minimum size=10pt] at (2.5,0.5) {};
    \node[gnode, minimum size=10pt] at (3.5,0.5) {};

    \draw (1.5,.5) node[cross=4pt] {};
    \draw (2.5,.5) node[cross=4pt] {};
    \draw (3.5,.5) node[cross=4pt] {};

    \node[below=4pt] at (0.5,0.5) {$\eps_1 - \eps_2$};
    \node[below=4pt] at (1.5,0.5) {$\eps_2 - \delta_1$};
    \node[below=4pt] at (2.5,0.5) {$\delta_1 - \eps_3$};
    \node[below=4pt] at (3.5,0.5) {$\eps_3 - \delta_2$};
  \end{tikzpicture}
\end{equation}

Odd reflections change the characterization of highest weights.  Let
us see how Verma modules are transformed.  Fix an ordered basis of
$\C^{m|n}$ and consider the Verma module $M(\lambda)$, with the
highest-weight vector $\ket{\Omega_\lambda}$.  Let $\alpha_s$ be an
odd root.  After the odd reflection about $\alpha_s$, the roles of the
raising operator $\CE_{s,s+1}$ and the lowering operator $\CE_{s+1,s}$
are exchanged, while all other lowering operators remain unchanged.
Consequently, the state
\begin{equation}
  \ket{\Omega_{\lambda'}}
  =
  \CE_{s+1,s} \ket{\Omega_\lambda}
\end{equation}
is annihilated by all elements of the new set of raising operators,
that is, it is a highest-weight state with respect to the new ordered
basis.  According to the PBW theorem~\eqref{eq:PBW}, the states of the
form
\begin{equation}
  x_1^{n_1} \dotsm x_{p-1}^{n_{p-1}} \CE_{s+1,s}^{n_p} \ket{\Omega_\lambda}
  =
  x_1^{n_1} \dotsm x_{p-1}^{n_{p-1}} \CE_{s,s+1}^{1-n_p} \ket{\Omega_{\lambda'}} \,,
\end{equation}
form a basis of the Fock space $V_\lambda$ for $M(\lambda)$, where
$(x_1, \dotsc, x_{p-1}, \CE_{s+1,s})$ is an ordered set of lowering
operators in the original ordered basis.  (Note that $n_p$ is either
$0$ or $1$.)  By the PBW theorem, we see that $M(\lambda)$ is the
Verma module $M(\lambda')$ with respect to the new ordered basis, with
\begin{equation}
  \label{eq:lambda'}
  \lambda' = \lambda - \alpha_s \,.
\end{equation}

In the spin chain, the highest weights of the Verma modules placed at
the spin sites are transformed by an odd reflection.  The weight of
each state of the spin chain remains the same, so the magnon numbers
must be transformed as
\begin{equation}
  \sum_{\ell=1}^L (\lambda^\ell)'
  - \sum_{r=1}^{m+n-1} M_r' \alpha_r'
  =
  \sum_{\ell=1}^L \lambda^\ell
  - \sum_{r=1}^{m+n-1} M_r \alpha_r \,,
  \quad
  \alpha_r' = \veps_{\sigma_s(r)} - \veps_{\sigma_s(r+1)} \,,
\end{equation}
or more explicitly,
\begin{equation}
  \label{eq:Mr'}
  M_r'
  =
  \begin{cases}
    L + M_{s-1} + M_{s+1} - M_s & (r = s) \,;
    \\
    M_r & (r \neq s) \,.
  \end{cases}
\end{equation}

The transformations of the highest weights and magnon numbers change
the Bethe equations.  Of course, this is merely a change in the
description of the spin chain states, so the solutions of the new
Bethe equations are in one-to-one correspondence with the solutions of
the original Bethe equations.  This correspondence is called the
fermionic duality generated by the odd reflection.

\subsubsection{Fermionic duality from string theory}

In the $\mathrm{D1}$--$\mathrm{D3}$--$\mathrm{NS5}$ duality frame that is S-dual to \eqref{eq:brane-4dCS}, the choice of
ordered basis of $\C^{m|n}$ is reflected in the ordering of NS5-branes
along $\R_X$; they are ordered as $\mathrm{NS5}_1$, $\mathrm{NS5}_2$,
$\dotsc$, $\mathrm{NS5}_{m+n}$ from left to right.  Therefore, the
string theory interpretation of the reflection about an odd root
$\alpha_s$ is clear: it swaps the positions of $\mathrm{NS5}_s$ and
$\mathrm{NS5}_{s+1}$.  We wish to understand the effect of the
exchange of positions on the highest weights and magnon numbers.

Recall that one of the boundary conditions for the BF theory that
emerges on the D3-branes creating the line defect $\CL^\ell$ is
specified by the parameters $\underline{\lambda}^\ell_i$, $i = 1$,
$\dotsc$, $m+n$.  The position of $\mathrm{D3}_i^\ell$ in $\C$ is
given by
\begin{equation}
  \zeta^\ell - (-1)^{[i]} \underline{\lambda}^\ell_i \,.
\end{equation}
Indeed, if we take $\underline{\lambda}_i^\ell$ to be of order
$\hbar^{-1}$, by the relation~\eqref{eq:lambda-zeta-mu} this quantity
coincides to order $\hbar^0$ with the twisted mass $\mu_i^\ell$, which
is identified with the classical location of $\mathrm{D3}_i^\ell$.
Swapping the positions of $\mathrm{NS5}_s$ and $\mathrm{NS5}_{s+1}$
also exchanges $\mathrm{D3}_s$ and $\mathrm{NS5}_{s+1}$ while keeping
their locations in $\C$ fixed.  Thus, $\{\underline{\lambda}_i^\ell\}$
are transformed to new values $\{(\underline{\lambda}^\ell_i)'\}$ such
that $(\underline{\lambda}^\ell_s)' = \underline{\lambda}^\ell_{s+1}$
and $(\underline{\lambda}^\ell_{s+1})' = \underline{\lambda}^\ell_s$
in the new ordered basis.  This simply means that we have
\begin{equation}
  (\underline{\lambda}^\ell)'
  =
  \sum_{i=1}^{m+n} (\underline{\lambda}^\ell)'_i \veps_{\sigma_s(i)}
  =
  \underline{\lambda}^\ell \,.
\end{equation}

Although $\underline{\lambda}^\ell$ is invariant under the odd
reflection, the Weyl vector $\rho$ is transformed to a new Weyl vector
$\rho'$.  Since $\rho$ is the half sum of even positive roots minus
the half sum of odd positive roots, from the
relation~\eqref{eq:OR-Phi+} between $\Phi^+$ and $\Phi^+_{\alpha_s}$
we see
\begin{equation}
  \rho'
  =
  \rho + \alpha_s \,.
\end{equation}
Then, the relation \eqref{eq:lambda-lambdat} between
$\underline{\lambda}^\ell$ and the highest weight $\lambda^\ell$ shows
that $\lambda^\ell$ is transformed to $(\lambda^\ell)'$ according to
the formula~\eqref{eq:lambda'}.

Exchanging the pairs $(\mathrm{NS5}_s, \mathrm{D3}_s)$ and
$(\mathrm{NS5}_{s+1}, \mathrm{D3}_{s+1})$ does not only transform the
highest weights, but also change the magnon numbers.  In the brane
picture, we can understand this phenomenon as creation and
annihilation of $\mathrm{D1}$-branes due to the Hanany--Witten effect
\cite{Hanany:1996ie}.%
\footnote{In general, Hanany--Witten processes for type IIA brane
  configurations for two-dimensional $\CN = (2,2)$ supersymmetric
  gauge theories classically suffer from ambiguities, which are only
  resolved if one takes brane bending into account or lifts the
  configurations to M-theory.~\cite{Hanany:1997vm}.  In the present
  case, such ambiguities do not arise because the relevant gauge group
  has the same number of fundamental and antifundamental chiral
  multiplets.}
In order to exchange the positions of the brane pairs, we first need
to move each $\mathrm{D1}$-brane between $\mathrm{NS5}_s$ and
$\mathrm{NS5}_{s+1}$ so that one of its end is attached to one of the
NS5-branes, say $\mathrm{NS5}_{s+1}$.  Then, we displace
$\mathrm{NS5}_{s+1}$ into the page and start moving it to the left.
At one point the D3-branes ending on $\mathrm{NS5}_{s+1}$ pass through
$\mathrm{NS5}_s$.  As a result, the $\mathrm{D1}$-branes ending on these
D3-branes are annihilated and a new $\mathrm{D1}$-brane is created on each of
those D3-branes that did not have $\mathrm{D1}$-brane ending on it.  In the case
in which $[s] = \bar0$, $[s+1] = \bar1$ and
$(M_{s-1}, M_s, M_{s+1}) = (1,2,1)$, the process looks as follows:
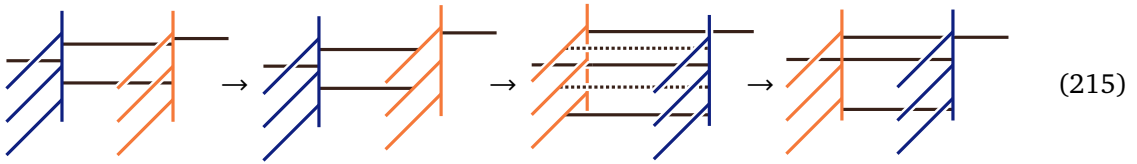
\begin{equation}
  \begin{tikzpicture}[scale=1.5]
    \begin{scope}[shift={(0,0.15)}]
      \draw[ZERO, very thick] (0.5,0.4) -- (1,0.4);
    \end{scope}

    \begin{scope}[shift={(1,-0.1)}]
      \draw[ZERO, very thick] (0,0.8) -- (1,0.8);
      \draw[ZERO, very thick] (0,0.45) -- (1,0.45);
    \end{scope}

    \begin{scope}[shift={(2,0.1)}]
      \draw[ZERO, very thick] (0,0.65) -- (0.5,0.65);
    \end{scope}

    \draw[white, line width=3pt] (0.5,0.3) -- (1,0.8);
    \draw[white, line width=3pt] (1.5,0) -- (2,0.5);
    \draw[white, line width=3pt] (1.5,0.3) -- (2,0.8);

    \draw[PLUS, very thick] (1,0) -- (1,1);
    \draw[MINUS, very thick] (2,0) -- (2,1);

    \begin{scope}[shift={(0,-0.3)}]
      \draw[PLUS, very thick] (0.5,0) -- (1,0.5);
      \draw[MINUS, very thick] (1.5,0) -- (2,0.5);
    \end{scope}

    \begin{scope}[shift={(0,0)}]
      \draw[PLUS, very thick] (0.5,0) -- (1,0.5);
      \draw[MINUS, very thick] (1.5,0) -- (2,0.5);
    \end{scope}

    \begin{scope}[shift={(0,0.3)}]
      \draw[PLUS, very thick] (0.5,0) -- (1,0.5);
      \draw[MINUS, very thick] (1.5,0) -- (2,0.5);
    \end{scope}
  \end{tikzpicture}
  \!\!\!\! \to \,
  \begin{tikzpicture}[scale=1.5]
    \begin{scope}[shift={(0,0.15)}]
      \draw[ZERO, very thick] (0.5,0.4) -- (1,0.4);
    \end{scope}

    \begin{scope}[shift={(1,-0.1)}]
      \draw[ZERO, very thick] (0,0.8) -- (0.9,0.8);
      \draw[ZERO, very thick] (0,0.45) -- (0.85,0.45);
    \end{scope}

    \draw[white, line width=3pt] (0.5,0.3) -- (1,0.8);

    \draw[PLUS, very thick] (1,0) -- (1,1);

    \begin{scope}[shift={(0,-0.3)}]
      \draw[PLUS, very thick] (0.5,0) -- (1,0.5);
    \end{scope}

    \begin{scope}[shift={(0,0)}]
      \draw[PLUS, very thick] (0.5,0) -- (1,0.5);
    \end{scope}

    \begin{scope}[shift={(0,0.3)}]
      \draw[PLUS, very thick] (0.5,0) -- (1,0.5);
    \end{scope}

    \begin{scope}[shift={(0.1,0.1)}]
      \begin{scope}[shift={(2,0.1)}]
        \draw[ZERO, very thick] (0,0.65) -- (0.5,0.65);
      \end{scope}

      \draw[MINUS, very thick] (2,0) -- (2,1);
      
      \begin{scope}[shift={(0,-0.3)}]
        \draw[MINUS, very thick] (1.5,0) -- (2,0.5);
      \end{scope}
      
      \begin{scope}[shift={(0,0)}]
        \draw[MINUS, very thick] (1.5,0) -- (2,0.5);
      \end{scope}
      
      \begin{scope}[shift={(0,0.3)}]
        \draw[MINUS, very thick] (1.5,0) -- (2,0.5);
      \end{scope}
    \end{scope}
  \end{tikzpicture}
  \!\!\!\! \to \,
  \begin{tikzpicture}[scale=1.5]
    \begin{scope}[shift={(-2.1,0.1)}]
      \begin{scope}[shift={(2,0.1)}]
        \draw[ZERO, very thick] (0,0.65) -- (1.5,0.65);
      \end{scope}
    \end{scope}

    \begin{scope}[shift={(-2.1,0.1)}]
      \draw[MINUS, very thick] (2,0) -- (2,1);
    \end{scope}

    \begin{scope}[shift={(1,-0.1)}]
      \draw[white, line width=3pt] (-1.3,0.8) -- (0,0.8);
      \draw[ZERO, very thick, densely dotted] (-1.3,0.8) -- (0,0.8);
      \draw[white, line width=3pt] (-1.35,0.45) -- (0,0.45);
      \draw[ZERO, very thick, densely dotted] (-1.35,0.45) -- (0,0.45);
      \draw[white, line width=3pt] (-1.3,0.2) -- (0,0.2);
      \draw[ZERO, very thick] (-1.3,0.2) -- (0,0.2);
    \end{scope}

    \begin{scope}[shift={(0,0.15)}]
      \draw[ZERO, very thick] (-0.6,0.4) -- (1,0.4);
    \end{scope}

    \draw[white, line width=3pt] (-0.6,0.4) -- (-0.4,0.6);

    \draw[white, line width=3pt] (0.5,0.3) -- (1,0.8);
    \draw[white, line width=3pt] (0.5,0) -- (1,0.5);
    \draw[white, line width=3pt] (0.5,-0.3) -- (1,0.2);

    \begin{scope}[shift={(0,-0.3)}]
      \draw[PLUS, very thick] (0.5,0) -- (1,0.5);
    \end{scope}

    \begin{scope}[shift={(0,0)}]
      \draw[PLUS, very thick] (0.5,0) -- (1,0.5);
    \end{scope}

    \begin{scope}[shift={(0,0.3)}]
      \draw[PLUS, very thick] (0.5,0) -- (1,0.5);
    \end{scope}

    \begin{scope}[shift={(-2.1,0.1)}]
      \begin{scope}[shift={(0,-0.3)}]
        \draw[MINUS, very thick] (1.5,0) -- (2,0.5);
      \end{scope}
    \end{scope}
    
    \draw[white, line width=3pt] (-0.15,0.55) -- (-0.05,0.55);
    \draw[ZERO, very thick] (-0.15,0.55) -- (-0.05,0.55);

    \draw[white, line width=3pt] (-0.2,0.5) -- (-0.11,0.59);

    \begin{scope}[shift={(-2.1,0.1)}]
      \begin{scope}[shift={(0,0)}]
        \draw[MINUS, very thick] (1.5,0) -- (2,0.5);
      \end{scope}

      \begin{scope}[shift={(0,0.3)}]
        \draw[MINUS, very thick] (1.5,0) -- (2,0.5);
      \end{scope}
    \end{scope}

    \draw[white, line width=3pt] (1,0.825) -- (1,0.9);
    \draw[PLUS, very thick] (1,0) -- (1,1);
  \end{tikzpicture}
  \!\!\!\! \to \,
  \begin{tikzpicture}[scale=1.5]
    \begin{scope}[shift={(1,-0.1)}]
      \draw[ZERO, very thick] (-1,0.2) -- (0,0.2);
    \end{scope}

    \begin{scope}[shift={(0,0.15)}]
      \draw[ZERO, very thick] (-0.5,0.4) -- (1,0.4);
    \end{scope}

    \begin{scope}[shift={(0,0.1)}]
      \draw[ZERO, very thick] (0,0.65) -- (1.5,0.65);
    \end{scope}

    \draw[white, line width=3pt] (-0.4,0.4) -- (-0.2,0.6);

    \draw[white, line width=3pt] (0.5,0.3) -- (1,0.8);
    \draw[white, line width=3pt] (0.5,0) -- (1,0.5);
    \draw[white, line width=3pt] (0.5,-0.3) -- (1,0.2);

    \begin{scope}[shift={(0,-0.3)}]
      \draw[PLUS, very thick] (0.5,0) -- (1,0.5);
    \end{scope}

    \begin{scope}[shift={(0,0)}]
      \draw[PLUS, very thick] (0.5,0) -- (1,0.5);
    \end{scope}

    \begin{scope}[shift={(0,0.3)}]
      \draw[PLUS, very thick] (0.5,0) -- (1,0.5);
    \end{scope}

    \begin{scope}[shift={(-2,0)}]
      \draw[MINUS, very thick] (2,0) -- (2,1);
    
      \begin{scope}[shift={(0,-0.3)}]
        \draw[MINUS, very thick] (1.5,0) -- (2,0.5);
      \end{scope}
    \end{scope}

    \begin{scope}[shift={(-2,0)}]
      \begin{scope}[shift={(0,0)}]
        \draw[MINUS, very thick] (1.5,0) -- (2,0.5);
      \end{scope}

      \begin{scope}[shift={(0,0.3)}]
        \draw[MINUS, very thick] (1.5,0) -- (2,0.5);
      \end{scope}
    \end{scope}

    \draw[white, line width=3pt] (1,0.825) -- (1,0.9);
    \draw[PLUS, very thick] (1,0) -- (1,1);
  \end{tikzpicture}
\end{equation}
The dotted lines indicate the annihilation of $\mathrm{D1}$-branes.  We see that
the numbers of $\mathrm{D1}$-branes between NS5-branes transform as in the
formula~\eqref{eq:Mr'}.

In the D2--D4--NS5 duality frame, the above manipulation is expected
to lead to an infrared duality of $\CN = (2,2)$ supersymmetric gauge
theories.  Indeed, there is a known duality transformation that sends
a theory with $\U(N_c)$ gauge group, $N_f$ fundamental chiral
multiplets and $N_a$ antifundamental chiral multiplets to a theory
with $\U(N_c')$ gauge group, $N_a$ fundamental chiral multiplets and
$N_f$ antifundamental chiral multiplets, plus mesons transforming in
the bifundamental representation of the flavor group
$\U(N_a) \times \U(N_f)$ \cite{Benini:2012ui}.  The rank of the dual
gauge group is $N_c' = \max(N_f, N_a) - N_c$.  This is consistent with
what we have found since $N_c = M_s$ and
$N_f = N_a = L + M_{s-1} + M_{s+1}$ in our case.  However, it appears
that the mesons are absent from our final brane configuration.
Fortunately, the mesons, being neutral under the gauge symmetry, do
not affect the Bethe equations.

\section{Bethe/gauge correspondence for compact superspin chains}
\label{sec:Bethe/gauge-compact}

The superspin chains that appear in the Bethe/gauge correspondence
discussed in the previous sections are noncompact, meaning that they
carry spins valued in infinite-dimensional representations of the
Yangian $Y(\glf(m|n))$.  Spin chains whose spins are valued in
finite-dimensional representations are said to be \emph{compact}.

In this section we discuss the Bethe/gauge correspondence for compact
rational $\glf(m|n)$ spin chains.  We will follow a line of reasoning
similar to our treatment of the noncompact case, but in the reverse
direction: we start with the construction of line defects for
finite-dimensional representations in four-dimensional Chern--Simons
theory, then identify their brane realization and apply dualities to
deduce the corresponding two-dimensional quiver gauge theories.

\subsection{Covariant and contravariant representations of
  $\glf(m|n)$}
\label{sec:cov-contra}

Finite-dimensional representations of $\glf(m|n)$ are most easily
discussed in the \emph{distinguished grading}, in which
\begin{equation}
  [i] =
  \begin{cases}
    \bar0 & (i \leq m) \,;
    \\
    \bar1 & (i > m) \,.
  \end{cases}
\end{equation}
For this reason, in this section we exclusively use the distinguished
grading.  We will write a weight
$\lambda = \sum_{i=1}^{m+n} \lambda_i \veps_i$ as
$(\lambda_1, \dotsc, \lambda_m|\lambda_{m+1}, \dotsc, \lambda_{m+n})$.

The Verma module $M(\lambda)$ with highest weight $\lambda$ contains a
unique maximal submodule.  In the distinguished grading, the
corresponding simple quotient module $L(\lambda)$ is
finite-dimensional if and only if
\begin{equation}
  \lambda_i - \lambda_{i+1} \in \Z_{\geq 0} \,,
  \quad
  i = 1, \dotsc, \dotsc, m+n-1 \,,
  \quad
  i \neq m \,,
\end{equation}
in other words, if and only if $(\lambda_1, \dotsc, \lambda_m)$ and
$(\lambda_{m+1}, \dotsc, \lambda_{m+n})$ are highest weights of
finite-dimensional irreducible representations of $\glf(m)$ and
$\glf(n)$, respectively.  Any finite-dimensional irreducible
representation of $\glf(m|n)$ is isomorphic to $L(\lambda)$ for some
$\lambda$.

We will consider two classes of finite-dimensional irreducible
representations of $\glf(m|n)$, called covariant representations and
contravariant representations.  Covariant representations appear in
tensor products of copies of the natural representation
\begin{equation}
  \C^{m|n} = L\bigl((1,0,\dotsc, 0|0, \dotsc, 0)\bigr) \,,
\end{equation}
whereas contravariant representations arise from tensor products of
copies of the dual representation%
\footnote{The dual $\pi^*$ of a representation $\pi$ is given by
  $\pi^* = \tau \circ \pi$, where $\tau(X) = -\strans{X}$ is the
  Chevalley automorphism.  The supertranspose $\strans{X}$ of $X$ is
  defined by $\strans{X}_{ij} = (-1)^{([i]+[j])[j]} X_{ji}$.  Our
  definition of supertranspose differs from a commonly used definition
  by a factor of $(-1)^{[i]+[j]}$.  With this definition, the quantum
  mechanical action~\eqref{eq:QM} is invariant under the natural
  action of $\GL(m|n) \times \GL(K|\Kb)$.}
\begin{equation}
  (\C^{m|n})^* = L\bigl((0,\dotsc, 0|0, \dotsc, 0, -1)\bigr) \,.
\end{equation}
Both covariant and contravariant representations are indexed by the
so-called $(m|n)$-hook partitions.

A partition $Y = (Y_1, \dotsc, Y_{l(Y)})$ of size $|Y|$ and length
$l(Y)$ is an $l(Y)$-tuple of positive integers such that
$Y_1 \geq \dotsb \geq Y_{l(Y)}$ and $Y_1 + \dotsb + Y_{l(Y)} = |Y|$.
It can be represented by a Young diagram with $l(Y)$ rows, with the
$\alpha$th row consisting of $Y_\alpha$ boxes.  The conjugate
partition $Y'$ has the Young diagram that is the transpose of the
Young diagram for $Y$.

A partition $Y$ is said to be \emph{$(m|n)$-hook} if $Y_{m+1} \leq n$.
If $Y$ is an $(m|n)$-hook partition, then $Y'$ is an $(n|m)$-hook
partition, $Y'_{n+1} \leq m$.  We let $\CH_{m|n}$ denote the set of all
$(m|n)$-hook partitions.

For an $(m|n)$-hook partition $Y$, we define the integral weight
\begin{equation}
  Y^\natural
  =
  (Y_1, \dotsc, Y_m|
  \langle Y'_1- m\rangle, \dotsc, \langle Y'_n - m\rangle) \,,
\end{equation}
where $\langle a\rangle = \max(0,a)$.  The even part of $Y^\natural$
is represented by the Young diagram formed by the first $m$ rows of
$Y$.  The Young diagram for the odd part of $Y^\natural$ is the
transpose of the remainder of $Y$, and its length is less than or
equal to $n$ by the $(m|n)$-hook condition.

Let $Y$ be an $(m|n)$-hook partition.  The \emph{covariant
  representation labeled by $Y$} is the highest-weight representation
$L(Y^\natural)$.  The \emph{contravariant representation labeled by
  $Y$} is the dual representation $L(Y^\natural)^*= L(\Yt^\natural)$.
Its highest weight $\Yt^\natural$ equals the minus of the lowest
weight of $L(Y^\natural)$ and is given by
\begin{equation}
  \Yt^\natural
  =
  (-\langle Y_m - n\rangle, \dotsc, -\langle Y_1- n\rangle|
  -Y'_n, \dotsc, -Y'_1\bigr) \,.
\end{equation}

\subsection{Line defects in covariant and contravariant
  representations}

Now we construct quantum mechanical systems whose Hilbert spaces are
covariant and contravariant representations of $\glf(m|n)$.  Coupled
to four-dimensional Chern--Simons theory with gauge group $\GL(m|n)$,
they describe line defects valued in these finite-dimensional
irreducible representations.

Let $K$, $\Kb$ be nonnegative integers and consider a pair of fields
\begin{align}
  \varphi &\in \Hom(\C^{K|\Kb}, \C^{m|n}) \,,
  \\
  \varphit &\in \Hom(\C^{m|n}, \C^{K|\Kb})
\end{align}
transforming in the bifundamental representations of
$\GL(m|n) \times \GL(K|\Kb)$.  Their components are $\Z_2$-graded, with
the grading given by
\begin{equation}
  [\varphi^i_\alpha] = [\varphit^\alpha_i] = [i] + [\alpha] \,.
\end{equation}
where $i$ and $\alpha$ are indices for $\C^{m|n}$ and $\C^{K|\Kb}$,
respectively.  The even components are bosonic and the odd ones are
fermionic.  The action of the theory is
\begin{equation}
  \label{eq:QM}
  \frac{1}{\hbar} \int_\R
  \str_{\C^{m|n}}(\varphi \, \rmd\varphit) \,.
\end{equation}

We will find it convenient to define
\begin{align}
  \eta^i_\alpha &= \frac{\iu}{\hbar} \varphi^i_\alpha \,,
  \\
  \chi_i^\alpha &= (-1)^{[\alpha]} \varphit_i^\alpha \,,
\end{align}
and
\begin{align}
  \chit^i_\alpha &= (-1)^{[i][\alpha] + [\alpha]} \eta^i_\alpha \,,
  \\
  \etat_i^\alpha &= (-1)^{[i][\alpha] + [\alpha]} \frac{\iu}{\hbar} \chi_i^\alpha \,.
\end{align}
Then, the canonical commutation relations read
\begin{equation}
  [\eta_\alpha^i, \chi_j^\beta]
  = [\chit_\alpha^i, \etat_j^\beta]
  = \delta^\beta_\alpha \delta_j^i \,.
\end{equation}

The theory has a $\GL(m|n) \times \GL(K|\Kb)$ global symmetry.  The
associated conserved charges are
\begin{align}
  q_{ij}
  &=
  \sum_{\alpha=1}^{K+\Kb}
  \chi_i^\alpha \eta_\alpha^j
  + (-1)^{[i]} c \delta_{ij} \,,
  \\
  Q_{\alpha\beta}
  &=
  -\sum_{i=1}^{m+n}
  (-1)^{([\alpha]+[\beta])[\alpha]} \chi_i^\beta \eta_\alpha^i
  + (-1)^{[\alpha]} C \delta_{\alpha\beta}
\end{align}
and satisfy the $\glf(m|n) \oplus \glf(K|\Kb)$ commutation relations:
\begin{align}
  [q_{ij}, q_{kl}]
  &= \delta_{jk} q_{il} - (-1)^{([i]+[j])([k]+[l])} \delta_{li} q_{kj} \,,
  \\
  [Q_{\alpha\beta}, Q_{\gamma\delta}]
  &= \delta_{\beta\gamma} Q_{\alpha\delta}
    - (-1)^{([\alpha]+[\beta])([\gamma]+[\delta])}
    \delta_{\delta\alpha} Q_{\gamma\beta} \,,
  \\
  [q_{ij}, Q_{\alpha\beta}] &= 0 \,.
\end{align}
The constants $c$ and $C$ account for the ambiguity in operator
ordering and will be fixed in a moment. Under
$\GL(m|n) \times \GL(K|\Kb)$, the sets of fields $\{\chi_i^\alpha\}$ and
$\{\chit^i_\alpha\}$ transform as the standard basis vectors for
$\C^{m|n} \otimes (\C^{K|\Kb})^*$ and $(\C^{m|n})^* \otimes \C^{K|\Kb}$,
respectively:
\begin{align}
  \label{eq:[q,varphi]}
  [q_{ij}, \chi_k^\alpha]
  &= \delta_{jk} \chi_i^\alpha
  = \sum_{l=1}^{m+n} (E_{ij})_{lk} \chi_l^\alpha \,,
  \\
  \label{eq:[Q,chi]}
  [Q_{\alpha\beta}, \chi_i^\gamma]
  &= -(-1)^{([\alpha]+[\beta])[\alpha]} \delta_{\alpha\gamma} \chi_i^\beta
  = \sum_{\delta=1}^{K+\Kb} (-\strans{E_{\alpha\beta}})_{\delta\gamma} \chi_i^\delta \,.
  \\
  [q_{ij}, \chit^k_\alpha]
  &= -(-1)^{([i]+[j])[i]} \delta_{ki} \chit^j_\alpha
  = \sum_{l=1}^{m+n} (-\strans{E_{ij}})_{lk} \chit^l_\alpha \,,
  \\
  [Q_{\alpha\beta}, \chit^i_\gamma]
  &= \delta_{\beta\gamma} \chit^i_\alpha
  = \sum_{\delta=1}^{K+\Kb} (E_{\alpha\beta})_{\delta\gamma} \chit^i_\delta \,.
\end{align}

For the construction of line defects we actually break the $\GL(K|\Kb)$
symmetry.  Let us gauge the Borel subgroup of
$\GL(K|\Kb)$.  We introduce an associated gauge field
\begin{equation}
  \CB
  =
  \sum_{\substack{\alpha,\beta = 1 \\ \alpha \leq \beta}}^{K+\Kb}
  \CB^{\alpha\beta} E_{\alpha\beta}
\end{equation}
and couple it to the theory by replacing the de Rham differential
$\rmd$ with $\rmd + \CB$.  For the moment we treat $\CB$ as a
background field and give it a diagonal value
\begin{equation}
  \label{eq:B-masses}
  \CB = \diag(b_1, \dotsc, b_{K+\Kb}) \,.
\end{equation}
In this background, the action becomes
\begin{equation}
  \frac{1}{\hbar} \int_\R \sum_{i=1}^{m+n} \sum_{\alpha=1}^{K+\Kb}
  (-1)^{[i]} (
  \varphi_\alpha^i \rmd\varphit^\alpha_i
  + b_\alpha \varphi_\alpha^i \varphit^\alpha_i) \,,
\end{equation}
and the $\GL(K|\Kb)$ symmetry is broken to the stabilizer of the gauge
field, which is generically the maximal torus.

The coupling to the gauge field does not affect the canonical
commutation relations, but modifies the Hamiltonian.  Before the
introduction of the gauge field, the theory was topological and the
Hamiltonian was zero.  It is now given by
\begin{equation}
  H
  =
  \iu\hbar \sum_{\alpha=1}^{K+\Kb} b_\alpha Q_{\alpha\alpha}
  \,.
\end{equation}
The Hilbert space of the theory is a $\Z_2$-graded Fock space
constructed from a vacuum state $\ket{0}$ by the action of the
creation operators.
The action of $\varphi_\alpha^i$ changes $H$ by $\iu\hbar b_\alpha$
while $\varphit^\alpha_i$ changes $H$ by $-\iu\hbar b_\alpha$.  Those
component fields that increase $\Re(\iu H/\hbar)$ are creation
operators, and those that decrease it are annihilation operators.%
\footnote{If the time axis is compactified to a circle of radius $1$,
  then the partition function involves trace twisted by
  $\exp(-\iu H/\hbar)$.  Creation operators should make this factor
  smaller.}
We can think of $\Re(\iu H/\hbar)$ as energy.

Suppose that we give the background value such that
\begin{equation}
  \label{eq:Im_b-order}
  0 < \Re b_{K+\Kb} < \Re b_{K+\Kb-1} < \dotsb < \Re b_1 \,.
\end{equation}
Then, $\chi^\alpha_i$ is a creation operator and $\eta_\alpha^i$ is an
annihilation operator.  Requiring the vacuum to be invariant under
(the maximal torus of) $\GL(K|\Kb)$, we find $c = C = 0$.  Let $\CF$ be
the corresponding Fock space.

The Fock space $\CF$ decomposes into tensor products of covariant
representations of $\glf(m|n)$ and contravariant representations of
$\glf(K|\Kb)$ \cite{MR0929437, MR1847665}:
\begin{equation}
  \CF
  =
  \bigoplus_{Y \in \CH_{m|n} \cap \CH_{K|\Kb}}
  L(Y^\natural_{m|n}) \otimes  L(Y^\natural_{K|\Kb})^*
  \,.
\end{equation}
(We use subscripts to distinguish weights for $\glf(m|n)$ and
$\glf(K|\Kb)$.)  For example, the first excited states take the form
\begin{equation}
  \sum_{i=1}^{m+n} \sum_{\alpha=1}^{K+\Kb} c^i_\alpha \chi^\alpha_i \ket{0}
\end{equation}
and span a subspace isomorphic to
\begin{equation}
  \C^{m|n} \otimes (\C^{K|\Kb})^* \,,
\end{equation}
as can be seen from the commutation relations~\eqref{eq:[q,varphi]}
and \eqref{eq:[Q,chi]}.

This Hilbert space is too large, and we need to reduce it to a single
covariant representation $L(Y^\natural_{m|n})$ of $\glf(m|n)$.  To do
so, we impose constraints that singles out the summand
$L(Y^\natural_{m|n}) \otimes L(Y^\natural_{K|\Kb})^*$ and further
projects it to the subspace of lowest-energy states.  Since the
raising operator $Q_{\alpha\beta}$, $\alpha < \beta$, changes the
energy $\Re(\iu H/\hbar)$ by $-\Re b_\alpha + \Re b_\beta < 0$, the
lowest-energy states have the highest weight with respect to
$\glf(K|\Kb)$.

We implement this projection by making $\CB$ dynamical.  The vacuum
expectation value of $\CB$ is given by the diagonal
matrix~\eqref{eq:B-masses}.  Let us add to the action the
Chern--Simons term
\begin{equation}
  - \iu \int_\R \underline{\Yt}^\natural_{K|\Kb}(\CB)
  \,.
\end{equation}
Then, the equations of motion for $\CB$ are
\begin{equation}
  Q_{\alpha\beta} = 0 \,,
  \quad
  \alpha < \beta
\end{equation}
and%
\footnote{The Weyl quantization of the classical expression of
  $Q_{\alpha\alpha}$ equals $Q_{\alpha\alpha} + (-1)^{[\alpha]} (m-n)/2$.}
\begin{equation}
  Q_{\alpha\alpha}
  =
  (\underline{\Yt}^\natural_{K|\Kb})_\alpha + \frac{m-n}{2} \,.
\end{equation}
The former equations restrict the Fock space to the subspace of states
that contains highest-weight vectors of covariant representations of
$\glf(K|\Kb)$:
\begin{equation}
  \bigoplus_{Y \in \CH_{m|n} \cap \CH_{K|\Kb}}
  L(Y^\natural_{m|n}) \otimes \ket{\Omega_{\Yt^\natural_{K|\Kb}}} \,.
\end{equation}
With the choice
\begin{equation}
  \Yt^\natural_{K|\Kb}
  =
  \underline{\Yt}^\natural_{K|\Kb}
  - \frac{m-n}{2} \sum_{\alpha=1}^{K+\Kb} \veps_\alpha \,,
\end{equation}
the second equation selects the highest weight $\Yt^\natural_{K|\Kb}$,
thereby reducing the Hilbert space to the covariant representation
$L(Y^\natural_{m|n}) \otimes \ket{\Omega_{\Yt^\natural_{K|\Kb}}}$
of $\glf(m|n)$.

In order to construct line defects in contravariant representations,
we take
\begin{equation}
  \Re b_{K+\Kb} < \Re b_{K+\Kb-1} < \dotsb < \Re b_1 < 0 \,.
\end{equation}
In this case, $\chit_i^\alpha$ is a creation operator and $\etat^i_\alpha$ is an
annihilation operator.  The corresponding Fock space $\CFt$ decomposes
as
\begin{equation}
  \CFt
  =
  \bigoplus_{Y \in \CH_{m|n} \cap \CH_{K|\Kb}}
  L(Y^\natural_{m|n})^* \otimes  L(Y^\natural_{K|\Kb})
  \,.
\end{equation}
Making $\CB$ dynamical and adding the Chern--Simons term
\begin{equation}
  - \iu \int_\R \underline{Y}^\natural_{K|\Kb}(\CB)
\end{equation}
to the action, we can reduce the Hilbert space to the contravariant
representation $L(Y^\natural)^*$, with
\begin{equation}
  Y^\natural_{K|\Kb}
  =
  \underline{Y}^\natural_{K|\Kb}
  - \frac{m-n}{2} \sum_{\alpha=1}^{K+\Kb} \veps_\alpha \,.
\end{equation}

\subsection{Brane construction of line defects}

The quantum mechanical system discussed above can be constructed with
D3-branes and D5-branes.  Let us remove from the brane system
\eqref{eq:brane-4dCS} the semi-infinite D3-branes and the F1-branes
stretched between the D5-branes, and instead introduce infinite
D3-branes $\mathrm{D3}_\alpha$, $\alpha = 1$, $\dotsc$, $K+\Kb$:
\begin{equation}
  \begin{array}{r@{\colon\ }c@{\ \times\ }c@{\ \times\ }c@{\ \times\ }c@{\ \times\ }c@{\ \times\ }c@{\ \times\ }c}
    \mathrm{Spacetime}
    & \R & S^1 & \C & \R_X & \R_Y
    & \R^2_{+\hbar} & \R^2_{-\hbar}
    \\
    \mathrm{D5}_i \ (i \leq m)
    & \R & S^1 & \C
    & \{X_i\} & \{Y_i\} & \R^2_{+\hbar} & \{0\}
    \\
    \mathrm{D5}_i \ (i > m)
    & \R & S^1 & \C
    & \{X_i\} & \{Y_i\} & \{0\} & \R^2_{-\hbar}
    \\
    \mathrm{D3}_\alpha \ (\alpha \leq K)
    & \R & \{y\} & \{\zeta\}
    & \{-\Re b_\alpha\} & \R_Y & \R^2_{+\hbar} & \{0\}
    \\
    \mathrm{D3}_\alpha \ (\alpha > K)
    & \R & \{y\} & \{\zeta\}
    & \{-\Re b_\alpha\} & \R_Y & \{0\} & \R^2_{-\hbar}
  \end{array}
\end{equation}
We claim that strings stretched between the D3-branes and the
D5-branes give rise to the quantum mechanical system in question.

The $K$ D3-branes $\mathrm{D3}_\alpha$, $\alpha \leq K$, and the $m$
D5-branes $\mathrm{D5}_i$, $i \leq m$, share the three-dimensional
spacetime $\R \times \R^2_{+\hbar}$.  Strings stretched between them
produce an $\CN = 4$ hypermultiplet in the bifundamental
representation of $\U(K) \times \U(m)$.  Let
\begin{align}
  \varphi^{00} &\in \Hom(\C^K, \C^m) \,,
  \\
  \varphit^{00} &\in \Hom(\C^m, \C^K)
\end{align}
be the scalar fields of this multiplet.  We are looking at the sector
of this theory that is invariant under the supercharge $Q$ for the
holomorphic--topological twist.  There is an $\Omega$-deformation
induced by the background RR two-form, and it has the effect of
localizing the hypermultiplet to the quantum mechanical model with
action \cite{Yagi:2014toa}
\begin{equation}
  \frac{1}{\hbar} \int_\R
  \tr_{\C^m}(\varphi^{00} \, \rmd\varphit^{00}) \,.
\end{equation}
Here we are using $\varphi^{00}$, $\varphit^{00}$ to denote the
one-dimensional fields that descend from the three-dimensional scalar
fields.

Similarly, from strings stretched between the $\Kb$ D3-branes
$\mathrm{D3}_\alpha$, $\alpha > K$, and the $n$ D5-branes
$\mathrm{D5}_i$, $i > m$, we get an $\CN = 4$ hypermultiplet in the
bifundamental representation of $\U(\Kb) \times \U(n)$ on the
three-dimensional spacetime $\R \times \R^2_{-\hbar}$.  By an
$\Omega$-deformation, the theory localizes to a quantum mechanical
system with action
\begin{equation}
  -\frac{1}{\hbar} \int_\R
  \tr_{\C^n}(\varphi^{11} \, \rmd\varphit^{11}) \,,
\end{equation}
where
\begin{align}
  \varphi^{11} &\in \Hom(\C^\Kb, \C^n) \,,
  \\
  \varphit^{11} &\in \Hom(\C^n, \C^\Kb) \,.
\end{align}

The branes $\mathrm{D3}_\alpha$, $\alpha \leq K$, and $\mathrm{D5}_i$,
$i > m$, intersect along the time axis $\R$, and from strings
stretched between them we get fermions
\begin{align}
  \varphi^{10} &\in \Hom(\C^K, \C^n) \,,
  \\
  \varphit^{10} &\in \Hom(\C^n, \C^K) \,.
\end{align}
They are described by the action
\begin{equation}
  \frac{1}{\hbar} \int_\R
  \tr_{\C^n}(\varphi^{10} \, \rmd\varphit^{10}) \,.
\end{equation}
This is the dimensional reduction of the two-dimensional chiral
fermions that arise from an intersection of D4-branes and D6-branes
\cite{Green:1996dd, Dijkgraaf:2007sw}.

In the same way, from strings stretched between $\mathrm{D3}_\alpha$,
$\alpha > K$, and $\mathrm{D5}_i$, $i \leq m$, we get fermionic fields
\begin{align}
  \varphi^{01} \in \Hom(\C^\Kb, \C^m) \,,
  \\
  \varphit^{01} \in \Hom(\C^m, \C^\Kb) \,,
\end{align}
described by the action
\begin{equation}
  -\frac{1}{\hbar} \int_\R
  \tr_{\C^m}(\varphi^{01} \, \rmd\varphit^{01}) \,.
\end{equation}

These four quantum mechanical systems can be combined into the single
quantum mechanical system described by the action~\eqref{eq:QM}, with
the fields
\begin{align}
  \varphi
  &=
  \begin{pmatrix}
    \varphi^{00} & \varphi^{01} \\
    \varphi^{10} & \varphi^{11}
  \end{pmatrix}
  \,,
  \\
  \varphit
  &=
  \begin{pmatrix}
    \varphit^{00} & \varphit^{01} \\
    \varphit^{10} & \varphit^{11}
  \end{pmatrix} \,.
\end{align}
The creation operator $\chi^\alpha_i$ adds a string stretched between
$\mathrm{D3}_\alpha$ and $\mathrm{D5}_i$.  The annihilation operator
$\eta_\alpha^i$ removes a string between them.

This quantum mechanical system is coupled to the four-dimensional
Chern--Simons theory that arises from the D5-branes and to the BF
theory that arises on the D3-branes.  As in the construction of line
defects valued in Verma modules, the boundary conditions on the BF
theory (at infinity, or at finite distance if we make the D3-branes
end on NS5-branes) breaks the $\GL(K|\Kb)$ gauge symmetry to a Borel
subgroup.  Which Borel subgroup is selected is determined by the
ordering of the D3-branes on $\R_X$.  For the ordering
\eqref{eq:Im_b-order} for $\{\Re b_\alpha\}$, it is the standard Borel
subgroup.

The situation in which there are no strings stretched between the
D3-branes and the D5-branes corresponds to the vacuum of the Fock
space $\CF$.  Here is how the vacuum looks like for $(m|n) = (1|2)$
and $(K|\Kb) = (2|1)$:
\begin{equation}
  \ket{0}
  =
  \begin{tikzpicture}[xscale=1.5, yscale=1.5]
    \draw[PLUS, very thick] (-2.3,0)
    node[below, black] {$\mathrm{D3}_1$} -- (-1.5,0.8);
    \draw[PLUS, very thick] (-1.8,0)
    node[below, black] {$\mathrm{D3}_2$} -- (-1,0.8);
    \draw[white, line width=3pt] (-1.3,0) -- (-0.5,0.8);
    \draw[MINUS, very thick] (-1.3,0)
    node[below, black] {$\mathrm{D3}_3$} -- (-0.5,0.8);
    \draw[white, line width=3pt] (0,0.25) -- (0,0);
    \draw[PLUS, very thick] (0,0)
    node[below, black] {$\mathrm{D5}_1$} -- (0,1);
    \draw[MINUS, very thick] (0.5,0)
    node[below, black] {$\mathrm{D5}_2$} -- (0.5,1);
    \draw[MINUS, very thick] (1,0)
    node[below, black] {$\mathrm{D5}_3$} -- (1,1);
  \end{tikzpicture}
\end{equation}

To project to a covariant representation $L(Y^\natural_{m|n})$
of $\glf(m|n)$, we fix the number of strings ending on each D3-brane.
Let us illustrate how this works with an example in which
$Y = (5,1,1)$.  For this choice of $Y$, we have $Y' = (3,1,1,1,1)$,
$Y^\natural_{m|n} = (5|2,0)$ and $\Yt^\natural_{K|\Kb} = (0,-4|-3)$.
The brane configuration for the highest-weight state of
$L(Y^\natural_{m|n}) \otimes
L(\Yt^\natural_{K|\Kb})$ is the following:
\begin{equation}
  \ket{\Omega_{Y^\natural_{m|n}}} \otimes \ket{\Omega_{\Yt^\natural_{K|\Kb}}}
  =
  \begin{tikzpicture}[xscale=1.5, yscale=1.5]
    \draw[ZERO, very thick]  (-1.1,0.7) -- (0,0.7);
    \draw[ZERO, very thick]  (-1.2,0.6) -- (0,0.6);
    \draw[ZERO, very thick]  (-1.3,0.5) -- (0,0.5);
    \draw[ZERO, very thick]  (-1.4,0.4) -- (0,0.4);
    \draw[white, line width=3pt] (-1.3,0) -- (-0.5,0.8);
    \draw[ZERO, very thick]  (-1.0,0.3) -- (0.5,0.3);
    \draw[ZERO, very thick]  (-1.1,0.2) -- (0.5,0.2);
    \draw[ZERO, very thick]  (-1.2,0.1) -- (0,0.1);
    \draw[PLUS, very thick] (-2.3,0)
    node[below, black] {$\mathrm{D3}_1$} -- (-1.5,0.8);
    \draw[PLUS, very thick] (-1.8,0)
    node[below, black] {$\mathrm{D3}_2$} -- (-1,0.8);
    \draw[MINUS, very thick] (-1.3,0)
    node[below, black] {$\mathrm{D3}_3$} -- (-0.5,0.8);
    \draw[white, line width=3pt] (0,0.15) -- (0,0.35);
    \draw[PLUS, very thick] (0,0)
    node[below, black] {$\mathrm{D5}_1$} -- (0,1);
    \draw[MINUS, very thick] (0.5,0)
    node[below, black] {$\mathrm{D5}_2$} -- (0.5,1);
    \draw[MINUS, very thick] (1,0)
    node[below, black] {$\mathrm{D5}_3$} -- (1,1);
    \node at (-0.6,-0.1) {$\mathrm{F1s}$};
  \end{tikzpicture}
\end{equation}
A string ending on $\mathrm{D5}_i$ from the left contributes $\veps_i$
to the $\glf(m|n)$ weight, and a string ending on $\mathrm{D3}_\alpha$
from the right contributes $-\veps_\alpha$ to the $\glf(K|\Kb)$ weight.
This configuration is the tensor product of two highest-weight vectors
because we cannot shorten any of the strings and stretch it between
another pair of D3-brane and D5-brane; by doing so we would get more
than one strings between $\mathrm{D3}_3$ and $\mathrm{D5}_1$, but that
is prohibited as such strings have necessarily coincident worldsheets
and are fermionic.  This brane diagram shows
\begin{equation}
  \ket{\Omega_{Y^\natural_{m|n}}} \otimes \ket{\Omega_{\Yt^\natural_{K|\Kb}}}
  = (\chi^3_2)^2 \chi^3_1 (\chi^2_1)^4 \ket{0} \,.
\end{equation}

The other vectors in
$L(Y^\natural_{m|n}) \otimes \ket{\Omega_{\Yt^\natural_{K|\Kb}}}$
can also be represented by brane configurations.  For example,
\begin{equation}
  \begin{split}
    q_{21} \ket{\Omega_{Y^\natural_{m|n}}} \otimes \ket{\Omega_{\Yt^\natural_{K|\Kb}}}
    &=
    \sum_{\alpha=1}^{3}
    \chi_2^\alpha \eta_\alpha^1
    (\chi^3_2)^2 \chi^3_1 (\chi^2_1)^4 \ket{0}
    \\
    &=
    -4 (\chi^3_2)^2 \chi^3_1 \chi_2^2 (\chi^2_1)^3 \ket{0}
    +
    (\chi^3_2)^3 (\chi^2_1)^4 \ket{0}
  \end{split}
\end{equation}
is a linear combination of two states, which one obtains from the
highest-weight vector by extending one of the strings to the right:
\begin{multline}
  q_{21} \ket{\Omega_{Y^\natural_{m|n}}} \otimes \ket{\Omega_{\Yt^\natural_{K|\Kb}}}
  =
  -4
  \begin{tikzpicture}[xscale=1.5, yscale=1.5]
    \draw[ZERO, very thick]  (-1.1,0.7) -- (0.5,0.7);
    \draw[ZERO, very thick]  (-1.2,0.6) -- (0,0.6);
    \draw[ZERO, very thick]  (-1.3,0.5) -- (0,0.5);
    \draw[ZERO, very thick]  (-1.4,0.4) -- (0,0.4);
    \draw[white, line width=3pt] (-1.3,0) -- (-0.5,0.8);
    \draw[ZERO, very thick]  (-1.0,0.3) -- (0.5,0.3);
    \draw[ZERO, very thick]  (-1.1,0.2) -- (0.5,0.2);
    \draw[ZERO, very thick]  (-1.2,0.1) -- (0,0.1);
    \draw[PLUS, very thick] (-2.3,0)
    node[below, black] {$\mathrm{D3}_1$} -- (-1.5,0.8);
    \draw[PLUS, very thick] (-1.8,0)
    node[below, black] {$\mathrm{D3}_2$} -- (-1,0.8);
    \draw[MINUS, very thick] (-1.3,0)
    node[below, black] {$\mathrm{D3}_3$} -- (-0.5,0.8);
    \draw[white, line width=3pt] (0,0.15) -- (0,0.35) (0,0.65) -- (0,0.75);
    \draw[PLUS, very thick] (0,0)
    node[below, black] {$\mathrm{D5}_1$} -- (0,1);
    \draw[MINUS, very thick] (0.5,0)
    node[below, black] {$\mathrm{D5}_2$} -- (0.5,1);
    \draw[MINUS, very thick] (1,0)
    node[below, black] {$\mathrm{D5}_3$} -- (1,1);
  \end{tikzpicture}
  \\
  +
  \begin{tikzpicture}[xscale=1.5, yscale=1.5]
    \draw[ZERO, very thick]  (-1.1,0.7) -- (0,0.7);
    \draw[ZERO, very thick]  (-1.2,0.6) -- (0,0.6);
    \draw[ZERO, very thick]  (-1.3,0.5) -- (0,0.5);
    \draw[ZERO, very thick]  (-1.4,0.4) -- (0,0.4);
    \draw[white, line width=3pt] (-1.3,0) -- (-0.5,0.8);
    \draw[ZERO, very thick]  (-1.0,0.3) -- (0.5,0.3);
    \draw[ZERO, very thick]  (-1.1,0.2) -- (0.5,0.2);
    \draw[ZERO, very thick]  (-1.2,0.1) -- (0.5,0.1);
    \draw[PLUS, very thick] (-2.3,0)
    node[below, black] {$\mathrm{D3}_1$} -- (-1.5,0.8);
    \draw[PLUS, very thick] (-1.8,0)
    node[below, black] {$\mathrm{D3}_2$} -- (-1,0.8);
    \draw[MINUS, very thick] (-1.3,0)
    node[below, black] {$\mathrm{D3}_3$} -- (-0.5,0.8);
    \draw[white, line width=3pt] (0,0.35) -- (0,0);
    \draw[PLUS, very thick] (0,0)
    node[below, black] {$\mathrm{D5}_1$} -- (0,1);
    \draw[MINUS, very thick] (0.5,0)
    node[below, black] {$\mathrm{D5}_2$} -- (0.5,1);
    \draw[MINUS, very thick] (1,0)
    node[below, black] {$\mathrm{D5}_3$} -- (1,1);
  \end{tikzpicture}
\end{multline}

The construction of a line defect in a contravariant representation of
$\glf(m|n)$ is analogous.  For a contravariant representation, the
D3-branes are placed to the right of the D5-branes.  For example, for
the same choice $(m|n) = (1|2)$, $(K|\Kb) = (2|1)$ and $Y = (5,1,1)$, we
have $\Yt^\natural_{m|n} = (-3|-1, -3)$ and
$Y^\natural_{K|\Kb} = (5,1|1)$, and the highest-weight vector is
represented by the configuration
\begin{equation}
  \begin{tikzpicture}[xscale=1.5, yscale=1.5]
    \draw[ZERO, very thick]  (1,0.7) -- (3.2,0.7);
    \draw[ZERO, very thick]  (1,0.1) -- (1.6,0.1);
    \draw[ZERO, very thick]  (1,0.6) -- (2.6,0.6);
    \draw[ZERO, very thick]  (0,0.5) -- (2,0.5);
    \draw[ZERO, very thick]  (0,0.4) -- (1.9,0.4);
    \draw[ZERO, very thick]  (0,0.3) -- (1.8,0.3);
    \draw[ZERO, very thick]  (0.5,0.2) -- (1.7,0.2);
    \draw[white, line width=3pt] (2.05,0.55) -- (2.3,0.8);
    \draw[white, line width=3pt] (2.65,0.65) -- (2.8,0.8);
    \draw[white, line width=3pt] (0.5,0.25) -- (0.5,0.55);
    \draw[white, line width=3pt] (1,0.15) -- (1,0.55);
    \draw[PLUS, very thick] (1.5,0)
    node[below, black] {$\mathrm{D3}_1$} -- (2.3,0.8);
    \draw[PLUS, very thick] (2,0)
    node[below, black] {$\mathrm{D3}_2$} -- (2.8,0.8);
    \draw[MINUS, very thick] (2.5,0)
    node[below, black] {$\mathrm{D3}_3$} -- (3.3,0.8);
    \draw[PLUS, very thick] (0,0)
    node[below, black] {$\mathrm{D5}_1$} -- (0,1);
    \draw[MINUS, very thick] (0.5,0)
    node[below, black] {$\mathrm{D5}_2$} -- (0.5,1);
    \draw[MINUS, very thick] (1,0)
    node[below, black] {$\mathrm{D5}_3$} -- (1,1);
  \end{tikzpicture}
\end{equation}

\subsection{Two-dimensional $\CN = (2,2)$ supersymmetric
  gauge theories}
\label{sec:finite-reps-2d}

Applying S-duality and T-duality on $S^1$ to the brane configurations
for a line defect, we obtain D2--D4--NS5 brane configurations which
describe two-dimensional $\CN = (2,2)$ supersymmetric field theories.
For a general choice of $(K|\Kb)$ and $Y$, the resulting theory does not
seem to admit a simple gauge theory description.

If we restrict to covariant representations with $K = 0$ and
$Y_{m+1} = 0$ and contravariant representations with $\Kb = 0$ and
$Y'_{n+1} = 0$, the two-dimensional theories are particularly nice.
Let us consider these cases.

For a covariant representation with $K = 0$ and $Y_{m+1} = 0$, the
relevant highest weights are
$Y^\natural_{m|n} = (Y_1, \dotsc, Y_m| 0, \dots, 0)$ and
$\Yt^\natural_{K|\Kb} = (-Y'_\Kb, \dots, -Y'_1)$.  The brane configuration
for the highest-weight vector of
$L(Y^\natural_{m|n}) \otimes L(\Yt^\natural_{K|\Kb})$ has one string
stretched between $\mathrm{NS5}_i$ and $\mathrm{D3}_{\Kb-\alpha+1}$ for each
$\alpha = 1$, $\dotsc$, $Y_i$. The following diagram depicts the
highest-weight vector for $(m|n) = (3|2)$, $(K|\Kb) = (0|4)$ and
$Y = (4,2,1)$, for which $Y' = (3,2,1,1)$,
$Y^\natural_{m|n} = (4,2,1|0,0)$ and
$\Yt^\natural_{K|\Kb} = (-1,-1,-2,-3)$:
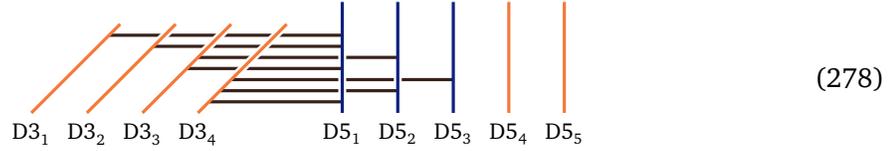
\begin{equation}
  \begin{tikzpicture}[xscale=1.5, yscale=1.5]
    \draw[ZERO, very thick]  (-2.1,0.7) -- (0,0.7);
    \draw[ZERO, very thick]  (-1.7,0.6) -- (0,0.6);
    \draw[ZERO, very thick]  (-1.3,0.5) -- (0.5,0.5);
    \draw[ZERO, very thick]  (-1.4,0.4) -- (0,0.4);
    \draw[ZERO, very thick]  (-1.0,0.3) -- (1,0.3);
    \draw[ZERO, very thick]  (-1.1,0.2) -- (0.5,0.2);
    \draw[ZERO, very thick]  (-1.2,0.1) -- (0,0.1);
    \draw[white, line width=3pt] (-1.65,0.65) -- (-1.5,0.8)
    (-1.25,0.55) -- (-1,0.8) (-0.95,0.35) -- (-0.5,0.8) (0,0.15) --
    (0,0.35) (0,0.45) -- (0,0.55) (0.5,0.25) -- (0.5,0.35);
    \draw[MINUS, very thick] (-2.8,0)
    node[below, black] {$\mathrm{D3}_1$} -- (-2,0.8);
    \draw[MINUS, very thick] (-2.3,0)
    node[below, black] {$\mathrm{D3}_2$} -- (-1.5,0.8);
    \draw[MINUS, very thick] (-1.8,0)
    node[below, black] {$\mathrm{D3}_3$} -- (-1,0.8);
    \draw[MINUS, very thick] (-1.3,0)
    node[below, black] {$\mathrm{D3}_4$} -- (-0.5,0.8);
    \draw[PLUS, very thick] (0,0)
    node[below, black] {$\mathrm{D5}_1$} -- (0,1);
    \draw[PLUS, very thick] (0.5,0)
    node[below, black] {$\mathrm{D5}_2$} -- (0.5,1);
    \draw[PLUS, very thick] (1,0)
    node[below, black] {$\mathrm{D5}_3$} -- (1,1);
    \draw[MINUS, very thick] (1.5,0)
    node[below, black] {$\mathrm{D5}_4$} -- (1.5,1);
    \draw[MINUS, very thick] (2,0)
    node[below, black] {$\mathrm{D5}_5$} -- (2,1);
  \end{tikzpicture}
\end{equation}

By moving the D3-branes past D5-branes, we can bring this
configuration to another configuration in which the D3-branes are
located between D5-branes and have no strings attached:%
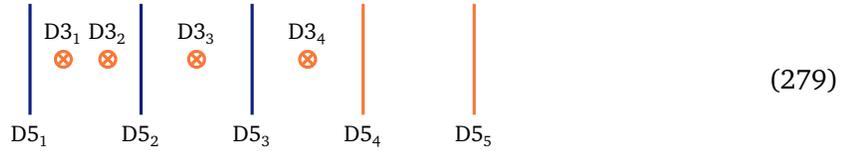
\footnote{An obstruction to generalize the present argument to more
  general covariant and contravariant representations is that we do not
  understand what happens when a D3-brane passes through a D5-brane of
  the same color in the presence of the RR two-form for
  $\Omega$-deformation.}
\begin{equation}
  \begin{tikzpicture}[xscale=3, yscale=1.5]
    \node[MINUS, very thick, gnode, minimum size=6pt] at (1.25,0.5) {};
    \draw (1.25,0.5) node[cross=3pt, MINUS, very thick] {};
    \node[above=2pt, black] at (1.25,0.5) {$\mathrm{D3}_4$};
    \node[MINUS, very thick, gnode, minimum size=6pt] at (0.75,0.5) {};
    \draw (0.75,0.5) node[cross=3pt, MINUS, very thick] {};
    \node[above=2pt, black] at (0.75,0.5) {$\mathrm{D3}_3$};
    \node[MINUS, very thick, gnode, minimum size=6pt] at (0.35,0.5) {};
    \draw (0.35,0.5) node[cross=3pt, MINUS, very thick] {};
    \node[above=2pt, black] at (0.35,0.5) {$\mathrm{D3}_2$};
    \node[MINUS, very thick, gnode, minimum size=6pt] at (0.15,0.5) {};
    \draw (0.15,0.5) node[cross=3pt, MINUS, very thick] {};
    \node[above=2pt, black] at (0.15,0.5) {$\mathrm{D3}_1$};
    \draw[PLUS, very thick] (0,0)
    node[below, black] {$\mathrm{D5}_1$} -- (0,1);
    \draw[PLUS, very thick] (0.5,0)
    node[below, black] {$\mathrm{D5}_2$} -- (0.5,1);
    \draw[PLUS, very thick] (1,0)
    node[below, black] {$\mathrm{D5}_3$} -- (1,1);
    \draw[MINUS, very thick] (1.5,0)
    node[below, black] {$\mathrm{D5}_4$} -- (1.5,1);
    \draw[MINUS, very thick] (2,0)
    node[below, black] {$\mathrm{D5}_5$} -- (2,1);
  \end{tikzpicture}
\end{equation}
The strings that were initially present get annihilated by the
Hanany--Witten transition.  The number of D3-branes between
$\mathrm{D5}_i$ and $\mathrm{D5}_{i+1}$ is equal to $Y_i - Y_{i+1}$.

If we stretch strings between D5-branes in this configuration, then by
the reverse Hanany--Witten moves we get a configuration for excited
states in
$L(Y^\natural_{m|n}) \otimes \ket{\Omega_{\Yt^\natural_{K|\Kb}}}$.  For
example, the configuration
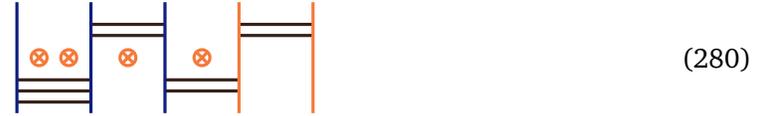
\begin{equation}
  \label{eq:HW-covariant}
  \begin{tikzpicture}[xscale=2, yscale=1.5]
    \draw[ZERO, very thick] (0,0.1) -- (0.5,0.1);
    \draw[ZERO, very thick] (0,0.2) -- (0.5,0.2);
    \draw[ZERO, very thick] (0,0.3) -- (0.5,0.3);
    \draw[ZERO, very thick] (0.5,0.7) -- (1,0.7);
    \draw[ZERO, very thick] (0.5,0.8) -- (1,0.8);
    \draw[ZERO, very thick] (1,0.2) -- (1.5,0.2);
    \draw[ZERO, very thick] (1.5,0.7) -- (2,0.7);
    \draw[ZERO, very thick] (1,0.3) -- (1.5,0.3);
    \draw[ZERO, very thick] (1.5,0.8) -- (2,0.8);
    \node[MINUS, very thick, gnode, minimum size=6pt] at (1.25,0.5) {};
    \draw (1.25,0.5) node[cross=3pt, MINUS, very thick] {};
    \node[MINUS, very thick, gnode, minimum size=6pt] at (0.75,0.5) {};
    \draw (0.75,0.5) node[cross=3pt, MINUS, very thick] {};
    \node[MINUS, very thick, gnode, minimum size=6pt] at (0.35,0.5) {};
    \draw (0.35,0.5) node[cross=3pt, MINUS, very thick] {};
    \node[MINUS, very thick, gnode, minimum size=6pt] at (0.15,0.5) {};
    \draw (0.15,0.5) node[cross=3pt, MINUS, very thick] {};
    \draw[PLUS, very thick] (0,0)
    -- (0,1);
    \draw[PLUS, very thick] (0.5,0)
    -- (0.5,1);
    \draw[PLUS, very thick] (1,0)
    -- (1,1);
    \draw[MINUS, very thick] (1.5,0)
    -- (1.5,1);
    \draw[MINUS, very thick] (2,0)
    -- (2,1);
  \end{tikzpicture}
\end{equation}
represents an excited state with weight
$(4,2,1|0,0) - 3 \alpha_1 - 2\alpha_2 - 2\alpha_3 - 2\alpha_4 =
(1,3,1|0,2)$.  One such state is represented by the configuration
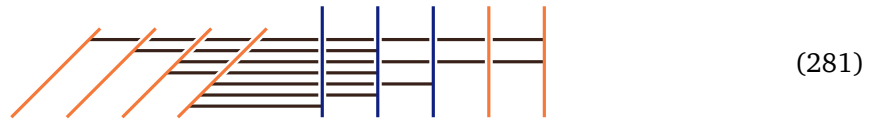
\begin{equation}
  \label{eq:HW-contravariant}
  \begin{tikzpicture}[xscale=1.5, yscale=1.5]
    \draw[ZERO, very thick]  (-2.1,0.7) -- (2,0.7);
    \draw[ZERO, very thick]  (-1.7,0.6) -- (0.5,0.6);
    \draw[ZERO, very thick]  (-1.3,0.5) -- (2,0.5);
    \draw[ZERO, very thick]  (-1.4,0.4) -- (0.5,0.4);
    \draw[ZERO, very thick]  (-1.0,0.3) -- (1,0.3);
    \draw[ZERO, very thick]  (-1.1,0.2) -- (0.5,0.2);
    \draw[ZERO, very thick]  (-1.2,0.1) -- (0,0.1);
    \draw[white, line width=3pt]
    (-1.65,0.65) -- (-1.5,0.8)
    (-1.25,0.55) -- (-1,0.8)
    (-0.95,0.35) -- (-0.5,0.8)
    (0,0.15) -- (0,0.75)
    (0.5,0.65) -- (0.5,0.75)
    (0.5,0.45) -- (0.5,0.55)
    (0.5,0.25) -- (0.5,0.35)
    (1,0.45) -- (1,0.75)
    (1.5,0.45) -- (1.5,0.75);
    \draw[MINUS, very thick] (-2.8,0)
    -- (-2,0.8);
    \draw[MINUS, very thick] (-2.3,0)
    -- (-1.5,0.8);
    \draw[MINUS, very thick] (-1.8,0)
    -- (-1,0.8);
    \draw[MINUS, very thick] (-1.3,0)
    -- (-0.5,0.8);
    \draw[PLUS, very thick] (0,0)
    -- (0,1);
    \draw[PLUS, very thick] (0.5,0)
    -- (0.5,1);
    \draw[PLUS, very thick] (1,0)
    -- (1,1);
    \draw[MINUS, very thick] (1.5,0)
    -- (1.5,1);
    \draw[MINUS, very thick] (2,0)
    -- (2,1);
  \end{tikzpicture}
\end{equation}

Similarly, for a contravariant representation with $\Kb = 0$ and
$Y'_{n+1} = 0$, we have
$\Yt^\natural_{m|n} \linebreak[1]= (0, \dotsc, 0|-Y'_n, \dotsc, -Y'_1)$ and
$Y^\natural_{K|\Kb} = (Y_1, \dotsc, Y_K)$, and the brane configuration
for the highest-weight vector of
$L(\Yt^\natural_{m|n}) \otimes L(Y^\natural_{K|\Kb})$ can be brought to
a configuration without any strings.  Take an example with
$(K|\Kb) = (3|0)$ and $Y = (2,2,1)$, for which $Y' = (3,2)$,
$\Yt^\natural_{m|n} = (0, 0, 0|-2, -3)$ and
$Y^\natural_{K|\Kb} = (2,2,1)$:
\begin{equation}
  \begin{tikzpicture}[xscale=2, yscale=1.5]
    \draw[ZERO, very thick] (0,0.5) -- (0.5,0.5);
    \draw[ZERO, very thick] (0.5,0.7) -- (1,0.7);
    \draw[ZERO, very thick] (1.5,0.7) -- (2,0.7);
    \draw[ZERO, very thick] (1,0.3) -- (1.5,0.3);
    \node[PLUS, very thick, gnode, minimum size=6pt] at (1.75,0.5) {};
    \draw (1.75,0.5) node[cross=3pt, PLUS, very thick] {};
    \node[PLUS, very thick, gnode, minimum size=6pt] at (1.35,0.5) {};
    \draw (1.35,0.5) node[cross=3pt, PLUS, very thick] {};
    \node[PLUS, very thick, gnode, minimum size=6pt] at (1.15,0.5) {};
    \draw (1.15,0.5) node[cross=3pt, PLUS, very thick] {};
    \draw[PLUS, very thick] (0,0)
    -- (0,1);
    \draw[PLUS, very thick] (0.5,0)
    -- (0.5,1);
    \draw[PLUS, very thick] (1,0)
    -- (1,1);
    \draw[MINUS, very thick] (1.5,0)
    -- (1.5,1);
    \draw[MINUS, very thick] (2,0)
    -- (2,1);
  \end{tikzpicture}
\end{equation}
This configuration represents an excited state with weight
$(0,0,0|-2,-3) - \alpha_1 - \alpha_2 - \alpha_3 - \alpha_4 \linebreak[1]=
(-1,0,0|-2,0)$.  The number of D3-branes between an adjacent pair of
D5-branes can be read off from $Y'$.

The dual D2--D4--NS5 brane configurations are of the type studied by
Hanany and Hori~\cite{Hanany:1997vm} and realize quiver gauge theories.
The quiver for the configuration~\eqref{eq:HW-covariant} is
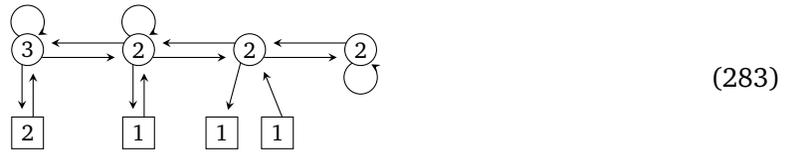
\begin{equation}
  \begin{tikzpicture}[scale=1.5]
    \draw[q->, shorten >=9pt] (1.37,0.56) --
    (0.5,0.56);
    \draw[q->, shorten >=9pt] (0.63,0.43) --
    (1.5,0.43);

    \draw[q->, shorten >=9pt] (2.37,0.56) --
    (1.5,0.56);
    \draw[q->, shorten >=9pt] (1.63,0.43) --
    (2.5,0.43);

    \draw[q->, shorten >=9pt] (3.37,0.56) --
    (2.5,0.56);
    \draw[q->, shorten >=9pt] (2.63,0.43) --
    (3.5,0.43);

    \draw[q->] (0.5,0.6) arc (270:-60:0.15)
    ;
    \draw[q->] (1.5,0.6) arc (270:-60:0.15)
    ;
    \draw[q->] (3.5,0.4) arc (-270:60:0.15);

    \draw[q->, shorten >=9pt] (0.45,0.5) --
    (0.45,-0.25);

    \draw[q->, shorten >=9pt] (0.55,-0.25) --
    (0.55,0.5);

    \draw[q->, shorten >=9pt] (1.45,0.5) --
    (1.45,-0.25);
    \draw[q->, shorten >=9pt] (1.55,-0.25) --
    (1.55,0.5);

    \draw[q->, shorten >=9pt] (2.45,0.5) --
    (2.25,-0.25);
    \draw[q->, shorten >=9pt] (2.85,-0.25) --
    (2.55,0.5);

    \node[gnode] at (0.5,0.5) {$3$};
    \node[gnode] at (1.5,0.5) {$2$};
    \node[gnode] at (2.5,0.5) {$2$};
    \node[gnode] at (3.5,0.5) {$2$};

    \node[fnode] at (0.5,-0.25) {$2$};
    \node[fnode] at (1.5,-0.25) {$1$};
    \node[fnode] at (2.25,-0.25) {$1$};
    \node[fnode] at (2.75,-0.25) {$1$};
\end{tikzpicture}
\end{equation}
and the quiver for the configuration~\eqref{eq:HW-contravariant} is
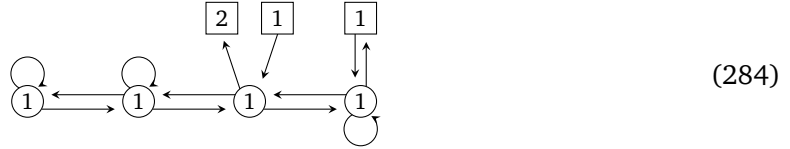
\begin{equation}
  \begin{tikzpicture}[scale=1.5]
    \draw[q->, shorten >=9pt] (1.37,0.56) -- (0.5,0.56);
    \draw[q->, shorten >=9pt] (0.63,0.43)
    -- (1.5,0.43);

    \draw[q->, shorten >=9pt] (2.37,0.56)
    -- (1.5,0.56);
    \draw[q->, shorten >=9pt] (1.63,0.43)
    -- (2.5,0.43);

    \draw[q->, shorten >=9pt] (3.37,0.56)
    -- (2.5,0.56);
    \draw[q->, shorten >=9pt] (2.63,0.43)
    -- (3.5,0.43);

    \draw[q->] (0.5,0.6) arc (270:-60:0.15);
    \draw[q->] (1.5,0.6) arc (270:-60:0.15);
    \draw[q->] (3.5,0.4) arc (-270:60:0.15);

    \draw[q->, shorten >=9pt] (2.45,0.5)
    -- (2.2,1.25);
    \draw[q->, shorten >=9pt] (2.8,1.25)
    -- (2.55,0.5);

    \draw[q->, shorten >=9pt] (3.55,0.5)
    -- (3.55,1.25);
    \draw[q->, shorten >=9pt] (3.45,1.25)
    -- (3.45,0.5);

    \node[gnode] at (0.5,0.5) {$1$};
    \node[gnode] at (1.5,0.5) {$1$};
    \node[gnode] at (2.5,0.5) {$1$};
    \node[gnode] at (3.5,0.5) {$1$};

    \node[fnode] at (2.25,1.25) {$2$};
    \node[fnode] at (2.75,1.25) {$1$};
    \node[fnode] at (3.5,1.25) {$1$};
\end{tikzpicture}
\end{equation}
The ranks of the gauge nodes are given by the numbers of F1-branes,
and the ranks of the flavor nodes are given by the numbers of
D3-branes.  There are $\CN = (4,4)$ cubic superpotential terms
involving adjoint chiral multiplets.

The flavor symmetry for the chiral multiplets charged under the $m$th
gauge node is doubled due to the lack of cubic superpotential term.
From the brane point of view, this is because a D4-brane between
$\mathrm{NS5}_m$ and $\mathrm{NS5}_{m+1}$ can be broken into half on
one of the NS5-branes which has the same color as the D4-brane:
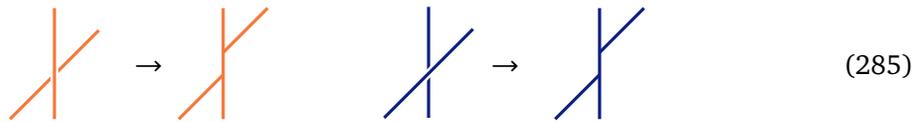
\begin{equation}
  \begin{tikzpicture}[xscale=1.5, yscale=1.5]
    \draw[MINUS, very thick] (-0.4,0) -- (0.4,0.8);
    \draw[white, line width=3pt] (0,0) -- (0,1);
    \draw[MINUS, very thick] (0,0) -- (0,1);
  \end{tikzpicture}
  \quad \to \
  \begin{tikzpicture}[xscale=1.5, yscale=1.5]
    \draw[MINUS, very thick] (-0.4,0) -- (-0,0.4);
    \draw[MINUS, very thick] (0,0.6) -- (0.4,1);
    \draw[MINUS, very thick] (0,0) -- (0,1);
  \end{tikzpicture}
  \qquad
  \qquad
  \begin{tikzpicture}[xscale=1.5, yscale=1.5]
    \draw[PLUS, very thick] (0,0) -- (0,1);
    \draw[white, line width=3pt] (-0.4,0) -- (0.4,0.8);
    \draw[PLUS, very thick] (-0.4,0) -- (0.4,0.8);
  \end{tikzpicture}
  \ \to \quad
  \begin{tikzpicture}[xscale=1.5, yscale=1.5]
    \draw[PLUS, very thick] (-0.4,0) -- (-0,0.4);
    \draw[PLUS, very thick] (0,0.6) -- (0.4,1);
    \draw[PLUS, very thick] (0,0) -- (0,1);
  \end{tikzpicture}
\end{equation}
Strings stretched between $\mathrm{D2}_m^{a_m}$, $a_m = 1$, $\dotsc$,
$M_m$, and one half of the D4-brane produce a fundamental chiral
multiplet for $\U(M_m)$, while strings between those D2-branes and the
other half of the D4-brane produce an antifundamental chiral
multiplet.

\subsection{Bethe/gauge correspondence for finite-dimensional
  representations}

Generalizing the above brane construction, we can obtain the
Bethe/gauge correspondence for spins valued in arbitrary
finite-dimensional representations of $\glf(m|n)$.  This is
essentially the correspondence proposed by
Nekrasov~\cite{Nekrasov:2018gne}.

Consider the rational $\glf(m|n)$ spin chain of length $L$, with the
$\ell$th spin takes values in finite-dimensional highest-weight
representations $L(\lambda^\ell)$.  For the moment, let us assume that
the highest weights are all integral and satisfy
\begin{equation}
  \label{eq:lambda-inequalities}
  \lambda_1^\ell \geq \dotsb \lambda_m^\ell
  \geq 0 \geq \lambda_{m+1}^\ell \geq \dotsb \geq \lambda_{m+n}^\ell \,.
\end{equation}
This is the case if all of the representations are of the type studied
in section~\ref{sec:finite-reps-2d}.  We define nonnegative integers
\begin{alignat}{2}
  K_r^\ell &= \lambda_r^\ell - \lambda_{r+1}^\ell \,,
  & \quad
  r &= 1, \dotsc, m-1,
  \\
  K_m^\ell &= \lambda_m^\ell \,,
  \\
  \Kb_r^\ell &= \lambda_r^\ell - \lambda_{r+1}^\ell \,,
  &\quad
  r &= m+1, \dotsc, m+n-1,
  \\
  \Kb_m^\ell &= -\lambda_{m+1}^\ell \,.
\end{alignat}
We look at a sector of fix magnon numbers $(M_1, \dotsc, M_{m+n-1})$.
The Bethe equations depend only on the highest weights and the magnon
numbers, so their form remain the same as in the case of Verma
modules.

The gauge theory corresponding to this magnon sector is similar to the
theory discussed in section~\ref{sec:gauge-side} and has the same
gauge symmetry.  The difference is that the chiral multiplets
$\Qsf_i$, $\Qsft_i$, $i = 1$, $\dotsc$, $m+n$, are replaced by chiral
multiplets
\begin{alignat}{3}
  \Rsf_r^\ell &\in \Hom(\C^{K_r^\ell},\C^{M_r}) \,,
  &\quad r &= 1, \dotsc, m \,,
  \\
  \Rsft_r^\ell &\in \Hom(\C^{M_r},\C^{K_r^\ell}) \,,
  & \quad r &= 1, \dotsc, m \,,
  \\
  \Ssf_r^\ell &\in \Hom(\C^{K_r^\ell},\C^{M_r}) \,,
  &\quad r &= m+1, \dotsc, m+n-1 \,,
  \\
  \Ssft_r^\ell &\in \Hom(\C^{M_r},\C^{K_r^\ell}) \,,
  & \quad r &= m+1, \dotsc, m+n-1 \,.
\end{alignat}
Letting
\begin{align}
  K_r &= \sum_{\ell=1}^L K_r^\ell \,,
  \\
  \Kb_r &= \sum_{\ell=1}^L \Kb_r^\ell \,,
\end{align}
we can combine them into chiral multiplets
$\Rsf_r \in \Hom(\C^{K_r}, \C^{M_r})$,
$\Rsft_r \in \Hom(\C^{M_r},\C^{K_r})$, $r = 1$, $\dotsc$, $m-1$, and 
$\Ssf_r \in \Hom(\C^{K_r}, \C^{M_r})$,
$\Ssft_r \in \Hom(\C^{M_r},\C^{K_r})$, $r = m+1$, $\dotsc$, $m+n-1$.

For $r < m$, the theory has the $\CN = (4,4)$ superpotential term
$\tr_{\C^{K_r}}(\Rsft_r \upphi_r \Rsf_r)$, and a flavor symmetry
$\U(K_r)$ act on $\Rsf_r$ and $\Rsft_r$.  For $r = m$, the cubic
superpotential is absent and two copies of $\U(K_m)$ act separately on
$\Rsf_m$ and $\Rsft_m$.  Similar statements hold for $\Ssf_r$ and
$\Ssft_r$.  Under $\U(1)_{\hbar}$, $\Rsf_r$ and $\Rsft_r$ have charge
$-1$ and $\Ssf_r$ and $\Ssft_r$ have charge $+1$.

The gauge and matter contents of the theory can be encoded in a
quiver.  For $(m|n) = (3,2)$, the quiver is
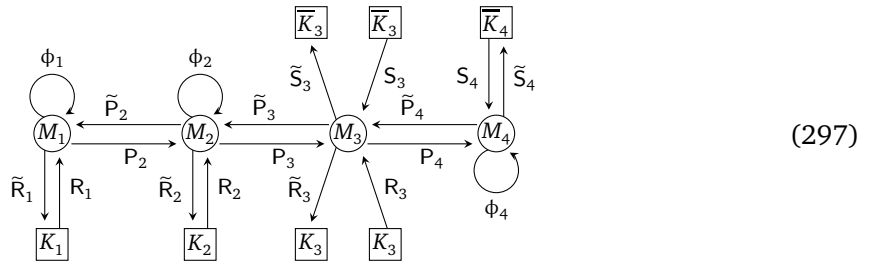
\begin{equation}
  \begin{tikzpicture}[scale=2]
    \draw[q->, shorten >=9pt] (1.37,0.56)
    -- node[above=-2pt]{$\Psft_2$} (0.5,0.56);
    \draw[q->, shorten >=9pt] (0.63,0.43)
    -- node[below=-2pt]{$\Psf_2$} (1.5,0.43);

    \draw[q->, shorten >=9pt] (2.37,0.56)
    -- node[above=-2pt]{$\Psft_3$} (1.5,0.56);
    \draw[q->, shorten >=9pt] (1.63,0.43)
    -- node[below=-2pt]{$\Psf_3$} (2.5,0.43);

    \draw[q->, shorten >=9pt] (3.37,0.56)
    -- node[above=-2pt]{$\Psft_4$} (2.5,0.56);
    \draw[q->, shorten >=9pt] (2.63,0.43)
    -- node[below=-2pt]{$\Psf_4$} (3.5,0.43);

    \draw[q->] (0.5,0.6) arc (270:-60:0.15) node[shift={(-4pt,22pt)}]{$\upphi_1$};
    \draw[q->] (1.5,0.6) arc (270:-60:0.15) node[shift={(-4pt,22pt)}]{$\upphi_2$};
    \draw[q->] (3.5,0.4) arc (-270:60:0.15) node[shift={(-4pt,-22pt)}]{$\upphi_4$};

    \draw[q->, shorten >=9pt] (0.45,0.5)
    -- node[left]{$\Rsft_1$} (0.45,-0.25);
    \draw[q->, shorten >=9pt] (0.55,-0.25)
    -- node[right]{$\Rsf_1$} (0.55,0.5);

    \draw[q->, shorten >=9pt] (1.45,0.5)
    -- node[left]{$\Rsft_2$} (1.45,-0.25);
    \draw[q->, shorten >=9pt] (1.55,-0.25)
    -- node[right]{$\Rsf_2$} (1.55,0.5);

    \draw[q->, shorten >=9pt] (2.45,0.5)
    -- node[left]{$\Rsft_3$} (2.2,-0.25);
    \draw[q->, shorten >=9pt] (2.8,-0.25)
    -- node[right]{$\Rsf_3$} (2.55,0.5);

    \draw[q->, shorten >=9pt] (2.45,0.5)
    -- node[left]{$\Ssft_3$} (2.2,1.25);
    \draw[q->, shorten >=9pt] (2.8,1.25)
    -- node[right]{$\Ssf_3$} (2.55,0.5);

    \draw[q->, shorten >=9pt] (3.55,0.5)
    -- node[right]{$\Ssft_4$} (3.55,1.25);
    \draw[q->, shorten >=9pt] (3.45,1.25)
    -- node[left]{$\Ssf_4$} (3.45,0.5);

    \node[gnode] at (0.5,0.5) {$M_1$};
    \node[gnode] at (1.5,0.5) {$M_2$};
    \node[gnode] at (2.5,0.5) {$M_3$};
    \node[gnode] at (3.5,0.5) {$M_4$};

    \node[fnode] at (0.5,-0.25) {$K_1$};
    \node[fnode] at (1.5,-0.25) {$K_2$};
    \node[fnode] at (2.25,-0.25) {$K_3$};
    \node[fnode] at (2.75,-0.25) {$K_3$};
    \node[fnode] at (2.25,1.25) {$\Kb_3$};
    \node[fnode] at (2.75,1.25) {$\Kb_3$};
    \node[fnode] at (3.5,1.25) {$\Kb_4$};
\end{tikzpicture}
\end{equation}
Furthermore, the theory admits a brane construction.  For example, for
$(M_1, M_2, M_3, M_4) = (3,2,2,2)$ and
$(K_1, K_2, K_3|\Kb_3, \Kb_4) = (2,1,1|1,1)$, the brane configuration for
the above quiver is
\begin{equation}
  \begin{tikzpicture}[xscale=2, yscale=1.5]
    \draw[ZERO, very thick] (0,0.1) -- (0.5,0.1);
    \draw[ZERO, very thick] (0,0.2) -- (0.5,0.2);
    \draw[ZERO, very thick] (0,0.3) -- (0.5,0.3);
    \draw[ZERO, very thick] (0.5,0.7) -- (1,0.7);
    \draw[ZERO, very thick] (0.5,0.8) -- (1,0.8);
    \draw[ZERO, very thick] (1,0.2) -- (1.5,0.2);
    \draw[ZERO, very thick] (1.5,0.7) -- (2,0.7);
    \draw[ZERO, very thick] (1,0.3) -- (1.5,0.3);
    \draw[ZERO, very thick] (1.5,0.8) -- (2,0.8);
    \node[MINUS, very thick, gnode, minimum size=6pt] at (0.75,0.5) {};
    \draw (0.75,0.5) node[cross=3pt, MINUS, very thick] {};
    \node[MINUS, very thick, gnode, minimum size=6pt] at (0.35,0.5) {};
    \draw (0.35,0.5) node[cross=3pt, MINUS, very thick] {};
    \node[MINUS, very thick, gnode, minimum size=6pt] at (0.15,0.5) {};
    \draw (0.15,0.5) node[cross=3pt, MINUS, very thick] {};
    \node[PLUS, very thick, gnode, minimum size=6pt] at (1.75,0.5) {};
    \draw (1.75,0.5) node[cross=3pt, PLUS, very thick] {};
    \node[PLUS, very thick, gnode, minimum size=6pt] at (1.06,0.5) {};
    \draw (1.06,0.5) node[cross=3pt, PLUS, very thick] {};
    \node[MINUS, very thick, gnode, minimum size=6pt] at (1.44,0.5) {};
    \draw (1.44,0.5) node[cross=3pt, MINUS, very thick] {};
    \draw[PLUS, very thick] (0,0)
    -- (0,1);
    \draw[PLUS, very thick] (0.5,0)
    -- (0.5,1);
    \draw[PLUS, very thick] (1,0)
    -- (1,1);
    \draw[MINUS, very thick] (1.5,0)
    -- (1.5,1);
    \draw[MINUS, very thick] (2,0)
    -- (2,1);
  \end{tikzpicture}
\end{equation}
Note that in order to realize the flavor symmetry
$\U(K_m)^2 \times \U(\Kb_m)^2$, each D4-brane between $\mathrm{NS5}_m$
and $\mathrm{NS5}_{m+1}$ needs to be brought to the NS5-brane of the
same color and broken into half.

We turn on mass parameters for the global symmetries in such a way
that higgsing give the following masses:
\begin{align}
  (\Rsf_r^\ell)^{a_r}{}_{l}&\colon
                             \sigma_r^{a_r} - \mu_{r,l}^\ell - \frac12 \hbar \,,
  \\
  (\Rsft_r^\ell)^{l}{}_{a_r}&\colon
    \mut_{r,l}^\ell - \sigma_r^{a_r} - \frac12 \hbar \,,
  \\
  (\Ssf_r^\ell)^{a_r}{}_{l}&\colon
    \sigma_r^{a_r} - \nu_{r,l}^\ell + \frac12 \hbar \,,
  \\
  (\Ssft_r^\ell)^{l}{}_{a_r}&\colon
    \nut_{r,l}^\ell - \sigma_r^{a_r} + \frac12 \hbar \,.
\end{align}
We necessarily have $\mu_{r,l}^\ell = \mut_{r,l}^\ell$ and
$\nu_{r,l}^\ell = \nut_{r,l}^\ell$ for $r \neq m$.

From these expressions for the masses, we see that for $r \leq m$, the
pair $(\Rsf_r^\ell, \Rsft_r^\ell)$ contributes to the
vacuum equations the factor
\begin{equation}
  \label{eq:vacuum-factor-1}
  \prod_{l=1}^{K_r^\ell}
  \frac{\sigma_r^{a_r} - \mu_{r,l}^\ell - \frac12 \hbar}
       {\sigma_r^{a_r} - \mut_{r,l}^\ell + \frac12 \hbar} \,,
\end{equation}
and for $r \geq m$, the pair $(\Ssf_r^\ell, \Ssft_r^\ell)$ contributes
the factor
\begin{equation}
  \label{eq:vacuum-factor-2}
  \prod_{l=1}^{\Kb_r^\ell}
  \frac{\sigma_r^{a_r} - \nu_{r,l}^\ell + \frac12 \hbar}
       {\sigma_r^{a_r} - \nut_{r,l}^\ell - \frac12 \hbar} \,.
\end{equation}

For the Bethe/gauge correspondence to exist, the above factors should
reproduce the factor
\begin{equation}
  \label{eq:Bethe-factor}
  \frac{\sigma_r^{a_r} - \zeta^\ell
        + (-1)^{[r+1]} \lambda_{r+1}^\ell \hbar - \frac12 c_r \hbar}
       {\sigma_r^{a_r} - \zeta^\ell
        + (-1)^{[r]} \lambda_r^\ell\hbar - \frac12 c_r \hbar}
\end{equation}
in the Bethe equations.  This is indeed possible if we identify the
parameters as
\begin{align}
  \label{eq:mu-lambda}
  \mu_{r,l}^\ell
  &=
  \mut_{r,l}^\ell
  =
  \zeta^\ell - (\lambda_r^\ell + \frac12 c_r + \frac12) \hbar + l\hbar \,,
  \\
  \label{eq:nu-lambda}
  \nu_{r,l}^\ell
  &=
  \nut_{r,l}^\ell
  =
  \zeta^\ell + (\lambda_{r+1}^\ell + \frac12 c_r - \frac12)\hbar + l\hbar \,.
\end{align}
We have
\begin{equation}
  \prod_{l=1}^{K_r^\ell}
  \frac{\sigma_r^{a_r} - \mu_{r,l}^\ell - \frac12 \hbar}
       {\sigma_r^{a_r} - \mut_{r,l}^\ell + \frac12 \hbar}
  =
  \frac{\sigma_r^{a_r} - \mu_{r,K_r^\ell}^\ell - \frac12 \hbar}
       {\sigma_r^{a_r} - \mu_{r,1}^\ell + \frac12 \hbar}
  =
  \frac{\sigma_r^{a_r} - \zeta^\ell
        + (\lambda_r^\ell - K_r^\ell) \hbar - \frac12 c_r \hbar}
       {\sigma_r^{a_r} - \zeta^\ell
        + \lambda_r^\ell\hbar - \frac12 c_r \hbar}
\end{equation}
for $r \leq m$ and
\begin{equation}
  \prod_{l=1}^{\Kb_r^\ell}
  \frac{\sigma_r^{a_r} - \nu_{r,l}^\ell + \frac12 \hbar}
       {\sigma_r^{a_r} - \nut_{r,l}^\ell - \frac12 \hbar}
  =
  \frac{\sigma_r^{a_r} - \nu_{r,1}^\ell + \frac12 \hbar}
  {\sigma_r^{a_r} - \nu_{r,\Kb_r^\ell}^\ell - \frac12 \hbar}
  =
  \frac{\sigma_r^{a_r} - \zeta^\ell
        - \lambda_{r+1}^\ell \hbar - \frac12 c_r \hbar}
       {\sigma_r^{a_r} - \zeta^\ell
        - (\lambda_{r+1}^\ell + \Kb_r^\ell)\hbar - \frac12 c_r \hbar}
\end{equation}
for $r \geq m$, so we obtain the factor~\eqref{eq:Bethe-factor} using
the definitions of $K_r^\ell$ and $\Kb_r^\ell$.

Now, let us consider the case in which the representations of the spin
variables are arbitrary finite-dimensional ones.  Even in this general
case, most of the above argument actually goes through, with the same
definitions of $K_r^\ell$ and $\Kb_r^\ell$ for $r \neq m$ and the
identifications~\eqref{eq:mu-lambda} and~\eqref{eq:nu-lambda}.  The
only place that fails is where we set $r = m$: if we choose
nonnegative integers $K_m^\ell$, $\Kb_m^\ell$ and write down the
product of the factors \eqref{eq:vacuum-factor-1} and
\eqref{eq:vacuum-factor-2} for $r = m$, we get
\begin{equation}
  \frac{\sigma_m^{a_m} - \zeta^\ell
        - \lambda_{m+1}^\ell \hbar - \frac12 c_m \hbar}
       {\sigma_m^{a_m} - \zeta^\ell
        + \lambda_m^\ell\hbar - \frac12 c_m \hbar}
  \biggl(
  \frac{\sigma_m^{a_m} - \zeta^\ell
        + (\lambda_m^\ell - K_m^\ell) \hbar - \frac12 c_m \hbar}
       {\sigma_m^{a_m} - \zeta^\ell
         - (\lambda_{m+1}^\ell + \Kb_m^\ell)\hbar - \frac12 c_m \hbar}
  \biggr) \,,
\end{equation}
whereas the Bethe equations do not contain the second fraction in the
parenthesis.

We can cancel this unwanted factor if we introduce additional chiral
multiplets
\begin{align}
  \Rsft_m^\ell &\in \Hom(\C^{M_m},\C^{K_m^\ell}) \,,
  \\
  \Ssf_m^\ell &\in \Hom(\C^{\Kb_m^\ell}, \C^{M_m})
\end{align}
and give them appropriate masses.  These chiral multiplets are
produced by semi-infinite D4-branes ending on $\mathrm{NS5}_m$ and
$\mathrm{NS5}_{m+1}$.

\section*{Acknowledgments}

We would like to thank Kevin Costello, Mykola Dedushenko and Philsang
Yoo for helpful discussions.  NI gratefully acknowledges support from
the Institute for Advanced Study and the National Science Foundation
under Grant No.\ PHY-1911298. The work of SFM is funded by the Natural
Sciences and Engineering Research Council of Canada (NSERC).  Research
at Perimeter Institute is supported in part by the Government of
Canada through the Department of Innovation, Science and Economic
Development Canada and by the Province of Ontario through the Ministry
of Colleges and Universities. Any opinions, findings, and conclusions
or recommendations expressed in this material are those of the authors
and do not necessarily reflect the views of the funding agencies.

\appendix
\section{Four-dimensional Chern--Simons theory with gauge supergroup
  from twisted string theory}
\label{sec:topstrings}

In this appendix we present an alternative construction of
four-dimensional Chern--Simons theory with gauge group $\GL(m|n)$,
using the framework of twists of superstring theory as developed in
\cite{Costello:2016mgj}. Twisted superstring theory refers to
superstring theory in a particular RR background where the bosonic
ghost for local supersymmetries may take a nonzero nilpotent vacuum
expectation value $Q$. When one considers D-branes in such
backgrounds, the coupling between D-branes and the bosonic ghost
dictates that $Q$ is added to the BRST differential of the worldvolume
theory \cite{Costello:2016mgj}. Therefore, the field content of
worldvolume theories of branes in twisted superstrings are naturally
$Q$-cohomology of that of the supersymmetric gauge theories one would
find in the absence of the additional RR background. As such, twisted
superstrings affords a useful framework for studying protected sectors
of supersymmetric gauge theories.

Costello and Li \cite{Costello:2016mgj} give conjectural descriptions
of such twists of superstrings in terms of topological strings. These
conjectures have passed several consistency checks
\cite{Costello:2016mgj, Costello:2018zrm, Costello:2020jbh} and have
been proven at the level of the free limit of the supergravity
approximation \cite{Saberi:2021weg}. Taking these conjectures as a
starting point, one can derive simple descriptions of twists of
worldvolume theories of D-branes using mathematical tools from the
study of topological strings. Though such calculations require
machinery from homological algebra, they have the benefit of
calculational ease. Tractable models of twisted worldvolume theories
can be determined from an $\Ext$-algebra computation, and the action
functional can be read off from an algebraic structure and trace on
the $\Ext$-algebra; no term-matching arguments involving the
Dirac--Born--Infeld action are required.

In this appendix, we work with field theory in the Batalin--Vilkovisky
(BV) formalism as articulated by \cite{MR2778558,
  Costello:2016vjw}. In particular, we freely make use of the language
of $L_{\infty}$-algebras. Much of the below is exposited elsewhere in
the literature. The construction of twisted supergravity and the
conjectural descriptions of twists of superstrings in terms of
topological strings are given in \cite{Costello:2016mgj}. Many of the
examples below are worked out in \cite{Raghavendran:2019zdq} where
more formal aspects of the framework are articulated and some
mathematical applications are discussed. We hope the exposition of
this appendix will have the simultaneous benefit of illustrating the
calculational utility of twisted superstrings, and making our
constructions parseable to more mathematically minded readers.

\subsection{Topological strings}
\label{sec:orga6d3a0f}

We begin with some recollections on topological strings. The
worldsheet theory of a topological string theory is a two-dimensional
oriented topological quantum field theory. Treating such theories via
the language of functorial field theory, the results of
\cite{MR2298823, MR2555928} tell us that such theories are determined
by the data of a Calabi--Yau category. Physically, we think of objects
of this category as D-branes in our topological string theory, and the
space of homomorphisms between two objects as the complex computing
BRST cohomology of the states of open strings stretched between the
branes. It is known that spaces of open string states have an algebra
structure, with respect to which the action for open string field
theory takes a simple form \cite{Witten:1992fb}. The data of a
Calabi--Yau category is exactly what is needed to make precise this
algebraic structure; this will be elaborated more on
subsection~\ref{open} below.

\begin{ex}\label{su3twist}
  Let $M$ be a symplectic four-manifold and $X$ a Calabi--Yau
  three-fold. The $\SU (3)$-invariant twist of type IIB string theory
  is given by the Calabi--Yau 5-category $\Fuk (M)\otimes \Coh
  (X)$. Here, $\Fuk (M)$ refers to the Fukaya category of $M$ and
  $\Coh (X)$ refers to the category of coherent sheaves on $X$. This
  describes a topological string theory that looks like a combination
  of the A-model into $M$ and the B-model into $X$.
\end{ex}

Here the terminology is meant to indicate that the above mixed A-B
model conjecturally arises from type IIB string theory in an RR
background in which the bosonic ghost takes a vacuum expectation value
given by an $\SU(3)$-invariant nilpotent element of the
ten-dimensional $\CN=(2,0)$ supersymmetry algebra.

\begin{rmk}

  Let us elaborate on our description of the A-model directions. For us the main relevant example will be when
  $M=\R^{2N}$. In this case, we will use a version of the Fukaya
  category that we will denote $\Fuk^0$ where we discard counts of
  pseudo-holomorphic discs with nonzero area. Explicitly, the objects
  in the category will consist of Lagrangians in $M$, and for two
  Lagrangians $L_1$, $L_2\subset \R^{2N}$ with clean intersection, we have that
  $\Hom_{\Fuk^0}(L_1, L_2)=\Omega^\bullet(L_1\cap L_2)$. This will
  suffice for our purposes as we will primarily care about
  perturbative phenomena on worldvolume theories of branes, so we may
  neglect worldsheet instantons.

 In addition to this restriction on the space of homomorphisms, this category does not include as objects, coisotropic A-branes. To the authors' knowledge, it is an open mathematical problem to construct a version of the Fukaya category that includes as objects such branes. Fortunately, we will not need to consider such branes in our analysis.
\end{rmk}

To a topological string theory, we may associate two field theories which are versions of open string field theory and closed string field theory respectively. The former recovers twists of worldvolume theories of branes in the physical string while the latter contains twists of supergravity.

\subsection{Topological open string field theory}
\label{open}

Let $\CC$ be an $A_\infty$-category and let $\CF\in\CC$ be an
object. Then $\Hom_{\CC}(\CF, \CF)$ is an $A_\infty$-algebra,
and skew-symmetrizing the $A_\infty$-operations yields an
$L_{\infty}$-algebra. Now suppose our $\CC$ is in fact a Calabi--Yau
$N$-category, and as such can be thought of as determining a
topological string theory. Then for any object $\CF\in \CC$, we
have an invariant pairing
$\tr\colon \Hom_{\CC}(\CF, \CF)\to \C[N]$.

In examples of interest, where $\CC$ is attached to a $2N$-manifold
thought of the target spacetime of our topological string,
$\Hom_{\CC}(\CF, \CF)$ will arise as sections of a natural
graded vector bundle over the support of $\CF$, the
$L_{\infty}$-structure maps will be given by polydifferential
operators, and the trace map will factor through integration over the
support of $\CF$. In such instances, the data of this
$L_\infty$-algebra and the trace pairing then determine the data of a
perturbative $\Z_2$-graded BV theory -- the space of fields of the
theory is $\Pi\Hom_{\CC}(\CF, \CF)$ and the action is given by
\begin{equation}
  S(\alpha)
  =
  \sum_{k \geq 1} \frac{1}{(k+1)!}
  \tr\bigl(\alpha \otimes \ell_k(\alpha^{\otimes k})\bigr) \,,
\end{equation}
where
$\ell_k\colon \Hom_{\CC}(\CF, \CF)^{\otimes k} \to \Hom_{\CC}(\CF,
\CF)$ are the $L_\infty$-structure maps. This theory is the
worldvolume theory of the D-brane $\CF$ in the topological string
theory determined by $\CC$. The conjectural descriptions of twists of
superstrings in terms of topological strings imply that for $\CC$
coming from a twist of a superstring theory, the worldvolume theory of
$\CF$ is a twist of the worldvolume theory of the corresponding brane
in the physical string theory.

\begin{ex}
  \label{holtopsix-dimensional11}
  Consider the $\SU (3)$-invariant twist of type IIB string theory on
  $\R^4 \times \C^3$ from example \ref{su3twist}, and consider a stack
  of $n$ D5-branes wrapping $\R^2\times \C^2$. As explained in the
  above example, this twist of type IIB string theory is described by
  the Calabi--Yau category $\CC = \Fuk^0(\R^4)\otimes \Coh (\C^3)$. The
  object describing our stack of branes is given by
  $(\R^2, \CO_{\C^2}^n)$. We have that
  \begin{equation}
    \begin{split}
      \Ext_{\CC}\bigl((\R^2, \CO_{\C^2}^n), (\R^2, \CO_{\C^2}^n)\bigr)
      &=
      \Hom_{\Fuk^0}(\R^2,\R^2)
      \otimes \Ext_{\Coh (\C^3)} (\CO_{\C^2}^n, \CO_{\C^2}^n) \\
      &=
      \Omega^\bullet (\R^2)
      \otimes \Ext_{\Coh (\C^3)} (\CO_{C^2}, \CO_{\C^2}) \otimes \glf(n) \\
      &=
      \Omega^\bullet(\R^2) \otimes
      \Omega^{0,\bullet}(\C^2)[\veps] \otimes \glf(n).
    \end{split}
  \end{equation}
  In the last step, we have used the following general result:
  \begin{lemma}
    Let $X$ be a Calabi--Yau manifold and let $Y\subset X$ be
    holomorphic. Then
    $\Ext_{\Coh (X)}(\CO_Y, \CO_Y) \linebreak[1] \cong \Omega^{0,\bullet}(Y,
    \wedge^{\bullet} N_{X/Y})$.
  \end{lemma}
  
  We can describe the $L_{\infty}$-structure as follows. There is an
  $L_\infty$-structure on
  $\Omega^{\bullet}(\R^{2})\otimes \Omega^{{0,\bullet}}(\C^{2})\otimes
  \glf(n)$ given by
  \begin{align}
    \ell_1
    &=
      \rmd \otimes 1_{\Omega^{0,\bullet}(\C^{2})} \otimes 1_{\glf(n)}
      + 1_{\Omega^{\bullet}(\R^{2})}\otimes\delb \otimes 1_{\glf(n)} \,,
    \\
    \ell_2
    &=
      \wedge\otimes \wedge\otimes [-,-]_{\glf(n)} \,,
    \\
    \ell_k
    &=
      0 \,, \quad k \geq 3 \,.
  \end{align}
  
  The $L_{\infty}$-structure on $\Omega^{\bullet}(\R^{4})\otimes
  \Omega^{{0,\bullet}}(\C^{2})[\veps]\otimes
  \glf(n)$ is given by the semidirect product
  \begin{equation}
    \bigl(\Omega^{\bullet}(\R^{4}) \otimes \Omega^{{0,\bullet}}(\C^{2})
    \otimes \glf(n) \bigr)
    \ltimes
    \veps\bigl(\Omega^{\bullet}(\R^{4}) \otimes \Omega^{{0,\bullet}}(\C^{2})
    \otimes \glf(n)\bigr)
  \end{equation}

  The trace pairing induced from the Calabi--Yau structure on $\CC$
  is given
  \begin{equation}
    \tr\colon
    \alpha \mapsto
    \int_{\R^{2}\times\C^{2|1}} \operatorname{Tr}(\alpha) \wedge \Omega \,,
  \end{equation}
  where $\Omega$ denotes the holomorphic volume form on $\C^2$ and
  $\operatorname{Tr}$ is the Killing form on $\glf(n)$. Thus, we find
  that the action of the theory is exactly
  \begin{equation}
    S(\alpha, \beta)
    =
    \int_{\R^2\times \C^{2|1}}
    \operatorname{Tr}\biggl(\frac{1}{2}\beta(d+\delb)\alpha
    + \frac{1}{6}\beta\wedge[\alpha, \alpha]\biggr)
    \wedge \Omega
  \end{equation}
  for
  \begin{align}
    \alpha
    &\in
      \Omega^\bullet(\R^2)\otimes \Omega^{0,\bullet}(\C^2)\otimes \glf(n) \,,
    \\
    \beta
    &\in
      \Omega^\bullet(\R^2) \otimes \Omega^{0,\bullet}(\C^{2}) \veps
      \otimes \glf(n) \,.
  \end{align}
  This is exactly the holomorphic--topological twist of
  six-dimensional $\CN = (1,1)$ super Yang--Mills theory, dubbed the
  rank $(1,1)$ partially holomorphic topological twist in
  \cite{Elliott:2020ecf}.
\end{ex}

\subsection{Topological closed string field theory}
\label{closed}

Let $Z$ be the worldsheet theory determined by the Calabi--Yau
category $\CC$. Naively, the closed string states of the theory
should be given by the local operators of the worldsheet theory,
$Z(S^1)$. However, the worldsheet theory in the physical string is
coupled to two-dimensional gravity -- closed string states should be
those local operators invariant under reparametrizations of the
worldsheet. Since the worldsheet theory is topological, Cartan's magic
formula tells us that small reparametrizations will act homotopically
trivially on the space of local operators. In the setting of
topological strings, there is a natural homotopy action of $S^1$ on
$Z(S^1)$ -- the closed string states will be the invariants
$Z(S^1)^{S^1}$. In terms of categorical data, this is computed by the
cyclic cochains of the category $\CC$, $\Cyc^\bullet (\CC)$. There
is a natural way to equip a shift of $\Cyc^{\bullet} (\CC)$ with an odd Poisson
tensor and an $L_\infty$-structure. In examples in which the
graded vector space underlying $\Cyc^\bullet (\CC)$ arises as the
space of sections of some graded vector bundle, this gives
$\Cyc^\bullet (\CC)$ the structure of a $\Z_2$-graded Poisson BV
theory. The constructions of the $L_\infty$- and shifted Poisson
structures in this generality are extraneous for our purposes -- we
will be focused on the following examples.

\begin{ex}
  Suppose $\CC = \Coh(X)$ with $X$ Calabi--Yau. Then the cyclic
  formality theorem \cite{MR2836399} tells us that there is an
  equivalence of $L_{\infty}$-algebras
  \begin{equation}
    \Cyc^\bullet (\CC) \cong \bigl(\PV^{\bullet,\bullet}(X)[\![t]\!] \,, \ell_1 = \delb+t\del \,, \ell_2 = \{-,-\} \bigr) \,,
  \end{equation}
  where $t$ is a parameter of degree $2$, $\del$ denotes the
  divergence operator, and $\{-,-\}$ denotes the Schouten bracket of
  polyvector fields. The Poisson tensor has Poisson kernel
  $(\del\otimes 1)\delta_{\Delta (X)}$, where
  $\Delta (X)\subset X\times X$ denotes the diagonal. This theory is
  Kodaira--Spencer gravity articulated as
  Bershadsky--Cecotti--Ooguri--Vafa theory studied by
  \cite{Bershadsky:1993cx, Costello:2012cy, Costello:2015xsa}.
\end{ex}

\begin{ex}
  Suppose $\CC = \Fuk (M)$ with $M$ being a symplectic manifold. The
  Hochschild (co)chains admit a description in terms of the quantum
  cohomology of the target. Together with the abstract
  $L_\infty$-structure and the $\Z_2$-graded Poisson structure, we
  expect that the result will be a version of the K\"ahler gravity
  \cite{Bershadsky:1994sr}. We will discard worldsheet instantons
  coming from the A-model directions of the twists of string theory we
  consider. Therefore, our ansatz will be that the closed string field
  theory for the A-model directions is described by the
  $L_\infty$-algebra $\Omega^\bullet (M)$ with $L_\infty$-structure
  given by $\ell_1 = \rmd$, $\ell_2 = \wedge$ and Poisson structure given
  by the wedge and integrate pairing. We will abusively continue to
  denote the closed string field theory in the A-model sans worldsheet
  instantons by $\Cyc^{\bullet}(\Fuk (M))$.
\end{ex}

\begin{ex}
  Putting the above two examples together, we can describe the closed
  string field theories for the twists of type IIB string theory we
  are interested in. The closed string field theory for the
  $\SU(3)$-invariant twist of type IIB string theory on
  $\R^4\times \C^3$ is given by the $L_\infty$-algebra
  $\Omega^\bullet (\R^4)\otimes \PV^{\bullet, \bullet}(\C^3)[\![t]\!]$
  with
  \begin{align}
    \ell_1
    &=
      \rmd\otimes 1_{\PV^{{\bullet,\bullet}}(\C^{3})[\![t]\!]}
      + 1_{{\Omega^{\bullet}(\R^{4})}}\otimes (\delb +t\del) \,,
    \\
    \ell_2
    &= \wedge\otimes \{-,-\} \,,
    \\
    \ell_k
    &=
      0 \,, \quad k \geq 3 \,.
  \end{align}
  The Poisson tensor is given by the Poisson kernel
  $(\del\otimes 1)\delta_{\Delta(\C^3)}\delta_{\Delta(\R^4)}$.
\end{ex}

\subsection{Closed--open map}

Given a Calabi--Yau category $\CC$ and an object $\CF\in \CC$, there
is always an $L_{\infty}$-map
\begin{equation}
  \Cyc^\bullet (\CC)
  \to
  \operatorname{CE}^\bullet\bigl(\Hom_{\CC}(\CF, \CF)\bigr) \,.
\end{equation}
Here, the target denotes Chevalley--Eilenberg cochains on the
$L_\infty$-algebra $\Hom_{\CC}(\CF, \CF)$; this is a model for Hamiltonian vector fields on
the formal moduli space describing fluctuations of the brane
$\CF$. This map takes a closed string field and produces a single trace-operator on the worldvolume theory of $\CF$, which
describes how the closed string field couples to the worldvolume
theory of $\CF$. We will wish to apply this to examples where
$\CC = \Fuk^0(\R^{10-2N})\otimes \Coh (\C^{N})$, and
$\CF = (\R^{5-N}, \CO_{\C^k}^n)$ for $k\leq N$. Then we have that
\begin{align}
  \Cyc^\bullet (\CC)
  &=
    \Omega^{\bullet}(\R^{10-2N})\otimes \PV^{0,\bullet}(\C^N)[\![t]\!] \,,
  \\
  \Hom_\CC(\CF, \CF)
  &=
    \Omega^\bullet(\R^{5-N}) \otimes
    \Omega^{0,\bullet}(\C^k)[\veps_{k+1},\dotsc, \veps_{N}] \otimes \glf(n) \,.
\end{align}

This map should be thought of as given by a sum of disk amplitudes
with boundary on the brane $\CF$ and with an arbitrary number of
marked points on the interior labeling closed string insertions. We
will only consider single closed string insertions of the form
$1\otimes \mu\in
\Omega^{\bullet}(\R^{10-2n}) \otimes \PV^{0,\bullet}(\C^N)$. In
particular the field does not depend on the A-twisted directions of
spacetime, or the parameter $t$. For such fields we have the following
explicit formula for the linear component of the closed--open map

\begin{equation}
  1 \otimes w_1^{a_1} \dotsm w_N^{a_N} \del_{w_1}^{b_1} \dotsm \del_{w_N}^{b_N}
  \mapsto
  I(\alpha) \,,
\end{equation}
where
\begin{multline}
  I(\alpha)
  =
  \frac{1}{(n+1)!}
  \int _{\R^{5-N}\times \C^{k|N-k}}
  \operatorname{Tr}(w_1^{a_1} \dotsm w_N^{a_N}
  \veps_{k+1}^{b_{k+1}} \dotsm \veps_{N}^{b_N}
  \\
  \times
  \del_{\veps_{k+1}}^{b_{k+1}} \alpha \wedge \dotsm
  \wedge \del_{\veps_N}^{b_N} \alpha \wedge
  \del_{w_1}^{b_1}\alpha \wedge \dotsm \wedge \del_{w_k}^{b_k} \alpha)
  \wedge \Omega \,.
\end{multline}

A version of this result including formulas for the deformation to all
orders in open string insertions is proved in
\cite{Costello:2015xsa}. It is worth emphasizing that deriving
formulas for this map at all orders is an extremely nontrivial problem
-- for $\mu\in \PV^{2,0}$ this is the content of the holomorphic
analogue of Kontsevich's theorem on deformation quantization.

\begin{ex}
  \label{ex:four-dimensionalcs}
  Consider the $\SU(3)$-invariant twist of type IIB string theory on
  $\R^4\times \C^3$. We fix once and for all coordinates $z$, $w_1$,
  $w_2$ on $\C^3$. We saw in the example above that a stack of $n$
  D5-branes wrapping $\R^2\times \C^2_{z, w_1}$ gives rise to a
  holomorphic--topological twist of six-dimensional $\CN = (1,1)$
  super Yang--Mills theory with gauge group $\U(n)$. Let us now
  consider what happens when we turn on a field
  $1\otimes w_1w_2 \in \Omega^\bullet
  (\R^4)\otimes\PV^{0,\bullet}(\C^3)$. Recall that the fields of the
  relevant twist of six-dimensional $\CN=(1,1)$ super Yang--Mills
  theory were given by
  $\Omega^\bullet(\R^2) \otimes \Omega^{0,\bullet}(\C^2_{z,
    w_1})[\veps]\otimes\glf(n)$.

  The image of the closed string field $w_1w_2$ under the closed
  open map becomes the functional
  \begin{equation}
    I(\alpha)
    =
    \int_{\R^2\times\C^{2|1}} \operatorname{Tr}(\alpha w_1\del_\veps\alpha \wedge\Omega)\,.
  \end{equation}
  Equivalently, this deforms the $L_\infty$-structure on
  $\Omega^\bullet (\R^2)\otimes \Omega^{0,\bullet}(\C^2)[\veps]\otimes
  \glf(n)$ so that
  $\ell_1 = \rmd \otimes 1_{\Omega^{0,\bullet}(\C^{2}[\veps])}\otimes
  1_{\glf(n)} \linebreak[1]+ 1_{\Omega^{\bullet}(\R^{2})}\otimes (\delb +
  w_1\del_\veps) \otimes 1_{\glf(n)}$. The differential
  $w_1\del_\veps$ has the effect of deforming the complex of fields of
  the theory into
  \begin{equation}
    \Omega^{\bullet}(\R^{2}) \otimes
    \Bigl(\Omega^{0,\bullet}(\C^{2})\veps
    \xrightarrow{w_{1}\del_{\veps}}
    \Omega^{0,\bullet}(\C^{2})\Bigr) \otimes \glf(n) \,.
  \end{equation}

  This is the Koszul resolution of the locus $w_{1}=0$, so is
  quasi-isomorphic to
  $\Omega^\bullet (\R^2)\otimes
  \Omega^{0,\bullet}(\C)\otimes\glf(n)$. This is exactly
  four-dimensional Chern--Simons theory as a $\Z_2$-graded BV theory.
\end{ex}

\begin{rmk}
  Note that the above construction differs slightly from the
  construction of four-dimensional Chern--Simons theory via
  $\Omega$-deformation in \cite{Costello:2018txb}. Conjecturally, the
  quadratic superpotential we have introduced should describe those
  components of the RR two-form used in \cite{Costello:2018txb} that
  are not exact for the twist we are performing. However, checking
  this explicitly is a difficult task. Moreover, at the level of field
  theory, the construction in~\cite{Costello:2018txb} came from
  subjecting the holomorphic--topological twist of six-dimensional
  $\CN=(1,1)$ super Yang--Mills theory with BV fields given by
  $\Omega^{\bullet}(\R^{4})\otimes \Omega^{0,\bullet}(\C)\otimes
  \glf(n)$ to a B-type $\Omega$-background along
  $\R^{2}\subset \R^{4}$. Such a construction involves replacing a
  factor of $\Omega^{\bullet}(\R^{2})$ with the Cartan model for
  $S^{1}$-equivariant cohomology of $\R^{2}$, which is given by the
  abelian dg Lie algebra
  $(\Omega^{\bullet}(\R^{2})[u]^{S^{1}}, \rmd +
  u \mathop{\iota}_{\del_{\theta}})$, where $u$ is an equivariant parameter, and
  $\del_{\theta}$ is the infinitesimal action of rotations. The
  localization theorem for equivariant cohomology tells us that for
  generic values of the equivariant parameter, the complex of fields
  of our theory is quasi-isomorphic to
  $\Omega^{\bullet}(\R^{2})\otimes
  \Omega^{0,\bullet}(\C)\otimes\glf(n)$. Note that the relation
  $u\CL_{\del_{\theta}}=[\rmd, u \mathop{\iota}_{\del_{\theta}}]$ coming from
  Cartan's magic formula tells us that the infinitesimal action of
  rotations on the fields of our theory is homotopically trivial.

  However, in the above construction we instead work with a more
  minimal twist of six-dimensional $\CN=(1,1)$ super Yang--Mills
  theory; the twist of the previous paragraph is gotten from deforming
  the differential on
  $\Omega^{\bullet}(\R^{2})\otimes
  \Omega^{0,\bullet}(\C^{2})[\veps]\otimes\glf(n)$ so that
  $\ell_{1} = \rmd \otimes 1_{\Omega^{0,\bullet}(\C)[\veps]} \otimes
  1_{\glf(n)} \linebreak[1]+1_{\Omega^{\bullet}(\R^{2})} \otimes (\delb+
  \veps\del_{w_{1}})\otimes 1_{\glf(n)}$. Instead of taking this
  further twist and working equivariantly along the topological plane
  that the $w_{1}$-plane becomes, we turned on a deformation coming
  from a quadratic superpotential. It is worth noting that there is a
  map from a twist of the four-dimensional $\CN=4$ superconformal
  algebra to the closed string sector of the $\SU(3)$-invariant twist
  of type IIB string theory -- the quadratic superpotential
  deformation considered above lies in the image of this
  map. Moreover, note that we have that
  $\CL_{w_{1}\del_{w_{1}}} = [\veps\del_{w_{1}},
  w_{1}\del_{{\veps}}]$; we see that the superconformal deformation
  also makes the complexified action of rotations exact for the
  B-twist supercharge. This appears to be part of a general pattern
  where a superconformal deformation of a holomorphic theory is
  equivalent to an $\Omega$-deformation of a further topological twist
  \cite{Oh:2019bgz, Saberi:2019fkq, Raghavendran:2019zdq}.
\end{rmk}

\begin{rmk}\label{6dtoplstrings}
  It is also interesting to consider the superpotential $w_{1}w_{2}$
  as a deformation of the entire topological string theory, that is,
  as a deformation of the category of branes. Morally, it should
  deform the category of coherent sheaves on $\C^{3}$ to the category
  of matrix factorizations for the superpotential $w_{1}w_{2}$; the
  B-model directions of the topological string are turned into a
  Landau--Ginzburg B-model. The category of matrix factorizations in
  this case can be described as the category of modules for the Jacobi
  algebra of the superpotential $w_{1}w_{2}$, which in this case is
  just the algebra $\C[z]$. Thus, we see that the $\SU(3)$-invariant
  twist of type IIB string theory localizes to a six-dimensional
  topological string theory on $\R^{4}\times\C$; this makes contact
  with the work of \cite{Ishtiaque:2018str}.
\end{rmk}

\subsection{Four-dimensional Chern--Simons theory with gauge
  supergroup from the $\SU (3)$-invariant twist of type IIB string
  theory}

In this section we will arrive at four-dimensional Chern--Simons
theory with gauge supergroup using the formalism developed in the
previous subsections. The calculation is essentially an easy corollary
of the examples therein.

We consider the $\SU (3)$-invariant twist of type IIB string theory on
$\R^4\times \C^3$ with a configuration of D-branes as in the following
table:
\begin{equation}  
  \begin{tabular}{rccccc}
    & $\R^2$ & $\R^2$ & $\C_{z}$ & $\C_{w_1}$ & $\C_{w_2}$\\
    \hline
    $n$ D5 & $\circ$ & $\times$ & $\times$ & $\times$ & $\circ$\\
    $m$ D5 & $\circ$ & $\times$ & $\times$ & $\circ$ & $\times$\\
  \end{tabular}
\end{equation}
A cross mark means that the D5-brane extends in that direction.  We
also turn on a closed string field given by the quadratic
superpotential $w_1w_2$. We arrive at a field theory description for
this system by first computing the open string field theory using the
techniques above, and then applying the closed--open map.

Let $\CF_1 = (\R^2, \CO^m_{\C^2_{z,w_1}})$,
$\CF_2 = (\R^2, \CO^n_{\C^2_{z,w_2}})$ denote the objects in the
categories of D-branes corresponding to the stacks of $n$ and $m$
D5-branes, respectively. We first wish to compute
$\Hom_{\CC}(\CF_1\oplus \CF_2, \CF_1\oplus \CF_2)$. Since
$\Hom$ commutes with direct sums, we have four summands:
\begin{itemize}
  \item From example \ref{ex:four-dimensionalcs}, we have that
    \begin{align}
      \Hom_{\CC}(\CF_1, \CF_1)
      &=
        \Omega^\bullet (\R^2) \otimes
        \Omega^{0, \bullet}(\C^2_{z, w_1})[\veps_{2}]\otimes \glf(m) \,,
      \\
      \Hom_{\CC}(\CF_2, \CF_2)
      &=
        \Omega^\bullet (\R^2) \otimes
        \Omega^{0,\bullet}(\C^2_{z, w_2})[\veps_{1}]\otimes \glf(n) \,.
    \end{align}
    Our convention above is that $\veps_{i}$ denotes a section of the
    normal bundle of $\C^{2}_{z,w_{j}}\subset\C^{3}$, where $i\neq j$.

    The trace pairing is given by
    \begin{equation}
      \tr(\alpha)
      =
      \int_{\R^2\times\C^{2|1}} \rmd z \, \rmd w_{j} \tr(\alpha) \,.
    \end{equation}

  \item We can compute, using free resolutions of the structure sheaves
    of the $w_{i}$-planes, that
    \begin{align}
      \Hom_{\CC}(\CF_1, \CF_2)
      &=
        \Omega^\bullet (\R^2) \otimes
        \Omega^{0,\bullet}(\C_z)\otimes \Hom (\C^m, \C^n)[-1] \,,
      \\
      \Hom_{\CC}(\CF_2, \CF_1)
      &=
        \Omega^\bullet (\R^2) \otimes
        \Omega^{0,\bullet}(\C_z) \otimes \Hom (\C^n, \C^m)[-1] \,.
    \end{align}

    Each of these are abelian $L_\infty$-algebras, with
    $\ell_1 = d\otimes 1_{\Omega^{0,\bullet}(\C_z)}\otimes 1_{\Hom} +
    1_{\Omega^{\bullet}(\R^{2})}\otimes \delb\otimes
    1_{\Hom}$. There is a natural trace pairing on the direct sum
    \begin{equation}
      \Omega^{\bullet}(\R^{2}) \otimes
      \Omega^{0,\bullet}(\C_{z}) \otimes T^{*}\Hom(\C^m,\C^n)[-1] \,.
    \end{equation}
    Letting $X$s denote fields valued in $\Hom(\C^m, \C^n)$ and $Y$s
    fields valued in the cotangent direction, the pairing is given by
    \begin{equation}
      \tr(X_{1}+Y_{1},X_{2}+Y_{2})
      =
      \int \rmd w \bigl(
      \operatorname{Tr}_{\C^n}(X_{1}Y_{2})
      - \operatorname{Tr}_{\C^m}(Y_{1}X_{2})\bigr) \,.
    \end{equation}
    The action functional induced by this pairing and abelian
    $L_{\infty}$-structure is exactly the BV action for a free
    hypermultiplet in the Kapustin twist. Restricted to fields of
    ghost number $1$, this recovers exactly the
    action~\eqref{eq:Kapustin-action}.
\end{itemize}

Thus we see that the entire space of open string states is given
by
\begin{equation}
  \CE
  =
  \Omega^\bullet (\R^2) \otimes \Omega^{0,\bullet}(\C_z) \otimes
  \left(
  \begin{gathered}
    \Omega^{0,\bullet}(\C_{w_1})[\veps_{2}]\otimes \glf(n) \\
    \oplus \\
    \Omega^{0,\bullet}(\C_{w_2})[\veps_{1}]\otimes \glf(m)\\
    \oplus \\
    T^*\Hom (\C^m, \C^n)[-1]
  \end{gathered}
  \right) .
\end{equation}

We now determine the $L_{\infty}$-structure. This is as usual gotten
by skew-symmetrizing the natural $A_{\infty}$-structure on
$\Hom_{\CC}(\CF_1\oplus \CF_2, \CF_1\oplus \CF_2)$. In terms of the
above direct summands, the $A_{\infty}$-structure is given in terms of
the following operations:
\begin{itemize}
\item
  $\Hom_{\CC}(\CF_{i}, \CF_{i}) \otimes \Hom_{\CC}(\CF_{i},
  \CF_{i})\to \Hom_{\CC}(\CF_{i}, \CF_{i})$ \,,
  \quad $A\otimes B\mapsto AB$ \,;

\item $\Hom_{\CC}(\CF_{i}, \CF_{j}) \otimes \Hom_{\CC}(\CF_{j}, \CF_{i})
  \to \Hom_{\CC}(\CF_{j}, \CF_{i})$ \,,
  \quad $Y\otimes X\mapsto YX$ \,;

\item
  $\Hom_{\CC}(\CF_{i}, \CF_{i}) \otimes \Hom_{\CC}(\CF_{j}, \CF_{i})
  \to \Hom_{\CC}(\CF_{j}, \CF_{i})$ \,,
  \quad $A\otimes X\mapsto AX$ \,,
\end{itemize}
where $i$, $j = 1,2$, $i \neq j$.

This induces the following $L_{\infty}$-structure:
\begin{itemize}
\item The first kind of $A_{\infty}$-operation above gives
  $L_{\infty}$-structures on $\Hom_{\CC}(\CF_{i},\CF_{i})$ with
  \begin{align}
    \ell_{1}
    &=
    \rmd \otimes 1_{\Omega^{0,\bullet}(\C^{2})[\veps_{j}]} \otimes 1_{\gf}
    +
    1_{\Omega^{\bullet}(\R^{2})}\otimes\delb\otimes 1_{{\gf}} \,,
    \\
    \ell_{2}
    &=
    \wedge\otimes [-,-]_{\gf} \,,
  \end{align}
  where $\gf= \glf(m)$ for $i=1$ and $\gf=\glf(n)$ for $i=2$. As
  explained in example \ref{holtopsix-dimensional11} these are
  precisely holomorphic--topological twists of the six-dimensional
  $\CN=(1,1)$ vector multiplets for $\glf(m)$ and $\glf(n)$.

\item There is a bracket
  \begin{multline}
    \bigl(\Omega^{\bullet}(\R^{2}) \otimes
    \Omega^{0,\bullet}(\C_{z}) \otimes
    T^{*}\Hom (\C^m,  \C^n)[-1]\bigr)^{\otimes 2}
    \\
    \to
    \Omega^{\bullet}(\R^{2}) \otimes \Omega^{0,\bullet}(\C_{z}) \otimes
    \left(
      \begin{gathered}
        \Omega^{0,\bullet}(\C_{w_{1}})[\veps_{2}]
        \otimes \glf(m)
        \\
        \oplus
        \\
        \Omega^{0,\bullet}(\C_{w_{2}})[\veps_{1}] \otimes \glf(n)
      \end{gathered}
    \right) ,
  \end{multline}
  explicitly given by wedging the form factors and taking the
  commutator of the matrices in $T^{*}\Hom(\C^m, \C^n)$.

  \item Finally, there is a bracket
    \begin{multline}
      \begin{aligned}
        \Omega^{\bullet}(\R^{2})\otimes \Omega^{0,\bullet}(\C_{z})&\otimes \left ( \begin{aligned}\Omega^{0,\bullet}(\C_{w_{1}})&[\veps_{2}] \otimes \glf(m)\\ & \oplus \\ \Omega^{0,\bullet}(\C_{w_{2}})&[\veps_{1}] \otimes \glf(n)\end{aligned} \right )\\ &\otimes  \\ \Omega^{\bullet}(\R^2)\otimes \Omega^{0,\bullet}(\C_{z}) & \otimes T^*\Hom(\C^m, \C^n)
      \end{aligned}
      \\
      \to
      \Omega^{\bullet}(\R^2) \otimes \Omega^{0,\bullet}(\C_{z})
      \otimes T^*\Hom(\C^m, \C^n)
    \end{multline}
    explicitly given by wedge product of forms and the natural action
    of $\glf(m)\oplus \glf(n)$ on $T^{*}\Hom (\C^m, \C^n)$.
\end{itemize}

The last of these brackets encodes the coupling between the
hypermultiplets in the Kapustin twist and the twist of the
six-dimensional $\CN=(1,1)$ vector multiplet. The second bracket,
which breaks the $\Z$-grading down to a $\Z_2$-grading, encodes an
extra gauge symmetry.

The open string field theory we have found can formally be regarded as
four-dimensional Chern--Simons theory on $\R^{2}\times \C_{z}$ for a
dg Lie superalgebra. We may schematically encode the above brackets by
writing the above dg Lie algebra as
\begin{equation}
  \left(
    \begin{array}{c|c}
      \Omega^{0,\bullet}(\C_{w_{1}})[\veps_{2}] \otimes \glf(m)
      & \Hom (\C^m, \C^n)[-1]
      \\ \hline
      \Hom (\C^n, \C^m)[-1]
      & \Omega^{0,\bullet}(\C_{w_{2}})[\veps_{1}] \otimes \glf(n)
    \end{array}
  \right) .
\end{equation}

Let us now analyze the effect of the closed string field $w_1w_2$. As
we saw before, the image of this closed string field under the closed
open map only affects the differential on the above dg Lie
superalgebra. Explicitly, the deformation looks like
\begin{multline}
  \left(
    \begin{array}{c|c}
      \bigl(\Omega^{0,\bullet}(\C_{w_{1}})\veps_{2}
      \xrightarrow{w_1\del_{\veps_2}} \Omega^{0,\bullet}(\C_{w_{1}})\bigr) \otimes \glf(m)
      & \Hom (\C^m, \C^n)[-1]
      \\ \hline
      \Hom(\C^n, \C^m)[-1]
      &
        \bigl(\Omega^{0,\bullet}(\C_{w_{2}})\veps_{1}
        \xrightarrow{w_2\del_{\veps_1}} \Omega^{0,\bullet}(\C_{w_{2}})\bigr) \otimes \glf(n)
    \end{array}\right)
  \\
  \cong
  \left(
    \begin{array}{c|c}
      \glf(m) & \Hom(\C^m, \C^n)[-1]
      \\ \hline
      \Hom(\C^n, \C^m)[-1] & \glf(n)
    \end{array}
  \right) .
\end{multline}

The remaining Lie brackets equip the above with with the structure of
the Lie superalgebra $\glf(m|n)$. Thus, we have found exactly
four-dimensional Chern--Simons theory for the Lie algebra $\glf(m|n)$
as claimed.

We note that the BRST transformations induced by the above Lie brackets are slightly different from those identified in the main body of the paper. This is an artifact of working with a particular model for the underlying $L_\infty$ algebra. For comparison, we explicate the BRST transformations below.

Note that the cochain complex underlying our $L_\infty$ algebra arises naturally as the totalization of a $\Z\times \Z/2$-graded cochain complex, where the fields valued in $\Hom (\C^m , \C^n)\oplus \Hom (\C^n\oplus \C^m)$ are placed in bidegree $(\bullet, 1)$. Though the lie brackets arising from the coupling of the hypermultiplets to the vector multiplets broke the $\Z$-grading down to a $\Z/2$-grading, these brackets are easily seen to preserve the above grading $\Z\times \Z/2$ grading.

We fix the following notation for components of our fields
\begin{equation}
\alpha_{ij} \in \Omega^i (\R) \otimes \Omega^{0,j}\otimes (\glf(m) \oplus \glf(n)), \ \ \ \ \ \beta_{ij} \in \Omega^i (\R) \otimes \Omega^{0,j}\otimes (\Hom(\C^m, \C^n) \oplus \Hom(\C^n, \C^m))
\end{equation}
and denote the corresponding linear operators the same way. The BRST variations determined by the then take the form

\begin{equation}
Q\alpha_{ij} = d\alpha_{(i-1)j} +\delb\alpha_{i(j-1)} + \sum_{a+c = i, b+ d = j} [\alpha_{ab},\alpha_{cd}] + \sum_{a+c = i, b+ d = j} [\beta_{ab},\beta_{cd}] 
\end{equation}

\begin{equation}
Q\beta_{ij} = d\beta_{(i-1)j} +\delb\beta_{i(j-1)} + \sum_{a+c = i, b+ d = j} [\alpha_{ab},\beta_{cd}]
\end{equation}

The brackets in these equations are the relevant brackets on the $L_\infty$ algebra we've identified. It would be interesting to construct an explicit $L_\infty$ equivalence between the BV complex we have identified and the $L_\infty$ algebra consisting of the fields $\CA, c, b, B$ and BRST transformations from section \ref{brsttrans}.

\begin{rmk}
  As in remark \ref{6dtoplstrings}, we can consider the effect of the
  quadratic superpotential as a deformation of the entire
  category. The result is a six-dimensional topological string theory
  on $\R^{4}\times \C$. The two stacks of D5-branes we have considered
  will localize to a stack of D4- and anti-D4-branes wrapping
  $\R^{2}\times \C$. This set up is very reminiscent of the
  topological strings construction of three-dimensional Chern--Simons
  theory with gauge supergroup of \cite{Vafa:2001qf} and should lend
  itself to a holographic realization of the Yangian of $\glf(m|n)$
  generalizing the analysis of \cite{Ishtiaque:2018str}
\end{rmk}

%\providecommand{\href}[2]{#2}\begingroup\raggedright
%\endgroup

\begin{thebibliography}{99}

\bibitem{Nekrasov:2009ui}
N.~A. Nekrasov and S.~L. Shatashvili, \emph{Quantum integrability and
  supersymmetric vacua},
  \href{https://doi.org/10.1143/PTPS.177.105}{\emph{Prog. Theor. Phys. Suppl.}
  {\bfseries 177} (2009) 105}
  [\href{https://arxiv.org/abs/0901.4748}{{\ttfamily 0901.4748}}].

\bibitem{Nekrasov:2009uh}
N.~A. Nekrasov and S.~L. Shatashvili, \emph{Supersymmetric vacua and {B}ethe
  ansatz}, \href{https://doi.org/10.1016/j.nuclphysbps.2009.07.047}{\emph{Nucl.
  Phys. B Proc. Suppl.} {\bfseries 192/193} (2009) 91}
  [\href{https://arxiv.org/abs/0901.4744}{{\ttfamily 0901.4744}}].

\bibitem{Nekrasov:2018gne}
N.~Nekrasov, \emph{Superspin chains and supersymmetric gauge theories},
  \href{https://doi.org/10.1007/jhep03(2019)102}{\emph{JHEP} {\bfseries 03}
  (2019) 102} [\href{https://arxiv.org/abs/1811.04278}{{\ttfamily
  1811.04278}}].

\bibitem{Nekrasov:2009rc}
N.~A. Nekrasov and S.~L. Shatashvili, \emph{Quantization of integrable systems
  and four dimensional gauge theories},  in \emph{X{VI}th {I}nternational
  {C}ongress on {M}athematical {P}hysics}, p.~265, World Sci. Publ.,
  Hackensack, NJ, (2010), \href{https://arxiv.org/abs/0908.4052}{{\ttfamily
  0908.4052}}, \href{https://doi.org/10.1142/9789814304634_0015}{DOI}.

\bibitem{Dorey:2011pa}
N.~Dorey, S.~Lee and T.~J. Hollowood, \emph{Quantization of integrable systems
  and a 2d/4d duality},
  \href{https://doi.org/10.1007/JHEP10(2011)077}{\emph{JHEP} {\bfseries 10}
  (2011) 077} [\href{https://arxiv.org/abs/1103.5726}{{\ttfamily 1103.5726}}].

\bibitem{Chen:2011sj}
H.-Y. Chen, N.~Dorey, T.~J. Hollowood and S.~Lee, \emph{A new 2d/4d duality via
  integrability}, \href{https://doi.org/10.1007/JHEP09(2011)040}{\emph{JHEP}
  {\bfseries 09} (2011) 040} [\href{https://arxiv.org/abs/1104.3021}{{\ttfamily
  1104.3021}}].

\bibitem{Gaiotto:2012xa}
D.~Gaiotto, L.~Rastelli and S.~S. Razamat, \emph{Bootstrapping the
  superconformal index with surface defects},
  \href{https://doi.org/10.1007/JHEP01(2013)022}{\emph{JHEP} {\bfseries 01}
  (2013) 022} [\href{https://arxiv.org/abs/1207.3577}{{\ttfamily 1207.3577}}].

\bibitem{Gadde:2013ftv}
A.~Gadde and S.~Gukov, \emph{2d index and surface operators},
  \href{https://doi.org/10.1007/JHEP03(2014)080}{\emph{JHEP} {\bfseries 03}
  (2014) 080} [\href{https://arxiv.org/abs/1305.0266}{{\ttfamily 1305.0266}}].

\bibitem{Gaiotto:2015usa}
D.~Gaiotto and S.~S. Razamat, \emph{{$\mathcal{N}=1$} theories of class
  {$\mathcal{S}_k$}},
  \href{https://doi.org/10.1007/JHEP07(2015)073}{\emph{JHEP} {\bfseries 07}
  (2015) 073} [\href{https://arxiv.org/abs/1503.05159}{{\ttfamily
  1503.05159}}].

\bibitem{Maruyoshi:2016caf}
K.~Maruyoshi and J.~Yagi, \emph{Surface defects as transfer matrices},
  \href{https://doi.org/10.1093/ptep/ptw151}{\emph{Prog. Theor. Exp. Phys.}
  (2016) 113B01, 52} [\href{https://arxiv.org/abs/1606.01041}{{\ttfamily
  1606.01041}}].

\bibitem{Yagi:2017hmj}
J.~Yagi, \emph{Surface defects and elliptic quantum groups},
  \href{https://doi.org/10.1007/JHEP06(2017)013}{\emph{JHEP} {\bfseries 06}
  (2017) 013} [\href{https://arxiv.org/abs/1701.05562}{{\ttfamily
  1701.05562}}].

\bibitem{Maruyoshi:2020cwy}
K.~Maruyoshi, T.~Ota and J.~Yagi, \emph{Wilson--'t {H}ooft lines as transfer
  matrices}, \href{https://doi.org/10.1007/jhep01(2021)072}{\emph{JHEP}
  {\bfseries 01} (2021) Paper No. 072, 30}
  [\href{https://arxiv.org/abs/2009.12391}{{\ttfamily 2009.12391}}].

\bibitem{Bullimore:2015lsa}
M.~Bullimore, T.~Dimofte and D.~Gaiotto, \emph{The {C}oulomb branch of 3d
  {$\mathcal{N}=4$} theories},
  \href{https://doi.org/10.1007/s00220-017-2903-0}{\emph{Comm. Math. Phys.}
  {\bfseries 354} (2017) 671}
  [\href{https://arxiv.org/abs/1503.04817}{{\ttfamily 1503.04817}}].

\bibitem{Braverman:2016pwk}
A.~Braverman, M.~Finkelberg and H.~Nakajima, \emph{{Coulomb branches of $3d$
  $\mathcal N=4$ quiver gauge theories and slices in the affine Grassmannian
  (with appendices by Alexander Braverman, Michael Finkelberg, Joel Kamnitzer,
  Ryosuke Kodera, Hiraku Nakajima, Ben Webster, and Alex Weekes)}},
  \href{https://arxiv.org/abs/1604.03625}{{\ttfamily 1604.03625}}.

\bibitem{Dedushenko:2020yzd}
M.~Dedushenko and D.~Gaiotto, \emph{{Correlators on the wall and
  $\mathfrak{sl}_n$ spin chain}},
  \href{https://arxiv.org/abs/2009.11198}{{\ttfamily 2009.11198}}.

\bibitem{Costello:2018txb}
K.~Costello and J.~Yagi, \emph{Unification of integrability in supersymmetric
  gauge theories}, {\emph{Adv. Theor. Math. Phys.} {\bfseries 24} (2020) 1931}
  [\href{https://arxiv.org/abs/1810.01970}{{\ttfamily 1810.01970}}].

\bibitem{Costello:2013zra}
K.~Costello, \emph{Supersymmetric gauge theory and the {Y}angian},
  \href{https://arxiv.org/abs/1303.2632}{{\ttfamily 1303.2632}}.

\bibitem{Costello:2013sla}
K.~Costello, \emph{Integrable lattice models from four-dimensional field
  theories},  in \emph{String-{M}ath 2013}, vol.~88 of \emph{Proc. Sympos. Pure
  Math.}, pp.~3--23, Amer. Math. Soc., Providence, RI, (2014),
  \href{https://arxiv.org/abs/1308.0370}{{\ttfamily 1308.0370}},
  \href{https://doi.org/10.1090/pspum/088/01483}{DOI}.

\bibitem{Costello:2017dso}
K.~Costello, E.~Witten and M.~Yamazaki, \emph{Gauge theory and integrability,
  {I}}, \href{https://doi.org/10.4310/ICCM.2018.v6.n1.a6}{\emph{ICCM Not.}
  {\bfseries 6} (2018) 46} [\href{https://arxiv.org/abs/1709.09993}{{\ttfamily
  1709.09993}}].

\bibitem{Nekrasov:2002qd}
N.~A. Nekrasov, \emph{Seiberg--{W}itten prepotential from instanton counting},
  \href{https://doi.org/10.4310/ATMP.2003.v7.n5.a4}{\emph{Adv. Theor. Math.
  Phys.} {\bfseries 7} (2003) 831}
  [\href{https://arxiv.org/abs/hep-th/0206161}{{\ttfamily hep-th/0206161}}].

\bibitem{Nekrasov:2003rj}
N.~A. Nekrasov and A.~Okounkov, \emph{Seiberg--{W}itten theory and random
  partitions},  in \emph{The unity of mathematics}, vol.~244 of \emph{Progr.
  Math.}, p.~525, Birkh{\"a}user Boston, Boston, MA, (2006),
  \href{https://arxiv.org/abs/hep-th/0306238}{{\ttfamily hep-th/0306238}},
  \href{https://doi.org/10.1007/0-8176-4467-9_15}{DOI}.

\bibitem{Nekrasov:2010ka}
N.~Nekrasov and E.~Witten, \emph{The {O}mega deformation, branes, integrability
  and {L}iouville theory},
  \href{https://doi.org/10.1007/JHEP09(2010)092}{\emph{JHEP} {\bfseries 09}
  (2010) 092} [\href{https://arxiv.org/abs/1002.0888}{{\ttfamily 1002.0888}}].

\bibitem{Yagi:2014toa}
J.~Yagi, \emph{{$\Omega$}-deformation and quantization},
  \href{https://doi.org/10.1007/JHEP08(2014)112}{\emph{JHEP} {\bfseries 08}
  (2014) 112} [\href{https://arxiv.org/abs/1405.6714}{{\ttfamily 1405.6714}}].

\bibitem{Luo:2014sva}
Y.~Luo, M.-C. Tan, J.~Yagi and Q.~Zhao, \emph{{$\Omega$}-deformation of
  {B}-twisted gauge theories and the 3d-3d correspondence},
  \href{https://doi.org/10.1007/JHEP02(2015)047}{\emph{JHEP} {\bfseries 02}
  (2015) 047} [\href{https://arxiv.org/abs/1410.1538}{{\ttfamily 1410.1538}}].

\bibitem{Orlando:2010uu}
D.~Orlando and S.~Reffert, \emph{Relating gauge theories via gauge/{B}ethe
  correspondence}, \href{https://doi.org/10.1007/JHEP10(2010)071}{\emph{JHEP}
  {\bfseries 10} (2010) 071, 30}
  [\href{https://arxiv.org/abs/1005.4445}{{\ttfamily 1005.4445}}].

\bibitem{MR2827177}
B.~Feigin, M.~Finkelberg, A.~Negut and L.~Rybnikov, \emph{Yangians and
  cohomology rings of {L}aumon spaces},
  \href{https://doi.org/10.1007/s00029-011-0059-x}{\emph{Selecta Math. (N.S.)}
  {\bfseries 17} (2011) 573} [\href{https://arxiv.org/abs/0812.4656}{{\ttfamily
  0812.4656}}].

\bibitem{Maulik:2012wi}
D.~Maulik and A.~Okounkov, \emph{Quantum groups and quantum cohomology},
  \href{https://doi.org/10.24033/ast}{\emph{Ast\'{e}risque} (2019) ix+209}
  [\href{https://arxiv.org/abs/1211.1287}{{\ttfamily 1211.1287}}].

\bibitem{Rimanyi:2021hzq}
R.~Rimanyi and L.~Rozansky, \emph{{New quiver-like varieties and Lie
  superalgebras}},  \href{https://arxiv.org/abs/2105.11499}{{\ttfamily
  2105.11499}}.

\bibitem{Ragoucy:2007kg}
E.~Ragoucy and G.~Satta, \emph{Analytical {B}ethe ansatz for closed and open
  {$\mathit{gl}(\mathcal{M}|\mathcal{N})$} super-spin chains in arbitrary
  representations and for any {D}ynkin diagram},
  \href{https://doi.org/10.1088/1126-6708/2007/09/001}{\emph{JHEP} {\bfseries
  09} (2007) 001} [\href{https://arxiv.org/abs/0706.3327}{{\ttfamily
  0706.3327}}].

\bibitem{Belliard:2008di}
S.~Belliard and E.~Ragoucy, \emph{The nested {B}ethe ansatz for `all' closed
  spin chains}, \href{https://doi.org/10.1088/1751-8113/41/29/295202}{\emph{J.
  Phys. A} {\bfseries 41} (2008) 295202, 33}
  [\href{https://arxiv.org/abs/0804.2822}{{\ttfamily 0804.2822}}].

\bibitem{Hori:2003ic}
K.~Hori, S.~Katz, A.~Klemm, R.~Pandharipande, R.~Thomas, C.~Vafa et~al.,
  \emph{Mirror symmetry}, vol.~1 of \emph{Clay Mathematics Monographs}.
  American Mathematical Society, Providence, RI; Clay Mathematics Institute,
  Cambridge, MA, 2003.

\bibitem{Uranga:1998vf}
A.~M. Uranga, \emph{Brane configurations for branes at conifolds},
  \href{https://doi.org/10.1088/1126-6708/1999/01/022}{\emph{JHEP} {\bfseries
  01} (1999) 022} [\href{https://arxiv.org/abs/hep-th/9811004}{{\ttfamily
  hep-th/9811004}}].

\bibitem{Hanany:1997vm}
A.~Hanany and K.~Hori, \emph{Branes and {$N=2$} theories in two dimensions},
  \href{https://doi.org/10.1016/S0550-3213(97)00754-2}{\emph{Nucl. Phys. B}
  {\bfseries 513} (1998) 119}
  [\href{https://arxiv.org/abs/hep-th/9707192}{{\ttfamily hep-th/9707192}}].

\bibitem{Hori:2011pd}
K.~Hori, \emph{Duality in two-dimensional {$(2,2)$} supersymmetric
  non-{A}belian gauge theories},
  \href{https://doi.org/10.1007/JHEP10(2013)121}{\emph{JHEP} {\bfseries 10}
  (2013) 121} [\href{https://arxiv.org/abs/1104.2853}{{\ttfamily 1104.2853}}].

\bibitem{Benini:2014mia}
F.~Benini, D.~S. Park and P.~Zhao, \emph{Cluster algebras from dualities of 2d
  {$\mathcal{N}=(2,2)$} quiver gauge theories},
  \href{https://doi.org/10.1007/s00220-015-2452-3}{\emph{Comm. Math. Phys.}
  {\bfseries 340} (2015) 47} [\href{https://arxiv.org/abs/1406.2699}{{\ttfamily
  1406.2699}}].

\bibitem{Hellerman:2011mv}
S.~Hellerman, D.~Orlando and S.~Reffert, \emph{String theory of the {O}mega
  deformation}, \href{https://doi.org/10.1007/JHEP01(2012)148}{\emph{JHEP}
  {\bfseries 01} (2012) 148} [\href{https://arxiv.org/abs/1106.0279}{{\ttfamily
  1106.0279}}].

\bibitem{Reffert:2011dp}
S.~Reffert, \emph{General {O}mega deformations from closed string backgrounds},
  \href{https://doi.org/10.1007/JHEP04(2012)059}{\emph{JHEP} {\bfseries 04}
  (2012) 059} [\href{https://arxiv.org/abs/1108.0644}{{\ttfamily 1108.0644}}].

\bibitem{Hellerman:2012zf}
S.~Hellerman, D.~Orlando and S.~Reffert, \emph{The omega deformation from
  string and {M}-theory},
  \href{https://doi.org/10.1007/JHEP07(2012)061}{\emph{JHEP} {\bfseries 07}
  (2012) 061} [\href{https://arxiv.org/abs/1204.4192}{{\ttfamily 1204.4192}}].

\bibitem{Ashwinkumar:2018tmm}
M.~Ashwinkumar, M.-C. Tan and Q.~Zhao, \emph{Branes and categorifying
  integrable lattice models},
  \href{https://doi.org/10.4310/atmp.2020.v24.n1.a1}{\emph{Adv. Theor. Math.
  Phys.} {\bfseries 24} (2020) 1}
  [\href{https://arxiv.org/abs/1806.02821}{{\ttfamily 1806.02821}}].

\bibitem{Ishtiaque:2020ufn}
N.~Ishtiaque and J.~Yagi, \emph{Disk, interval, point: on constructions of
  quantum field theories with holomorphic action functionals},
  \href{https://doi.org/10.1007/jhep06(2020)180}{\emph{JHEP} {\bfseries 06}
  (2020) 180} [\href{https://arxiv.org/abs/2002.10488}{{\ttfamily
  2002.10488}}].

\bibitem{Mikhaylov:2014aoa}
V.~Mikhaylov and E.~Witten, \emph{Branes and supergroups},
  \href{https://doi.org/10.1007/s00220-015-2449-y}{\emph{Comm. Math. Phys.}
  {\bfseries 340} (2015) 699}
  [\href{https://arxiv.org/abs/1410.1175}{{\ttfamily 1410.1175}}].

\bibitem{Yamazaki:2019prm}
M.~Yamazaki, \emph{New {T}-duality for {C}hern--{S}imons theory},
  \href{https://doi.org/10.1007/jhep12(2019)090}{\emph{JHEP} {\bfseries 12}
  (2019) 090} [\href{https://arxiv.org/abs/1904.04976}{{\ttfamily
  1904.04976}}].

\bibitem{Witten:2011zz}
E.~Witten, \emph{Fivebranes and knots},
  \href{https://doi.org/10.4171/QT/26}{\emph{Quantum Topol.} {\bfseries 3}
  (2012) 1} [\href{https://arxiv.org/abs/1101.3216}{{\ttfamily 1101.3216}}].

\bibitem{Kapustin:2006hi}
A.~Kapustin, \emph{{Holomorphic reduction of $\mathcal{N} = 2$ gauge theories,
  Wilson--'t Hooft operators, and S-duality}},
  \href{https://arxiv.org/abs/hep-th/0612119}{{\ttfamily hep-th/0612119}}.

\bibitem{Kapustin:2009cd}
A.~Kapustin and N.~Saulina, \emph{Chern-{S}imons-{R}ozansky-{W}itten
  topological field theory},
  \href{https://doi.org/10.1016/j.nuclphysb.2009.07.006}{\emph{Nucl. Phys. B}
  {\bfseries 823} (2009) 403}
  [\href{https://arxiv.org/abs/0904.1447}{{\ttfamily 0904.1447}}].

\bibitem{MR4050665}
Y.~Su and R.~B. Zhang, \emph{Mixed cohomology of {L}ie superalgebras},
  \href{https://doi.org/10.1016/j.jalgebra.2019.11.036}{\emph{J. Algebra}
  {\bfseries 549} (2020) 1} [\href{https://arxiv.org/abs/1902.10627}{{\ttfamily
  1902.10627}}].

\bibitem{Ikeda:1993fh}
N.~Ikeda, \emph{Two-dimensional gravity and nonlinear gauge theory},
  \href{https://doi.org/10.1006/aphy.1994.1104}{\emph{Ann. Physics} {\bfseries
  235} (1994) 435} [\href{https://arxiv.org/abs/hep-th/9312059}{{\ttfamily
  hep-th/9312059}}].

\bibitem{Schaller:1994es}
P.~Schaller and T.~Strobl, \emph{Poisson structure induced (topological) field
  theories}, \href{https://doi.org/10.1142/S0217732394002951}{\emph{Modern
  Phys. Lett. A} {\bfseries 9} (1994) 3129}
  [\href{https://arxiv.org/abs/hep-th/9405110}{{\ttfamily hep-th/9405110}}].

\bibitem{Cattaneo:2003dp}
A.~S. Cattaneo and G.~Felder, \emph{Coisotropic submanifolds in {P}oisson
  geometry and branes in the {P}oisson sigma model},
  \href{https://doi.org/10.1007/s11005-004-0609-7}{\emph{Lett. Math. Phys.}
  {\bfseries 69} (2004) 157}
  [\href{https://arxiv.org/abs/math/0309180}{{\ttfamily math/0309180}}].

\bibitem{Kontsevich:1997vb}
M.~Kontsevich, \emph{Deformation quantization of {P}oisson manifolds},
  \href{https://doi.org/10.1023/B:MATH.0000027508.00421.bf}{\emph{Lett. Math.
  Phys.} {\bfseries 66} (2003) 157}
  [\href{https://arxiv.org/abs/q-alg/9709040}{{\ttfamily q-alg/9709040}}].

\bibitem{Cattaneo:1999fm}
A.~S. Cattaneo and G.~Felder, \emph{A path integral approach to the
  {K}ontsevich quantization formula},
  \href{https://doi.org/10.1007/s002200000229}{\emph{Comm. Math. Phys.}
  {\bfseries 212} (2000) 591}
  [\href{https://arxiv.org/abs/math/9902090}{{\ttfamily math/9902090}}].

\bibitem{MR3305442}
P.~Batakidis, \emph{Reduction algebra and differential operators on {L}ie
  groups}, \href{https://doi.org/10.1007/s13366-013-0157-3}{\emph{Beitr.
  Algebra Geom.} {\bfseries 56} (2015) 175}.

\bibitem{MR2504434}
A.~S. Cattaneo and C.~Torossian, \emph{Quantification pour les paires
  sym\'{e}triques et diagrammes de {K}ontsevich},
  \href{https://doi.org/10.24033/asens.2082}{\emph{Ann. Sci. \'{E}c. Norm.
  Sup\'{e}r. (4)} {\bfseries 41} (2008) 789}.

\bibitem{MR3118583}
A.~S. Cattaneo, C.~A. Rossi and C.~Torossian, \emph{Biquantization of symmetric
  pairs and the quantum shift},
  \href{https://doi.org/10.1016/j.geomphys.2013.07.001}{\emph{J. Geom. Phys.}
  {\bfseries 74} (2013) 211}.

\bibitem{doi:10.1142/9789812798268_0007}
F.~Woynarovich, \emph{Low-energy excited states in a hubbard chain with on-site
  attraction}, {\emph{J. Phys. C} {\bfseries 16} (1983) 6593}.

\bibitem{Essler:1992he}
F.~H.~L. Essler and V.~E. Korepin, \emph{{Higher conservation laws and
  algebraic Bethe ansatze for the supersymmetric $t$-$J$ model}},
  \href{https://doi.org/10.1103/PhysRevB.46.9147}{\emph{Phys. Rev. B}
  {\bfseries 46} (1992) 9147}.

\bibitem{Tsuboi:1998ne}
Z.~Tsuboi, \emph{Analytic {B}ethe ansatz and functional equations associated
  with any simple root systems of the {L}ie superalgebra
  {${\mathit{sl}}(r+1|s+1)$}},
  \href{https://doi.org/10.1016/S0378-4371(97)00625-0}{\emph{Phys. A}
  {\bfseries 252} (1998) 565}
  [\href{https://arxiv.org/abs/0911.5387}{{\ttfamily 0911.5387}}].

\bibitem{MR2074080}
F.~G\"{o}hmann and A.~Seel, \emph{A note on the {B}ethe ansatz solution of the
  supersymmetric {$t$}-{$J$} model},
  \href{https://doi.org/10.1023/B:CJOP.0000010530.54520.12}{\emph{Czechoslovak
  J. Phys.} {\bfseries 53} (2003) 1041}
  [\href{https://arxiv.org/abs/cond-mat/0309138}{{\ttfamily
  cond-mat/0309138}}].

\bibitem{MR2042980}
F.~G\"{o}hmann and A.~Seel, \emph{Algebraic {B}ethe ansatz for the {${\rm
  gl}(1|2)$} generalized model. {II}. {T}he three gradings},
  \href{https://doi.org/10.1088/0305-4470/37/8/001}{\emph{J. Phys. A}
  {\bfseries 37} (2004) 2843}
  [\href{https://arxiv.org/abs/cond-mat/0309135}{{\ttfamily
  cond-mat/0309135}}].

\bibitem{Volin:2010xz}
D.~Volin, \emph{String hypothesis for
  {$\mathfrak{gl}(\mathfrak{n}|\mathfrak{m})$} spin chains: a particle/hole
  democracy}, \href{https://doi.org/10.1007/s11005-012-0570-9}{\emph{Lett.
  Math. Phys.} {\bfseries 102} (2012) 1}
  [\href{https://arxiv.org/abs/1012.3454}{{\ttfamily 1012.3454}}].

\bibitem{Hanany:1996ie}
A.~Hanany and E.~Witten, \emph{Type {IIB} superstrings, {BPS} monopoles, and
  three-dimensional gauge dynamics},
  \href{https://doi.org/10.1016/S0550-3213(97)00157-0}{\emph{Nucl. Phys. B}
  {\bfseries 492} (1997) 152}
  [\href{https://arxiv.org/abs/hep-th/9611230}{{\ttfamily hep-th/9611230}}].

\bibitem{Benini:2012ui}
F.~Benini and S.~Cremonesi, \emph{Partition functions of {$\mathcal{N}=(2,2)$}
  gauge theories on {$S^2$} and vortices},
  \href{https://doi.org/10.1007/s00220-014-2112-z}{\emph{Comm. Math. Phys.}
  {\bfseries 334} (2015) 1483}
  [\href{https://arxiv.org/abs/1206.2356}{{\ttfamily 1206.2356}}].

\bibitem{MR0929437}
A.~Brini, A.~Palareti and A.~G.~B. Teolis, \emph{Gordan-{C}apelli series in
  superalgebras}, \href{https://doi.org/10.1073/pnas.85.5.1330}{\emph{Proc.
  Nat. Acad. Sci. U.S.A.} {\bfseries 85} (1988) 1330}.

\bibitem{MR1847665}
S.-J. Cheng and W.~Wang, \emph{Howe duality for {L}ie superalgebras},
  \href{https://doi.org/10.1023/A:1017594504827}{\emph{Compositio Math.}
  {\bfseries 128} (2001) 55}
  [\href{https://arxiv.org/abs/math/0008093}{{\ttfamily math/0008093}}].

\bibitem{Green:1996dd}
M.~B. Green, J.~A. Harvey and G.~Moore, \emph{{$I$}-brane inflow and anomalous
  couplings on {D}-branes},
  \href{https://doi.org/10.1088/0264-9381/14/1/008}{\emph{Classical Quantum
  Gravity} {\bfseries 14} (1997) 47}
  [\href{https://arxiv.org/abs/hep-th/9605033}{{\ttfamily hep-th/9605033}}].

\bibitem{Dijkgraaf:2007sw}
R.~Dijkgraaf, L.~Hollands, P.~Su\l~kowski and C.~Vafa, \emph{Supersymmetric
  gauge theories, intersecting branes and free fermions},
  \href{https://doi.org/10.1088/1126-6708/2008/02/106}{\emph{JHEP} {\bfseries
  02} (2008) 106} [\href{https://arxiv.org/abs/0709.4446}{{\ttfamily
  0709.4446}}].

\bibitem{Costello:2016mgj}
K.~Costello and S.~Li, \emph{{Twisted supergravity and its quantization}},
  \href{https://arxiv.org/abs/1606.00365}{{\ttfamily 1606.00365}}.

\bibitem{Costello:2018zrm}
K.~Costello and D.~Gaiotto, \emph{{Twisted holography}},
  \href{https://arxiv.org/abs/1812.09257}{{\ttfamily 1812.09257}}.

\bibitem{Costello:2020jbh}
K.~Costello and N.~M. Paquette, \emph{Twisted supergravity and {K}oszul
  duality: a case study in {$\rm AdS_3$}},
  \href{https://doi.org/10.1007/s00220-021-04065-3}{\emph{Comm. Math. Phys.}
  {\bfseries 384} (2021) 279}
  [\href{https://arxiv.org/abs/2001.02177}{{\ttfamily 2001.02177}}].

\bibitem{Saberi:2021weg}
I.~Saberi and B.~R. Williams, \emph{{Twisting pure spinor superfields, with
  applications to supergravity}},
  \href{https://arxiv.org/abs/2106.15639}{{\ttfamily 2106.15639}}.

\bibitem{MR2778558}
K.~Costello, \emph{Renormalization and effective field theory}, vol.~170 of
  \emph{Mathematical Surveys and Monographs}. American Mathematical Society,
  Providence, RI, 2011,
  \href{https://doi.org/10.1090/surv/170}{10.1090/surv/170}.

\bibitem{Costello:2016vjw}
K.~Costello and O.~Gwilliam, \emph{Factorization algebras in quantum field
  theory. {V}ol. 1}, vol.~31 of \emph{New Mathematical Monographs}. Cambridge
  University Press, Cambridge, 2017,
  \href{https://doi.org/10.1017/9781316678626}{10.1017/9781316678626}.

\bibitem{Raghavendran:2019zdq}
S.~Raghavendran and P.~Yoo, \emph{{Twisted S-duality}},
  \href{https://arxiv.org/abs/1910.13653}{{\ttfamily 1910.13653}}.

\bibitem{MR2298823}
K.~Costello, \emph{Topological conformal field theories and {C}alabi-{Y}au
  categories}, \href{https://doi.org/10.1016/j.aim.2006.06.004}{\emph{Adv.
  Math.} {\bfseries 210} (2007) 165}
  [\href{https://arxiv.org/abs/math/0412149}{{\ttfamily math/0412149}}].

\bibitem{MR2555928}
J.~Lurie, \emph{On the classification of topological field theories},  in
  \emph{Current developments in mathematics, 2008}, pp.~129--280, Int. Press,
  Somerville, MA, (2009), \href{https://arxiv.org/abs/0905.0465}{{\ttfamily
  0905.0465}}.

\bibitem{Witten:1992fb}
E.~Witten, \emph{Chern--{S}imons gauge theory as a string theory},  in
  \emph{The {F}loer memorial volume}, vol.~133 of \emph{Progr. Math.},
  pp.~637--678, Birkh\"{a}user, Basel, (1995),
  \href{https://arxiv.org/abs/hep-th/9207094}{{\ttfamily hep-th/9207094}}.

\bibitem{Elliott:2020ecf}
C.~Elliott, P.~Safronov and B.~R. Williams, \emph{{A Taxonomy of Twists of
  Supersymmetric Yang--Mills Theory}},
  \href{https://arxiv.org/abs/2002.10517}{{\ttfamily 2002.10517}}.

\bibitem{MR2836399}
T.~Willwacher, \emph{Formality of cyclic chains},
  \href{https://doi.org/10.1093/imrn/rnq196}{\emph{Int. Math. Res. Not. IMRN}
  (2011) 3939}.

\bibitem{Bershadsky:1993cx}
M.~Bershadsky, S.~Cecotti, H.~Ooguri and C.~Vafa, \emph{Kodaira-{S}pencer
  theory of gravity and exact results for quantum string amplitudes},
  {\emph{Comm. Math. Phys.} {\bfseries 165} (1994) 311}
  [\href{https://arxiv.org/abs/hep-th/9309140}{{\ttfamily hep-th/9309140}}].

\bibitem{Costello:2012cy}
K.~J. Costello and S.~Li, \emph{{Quantum BCOV theory on Calabi-Yau manifolds
  and the higher genus B-model}},
  \href{https://arxiv.org/abs/1201.4501}{{\ttfamily 1201.4501}}.

\bibitem{Costello:2015xsa}
K.~Costello and S.~Li, \emph{{Quantization of open-closed BCOV theory, I}},
  \href{https://arxiv.org/abs/1505.06703}{{\ttfamily 1505.06703}}.

\bibitem{Bershadsky:1994sr}
M.~Bershadsky and V.~Sadov, \emph{Theory of {K}\"{a}hler gravity},
  \href{https://doi.org/10.1142/S0217751X96002157}{\emph{Internat. J. Modern
  Phys. A} {\bfseries 11} (1996) 4689}
  [\href{https://arxiv.org/abs/hep-th/9410011}{{\ttfamily hep-th/9410011}}].

\bibitem{Oh:2019bgz}
J.~Oh and J.~Yagi, \emph{Chiral algebras from {$\Omega$}-deformation},
  \href{https://doi.org/10.1007/JHEP08(2019)143}{\emph{JHEP} {\bfseries 08}
  (2019) 143} [\href{https://arxiv.org/abs/1903.11123}{{\ttfamily
  1903.11123}}].

\bibitem{Saberi:2019fkq}
I.~Saberi and B.~R. Williams, \emph{{Superconformal algebras and holomorphic
  field theories}},  \href{https://arxiv.org/abs/1910.04120}{{\ttfamily
  1910.04120}}.

\bibitem{Ishtiaque:2018str}
N.~Ishtiaque, S.~F. Moosavian and Y.~Zhou, \emph{Topological holography: the
  example of the {D}2-{D}4 brane system},
  \href{https://doi.org/10.21468/scipostphys.9.2.017}{\emph{SciPost Phys.}
  {\bfseries 9} (2020) 017} [\href{https://arxiv.org/abs/1809.00372}{{\ttfamily
  1809.00372}}].

\bibitem{Vafa:2001qf}
C.~Vafa, \emph{{Brane/anti-brane systems and $U(N|M)$ supergroup}},
  \href{https://arxiv.org/abs/hep-th/0101218}{{\ttfamily hep-th/0101218}}.

\end{thebibliography}
\end{document}